\documentclass[amssymb,prb,twocolumn,superscriptaddress,floats,showkeys,amsmath]{revtex4-2}

\usepackage{graphicx}
\usepackage{epstopdf}
\usepackage{hyperref}
\hypersetup{colorlinks,citecolor=blue, filecolor=blue ,linkcolor=blue , urlcolor=blue, pdftex}

\def \FUW{University of Warsaw, Faculty of Physics, 02-093 Warsaw, Poland}
\def \Watanabe{Research Center for Electronic and Optical Materials, National Institute for Materials Science, 1-1 Namiki, Tsukuba 305-0044, Japan}
\def \Taniguchi{Research Center for Materials Nanoarchitectonics, National Institute for Materials Science,  1-1 Namiki, Tsukuba 305-0044, Japan}
\def \Sapienza{Physics Department, Sapienza University of Rome, 00185 Rome, Italy}
\def \CNR{Institute for Photonics and Nanotechnologies, National Research Council (CNR-IFN), 00133 Rome, Italy}
\def \LNCMI{Laboratoire National des Champs Magnétiques Intenses, CNRS-UGA-UPS-INSA-EMFL, 38042 Grenoble, France}

\def \Centera{CENTERA, CEZAMAT, Warsaw University of Technology, Warsaw, Poland}

\def \Prague{Department of Inorganic Chemistry, University of Chemistry and Technology Prague, Technická 5, 166 28 Prague 6, Czech Republic}
\def \Singapur{Institute for Functional Intelligent Materials, National University of Singapore, Singapore 117544, Singapore}
\def \PhyUCF{Department of Physics, University of Central Florida, Orlando, Florida 32816, USA}
\def \EngUCF{Department of Electrical and Computer Engineering, University of Central Florida, Orlando, Florida 32816, USA}

\begin{document}

\title{Extremely high excitonic $g$-factors in 2D crystals \\ by alloy-induced admixing of band states}

\author{Katarzyna Olkowska-Pucko}
\email{katarzyna.olkowska-pucko@fuw.edu.pl}
\affiliation{\FUW}
\author{Tomasz Woźniak}
\email{tomasz.wozniak@fuw.edu.pl}
\affiliation{\FUW}
\author{Elena Blundo}
\email{elena.blundo@uniroma1.it}
\affiliation{\Sapienza}
\author{Natalia Zawadzka}
\affiliation{\FUW}
\author{Łucja~Kipczak}
\affiliation{\FUW}
\author{Paulo~E.~{Faria~Junior}}
\affiliation{\PhyUCF}
\affiliation{\EngUCF}
\author{Jan Szpakowski}
\affiliation{\FUW}
\author{Grzegorz~Krasucki}
\affiliation{\FUW}
\author{Salvatore~Cianci}
\affiliation{\Sapienza}
\author{Diana~Vaclavkova}
\affiliation{\LNCMI}
\author{Dipankar~Jana}
\affiliation{\LNCMI}
\author{Piotr~Kapu\'sci\'nski}
\affiliation{\LNCMI}
\author{Amit~Pawbake}
\affiliation{\LNCMI}
\author{Shalini Badola}
\affiliation{\LNCMI}
\author{Magdalena Grzeszczyk}
\affiliation{\FUW}
\affiliation{\Singapur}
\author{Daniele Cecchetti}
\affiliation{\CNR}
\author{Giorgio Pettinari}
\affiliation{\CNR}
\author{Igor Antoniazzi}
\affiliation{\FUW}
\author{Zden\v{e}k Sofer}
\affiliation{\Prague}
\author{Iva Plutnarová}
\affiliation{\Prague}
\author{Kenji~Watanabe}
\affiliation{\Watanabe}
\author{Takashi Taniguchi}
\affiliation{\Taniguchi}
\author{Clement~Faugeras}
\affiliation{\LNCMI}
\author{Marek Potemski}
\affiliation{\FUW}
\affiliation{\LNCMI}
\affiliation{\Centera}
\author{Adam Babi\'nski}
\affiliation{\FUW}
\author{Antonio Polimeni}
\affiliation{\Sapienza}
\author{Maciej R. Molas}
\email{maciej.molas@fuw.edu.pl}
\affiliation{\FUW}

\begin{abstract}
Monolayers (MLs) of semiconducting transition metal dichalcogenides (\mbox{S-TMDs}) emit light very efficiently and display rich spin-valley physics, with gyromagnetic ($g$-) factors of about -4. 
Here, we investigate how these properties can be tailored by alloying. 
Magneto-optical spectroscopy is used to reveal the peculiar properties of excitonic complexes in Mo$_{x}$W$_{1-x}$Se$_2$ MLs with different Mo and W concentrations. 
We show that the alloys feature extremely high $g$-factors for neutral excitons, that change gradually with the composition up to reaching values of the order of -10 for $x \approx 0.2$. 
First-principles calculations corroborate the experimental findings and provide evidence that alloying in S-TMDs results in a non-trivial band structure engineering, being at the origin of the high $g$-factors. 
The theoretical framework also suggests a higher strain sensitivity of the alloys, making them promising candidates for tailor-made optoelectronic devices.

\end{abstract}

\maketitle


\textit{Introduction---}{Semiconducting transition metal dichalcogenides (\mbox{S-TMDs}) monolayers (MLs), such as MoS$_2$, MoSe$_2$, WS$_2$, WSe$_2$, and MoTe$_2$, exhibit several intriguing optical properties}~\cite{Mak2010, Cao2012, Wang2014, Arora2015W, aroramose2, Tonndorf2013, Lezama2015, Koperski2017, molasNanoscale, Molas2017, Liu2020, Grzeszczyk2021}. 
In particular, the optical response of S-TMDs is dominated by excitonic effects~\cite{Koperski2017, Cadiz2017, Malic2018, Taghizadeh2019, Gerber2019}, and, moreover, 
the MLs can be divided into two subgroups: bright MLs with a bright optically-active exciton ground state (MoSe$_2$ and MoTe$_2$), and darkish MLs whose exciton ground state is spin-forbidden and thus optically-dark (MoS$_2$, WS$_2$, and WSe$_2$)~\cite{Robert2016, Molas2017, Koperski2017, Zhang2017, Robert2017, Molas2019, LiuGate2019, Lu_2020, Arora2020, Liu2020, He2020, Robert2020, Zinkiewicz2020, Zinkiewicz2021, Kapuscinski2021, Kipczak2024}.

In S-TMD MLs, the lowest energy optically-active excitons (referred to as A excitons) have energies that span the spectral range from~1.15 eV for MoTe$_2$ ~\cite{Lezama2015} up to ~2.1 eV for WS$_2$~\cite{Zinkiewicz2021}.
Continuous energy tuning within this range can be achieved by mixing two transition metal atoms ($i.e.$, Mo and W) or two chalcogen atoms ($e.g.$, S and Se), which induces the creation of S-TMD alloys with adjustable stoichiometry ratio ($e.g.$, Mo$_x$W$_{1-x}$Se$_2$ and WS$_{2x}$Se$_{2-2x}$)~\cite{Zhao2018}.
Such compositional tuning opens new possibilities for absorbing and emitting light at desired wavelengths, which is crucial for optoelectronics applications~\cite{Xie2015, Yucheng2018, Yumei2020, BIKEROUIN2022, Sumaira2022}. 
The opportunities enabled by alloying can be further exploited in fields such as valleytronics~\cite{Wu2021}, which employs local energy extrema, $i.e.$, the so-called valleys, to encode and process binary information, with applications for quantum computers~\cite{Schaibley2016, Mak2018, Pal2023,Vitale2018,alice_and_bob}.

Bright A excitons in S-TMD ML are formed by an electron-hole pair originating from the inequivalent $\pm$K valleys of the hexagonal Brillouin zone, and are associated selectively to $\sigma^\pm$ circularly polarized light~\cite{Kormányos_2015, Koperski2017, Wang2018}.
The energy degeneracy of excitons at the $\pm$K valleys can be lifted by applying a magnetic field perpendicular to the ML plane giving rise to the excitonic Zeeman effect described by the excitonic Landé $g$-factor ($g$)~\cite{Koperski2019, Wozniak2020}. 
	In an external field ($B$), the $g$-factor value is related to the energy splitting between excitons from the $\pm$K points with $\sigma^\pm$ polarizations, $i.e.$, $\Delta E(B) = E_{\sigma^+}(B) - E_{\sigma^-}(B)=g\,\mu_B B,$ ($\mu_B$ is the Bohr magneton).
	Theoretically, the magnetic field couples to both the spin and orbital parts of the Bloch states, and thus the excitonic $g$-factor reflects the energy difference between the conduction (CB) and valence (VB) bands in both the $\pm$K valleys~\cite{Ivchenko2005, Koperski2019, Wozniak2020}.

For S-TMD MLs, the $g$-factors are approximately equal to $-4$~\cite{Robert2017, Cadiz2017, Stier2018, Koperski2019, Goryca2019, Arora2020,  LiuGate2019, Robert2020, Förste2020, Jadczak_2021, Zinkiewicz2021} for the A excitons.
Moreover, while a significant variation in the ML dielectric environment affects such physical quantities as the exciton binding energies and therefore the diamagnetic shift of excitons~\cite{Komsa2012, Lin2014, Latini2015, Andersen2015, Stier2016}, the $g$-factor value stays insensitive~\cite{Stier2016, Faria2023}.
For the WSSe ML with almost 50/50 S/Se ratio, the excitonic $g$-factor is close to $-4$~\cite{Pucko_2023}, similarly to the WS$_2$ and WSe$_2$ MLs.
On the other hand, investigating the $g$-factors of MLs of Mo-W alloys could be of particular interest, since these alloys combine bright and darkish MLs. 

\begin{figure}[t]
	\centering
	\includegraphics[width=1\linewidth]{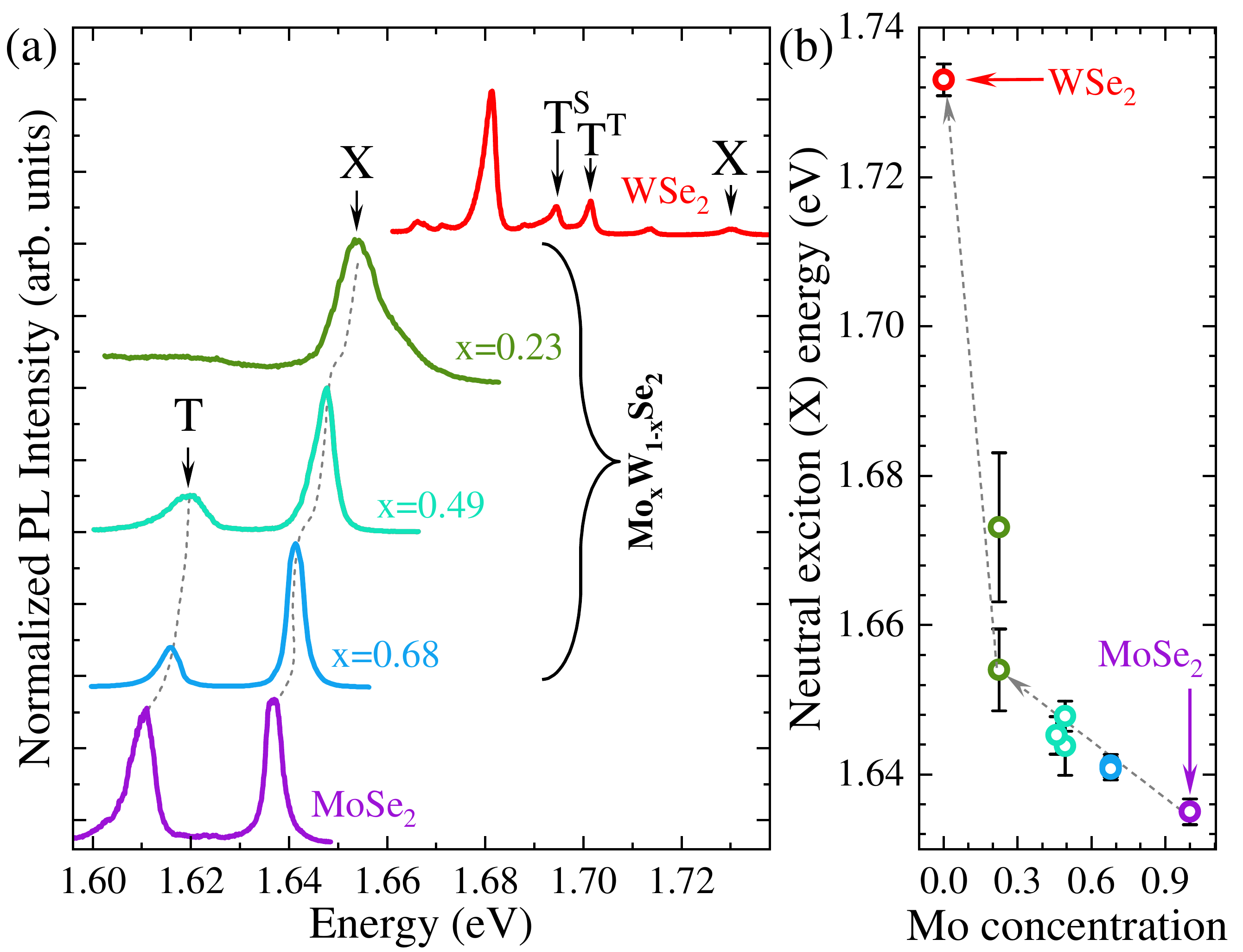}
	\caption{(a) Low-temperature ($T$=10~K) example PL spectra of MoSe$_2$, WSe$_2$, and Mo$_x$W$_{1-x}$Se$_2$ MLs encapsulated in hBN flakes. (b) Extracted energy of neutral exciton (X) as a function of molybdenum (Mo) concentration. The dashed lines are guided to the eye.} 
	\label{fig:parents}
\end{figure}


In this Letter, we investigate the magneto-optical properties of excitonic complexes in Mo$_{x}$W$_{1-x}$Se$_2$ MLs encapsulated in hexagonal BN (hBN) with different ratios of Mo and W atoms and compare the results with those achieved for WSe$_2$ and MoSe$_2$ MLs.
Using low-temperature ($T$=10~K) photoluminescence (PL) experiments carried out in external out-of-plane magnetic fields up to 30~T, we extract the $g$-factors of the neutral (X) and charged (T) excitons.
While the measured T $g$-factors for Mo$_{x}$W$_{1-x}$Se$_2$ MLs vary from around -3.5 to almost -5 and are thus comparable to those of MoSe$_2$ and WSe$_2$ MLs, the $g$-factors for the X transitions change gradually from about -4 for MoSe$_2$, to about -4.5  when the Mo concentration is $\sim 85\%$
	then -6 when the Mo concentration is $\sim 70 \%$, to about -7 when it reaches $\sim 50 \%$, and even up to about -10 for Mo concentrations of $\sim 20 \%$, to then go back to -4 in WSe$_2$.
This striking tunability of the $g$-factor is verified by first principles calculations of the band structures and angular momenta of MoSe$_2$ and WSe$_2$ MLs and their alloys.
The calculated values of the $g$-factors show a trend similar to the experimental ones and also reveal an additional increase and decrease under the application of compressive or tensile biaxial strains, respectively.
Our studies indicate that the alloying of S-TMD MLs is an efficient mechanism to enhance the $g$-factors of neutral excitons, up to values that have only been observed for interlayer excitons in S-TMDs heterostructures (HSs) with nearly 0$^\circ$ or 60$^\circ$ twist angles so far~\cite{Seyler2019, Liu2021, Jiang2021, Brotons2021, Blundo2024}. 
Due to the much simpler fabrication process of MLs compared to S-TMD HSs with specific twist angles, alloy MLs open new avenues as potential candidates for valleytronic and quantum devices~\cite{SJPrado_2004}.\\

\begin{figure}[htbp]
	\centering
	\includegraphics[width=1\linewidth]{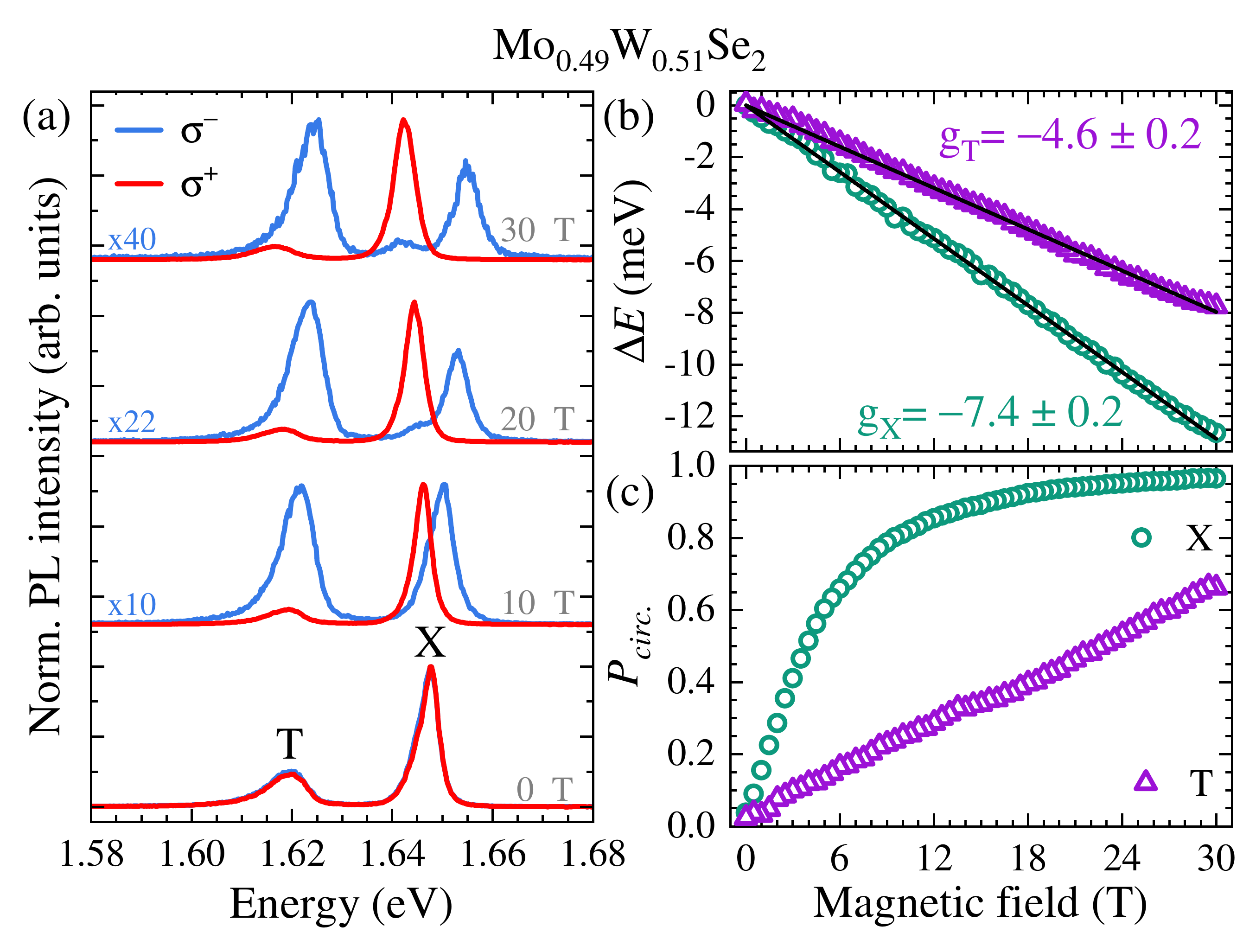}
	\caption{(a) Helicity-resolved PL spectra of an hBN-encapsulated Mo$_{0.49}$W$_{0.51}$Se$_2$ ML at $T$=4.2~K measured at selected values of the applied out-of-plane magnetic field. The red (blue) color corresponds to the $\sigma^+$ ($\sigma^-$) polarized spectra. The measurements were performed  under excitation energy of 2.41~eV and power of 2.5~$\mu$W. The $\sigma^+$-polarized spectra were normalized to the X intensity, while the $\sigma^-$-polarized spectra were multiplied by scaling factors to make them better visible. The spectra are vertically shifted for clarity.  (b) Magnetic-field evolutions of the energy differences ($\Delta E$) between the two circularly polarized split components of the X and T transitions. The solid lines represent fits according to the equation described in the text.}
	\label{fig:mag_PL}
\end{figure}


\textit{Investigated~samples---}A series of hBN-encapsulated Mo$_{x}$W$_{1-x}$Se$_2$ MLs with five different Mo/W ratios, plus the parent compounds MoSe$_2$ and WSe$_2$, was prepa
 by exfoliation and dry transfer techniques, see Methods section of the Supplemental Material (SM)~\cite{SM} for details.
The precise stoichiometry of each crystal was measured by scanning electron microscopy (SEM) with energy dispersive X-ray analysis (EDX). 
Each alloyed crystal was measured at multiple points, giving the following average Mo concentrations ($x$): 
	$(22.5 \pm 1.9) \% $, $(45.8 \pm 1.7) \% $, $(49.40 \pm 0.81) \% $, $(67.79 \pm 0.94) \% $, and $(85.41 \pm 0.63) \% $.
We further performed SEM-EDX mapping measurements to verify the crystal homogeneity, see Sec.\ S1 of the SM~\cite{SM}.
The crystal with 49 \% Mo concentration was grown by chemical vapor transport, while the other crystals were purchased from HQ graphene.
Our experiments show very consistent results between grown and purchased samples, with the grown sample being characterized by a higher homogeneity.\\

\textit{Optical properties of the alloys---}Low-temperature example PL spectra of MoSe$_2$, WSe$_2$ MLs and of the Mo$_{x}$W$_{1-x}$Se$_2$ alloys are shown in Fig.~\ref{fig:parents}(a).  
The optical properties of the MoSe$_2$ and WSe$_2$ MLs originates from the bright~\cite{Robert2017} and dark~\cite{Arora2015W, Molas2017, Molas2019, Zinkiewicz2020, Robert2020} characters of the ground excitonic states, respectively.
The PL spectrum of the MoSe$_2$ ML~\cite{Robert2017, Lu_2020, Robert2020, Molas2019Energy, Koperski2019, Oreszczuk2023} consists of only two emission lines, while the WSe$_2$ spectrum is rather complex, since it is composed of several lines~\cite{Courtade2017,Robert2017, Li2018, Chen2018, Barbone2018, Paur2019, Li2019, Lireplica2019, Li2019momentum, Molas2019, LiuValley, Liu2020, He2020, Arora2020, Robert2021, Robert2021PRL, Zinkiewicz2022}.
Following previous works, the X and T lines can be ascribed to the neutral and charged exciton~\cite{Koperski2017}, respectively, see Sec.\ S2 of SM for details~\cite{SM}.
The PL spectra of the Mo$_{0.85}$W$_{0.15}$Se$_2$, Mo$_{0.68}$W$_{0.32}$Se$_2$ and Mo$_{0.49}$W$_{0.51}$Se$_2$ MLs are similar to those of MoSe$_2$ ML, $i.e.$, two emission lines can be distinguished.
A small energy shift of the X and T lines is apparent in the transition from MoSe$_2$ ML to alloys with decreasing Mo concentration up to 46 \%, and, thus, by analogy, we attributed them correspondingly to the neutral and charged excitons, see Sec.\ S2 of SM for details~\cite{SM}.
The PL spectra measured on the Mo$_{0.85}$W$_{0.15}$Se$_2$, Mo$_{0.68}$W$_{0.32}$Se$_2$, Mo$_{0.49}$W$_{0.51}$Se$_2$, and Mo$_{0.46}$W$_{0.54}$Se$_2$ (the latter is not shown in Fig. \ref{fig:parents}) MLs indicate that the quality of the investigated MLs is very high similar to that of the MoSe$_2$ ML (the X linewidths are about 4--5~meV) see Sec.\ S3 of SM for details~\cite{SM}.
This suggests that while the overall arrangement of Mo and W atoms in these MLs is random, the Mo/W ratio stays at a fixed level at least in the spatial area comparable to the diameter of the excitation spot ($i.e.$, 1~$\mu$m).
This contrasts to the PL spectrum of the Mo$_{0.23}$W$_{0.77}$Se$_2$ ML, where the X linewidth reaches the value of about 11~meV. 
This points to a lower quality of the Mo$_{0.23}$W$_{0.77}$Se$_2$ ML, probably associated with a larger variation of the Mo/W ratio below the spatial area of excitation ($i.e.$, higher disorder of the atoms at the nanoscale), as reported for WSSe MLs~\cite{Pucko_2023}.
We performed a detailed analysis of the temperature- and power-dependent PL spectra measured on MoWSe$_2$ MLs, as discussed in Sec.\ S2 of SM that confirm the attributions of the X and T lines~\cite{SM}.

Indeed, the alloy composition can efficiently tune the exciton energy.
Fig.~\ref{fig:parents}(b) presents the neutral exciton energy as a function of the Mo composition in MoWSe$_2$ MLs extracted from all studied samples. 
As the Mo composition increases from the WSe$_2$ side, the neutral exciton (X) energies demonstrate two distinct trends. Initially, the X energies decrease significantly, with a pronounced drop observed for the Mo$_{0.23}$W$_{0.77}$Se$_2$ ML. 
This indicates that even small variations in Mo concentration (below 23\%) lead to substantial changes in exciton energies. Beyond this critical point, the exciton energies vary at a much lower rate.
A similar trend of the X energy has previously been reported for MoWSe$_2$ MLs~\cite{zhang2014, Wang2015, Catanzaro2024,Brito_2025}.

\begin{figure}[htbp]
	\centering
	\includegraphics[width=1\linewidth]{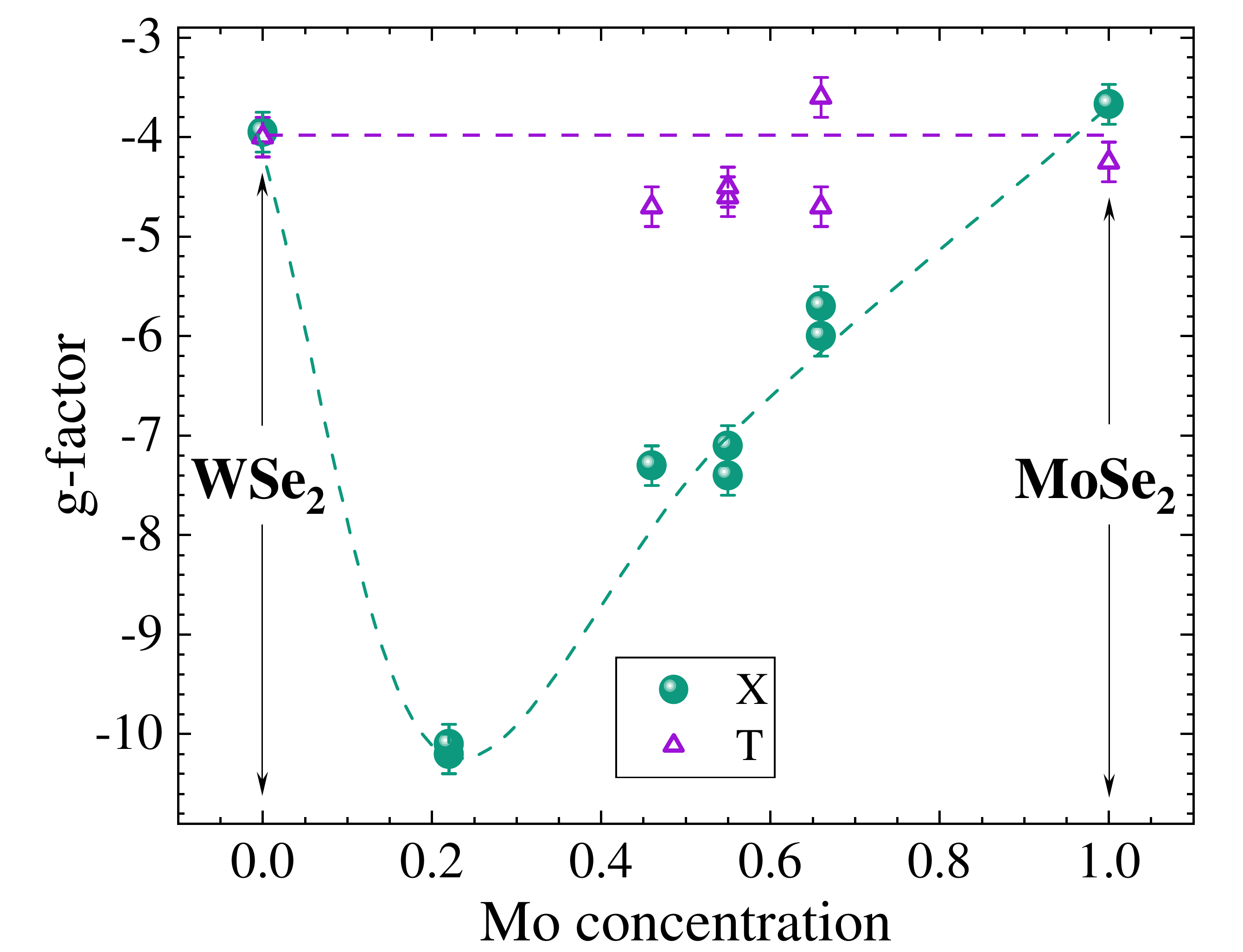}
	\caption{Experimental values of $g$-factors extracted for the neutral and charged excitons measured on the MoSe$_2$, WSe$_2$, and MoWSe$_2$ MLs with different Mo concentration. The dashed curves are guide to the eye.}
	\label{fig:gfactor_exp}
\end{figure}

Slight variations in the exciton energies between different works can be attributed to concentration inhomogeneities and possible strains induced during the sample preparation. Consistently, by probing different MLs exfoliated from the same crystal, we observed X energy variations of almost 4~meV \cite{Blundo2020,Cianci2024}. 
A larger variability, up to almost 20 meV, was only observed for Mo$_{0.23}$W$_{0.77}$Se$_2$ MLs, $i.e.$ for the crystal characterized by a larger inhomogeneity, as also deduced by the broader lineshape and by SEM-EDX measurements. Furthermore, for small Mo concentrations, the energy is highly dependent on the concentration, causing the observed energy variability.\\


\textit{Magneto-optical response---}~We now focus on the effects of out-of-plane magnetic fields on the alloyed MLs.
Fig.~\ref{fig:mag_PL}(a) shows the helicity-resolved PL spectra of a hBN-encapsulated Mo$_{0.49}$W$_{0.51}$Se$_2$ ML in magnetic fields up to 30 T.
Indeed, the application of the magnetic field results in a clearly observable Zeeman effect, which manifests itself as a splitting of the two counter-circularly-polarized components ($\sigma\pm$).

\begin{figure*}[t]
	\centering
	\includegraphics[width=1\linewidth]{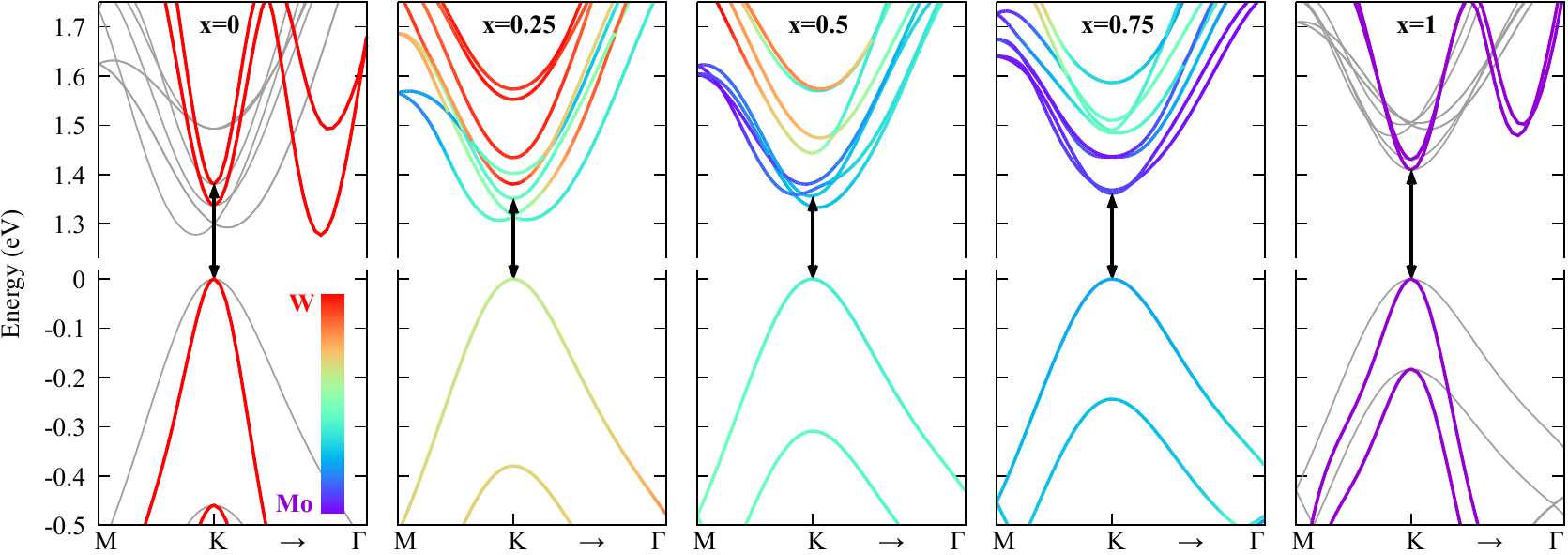}
	\caption{Electronic band structures calculated for the studied Mo$_x$W$_{1-x}$Se$_2$ MLs with $a_0(x)$. The color code reveals the wavefunction localization at the W and Mo atomic sites.
		Band structures using the 2$\times$2 supercells for pure MoSe$_2$ and WSe$_2$ crystals are shown with thin grey lines for comparison. 
		The lowest energy interband transitions with circular polarization are indicated with vertical arrows.
		The unfolded band structures are shown in Fig. S7.2 of the SM~\cite{SM}.
	}
	\label{fig:band}
\end{figure*}

The X and T bands are fitted by Lorentzian functions to extract the Zeeman splitting ($\Delta E$, Fig.~\ref{fig:mag_PL}(b)) and to derive information on the intensity to calculate the degree of circular polarization, shown in SM~\cite{SM} Sec. S3).
Indeed, the PL spectra reveal a large intensity variation for the $\sigma^+$ and $\sigma^-$ components when increasing the field. For both the X and T bands, the lowest energy transitions ($\sigma^+$) gain intensity, whereas the highest energy transitions ($\sigma^-$) lose intensity with increasing field.

From the Zeeman splitting data shown in Fig.~\ref{fig:mag_PL}(b), we can quantify the $g$-factor through linear fits: $g_\textrm{X}=-7.4\pm0.2$ and $g_\textrm{T}=-4.6\pm0.2$.
Although the $g$-factor for the trion is similar to those of the MoSe$_2$ and WSe$_2$ MLs (about $-4.0$ -- $-4.3$), the $g$-factor for the X line is almost two times larger than for the parent MLs (around $-3.7$ -- $-4.0$, see Sec.\ S3 of SM for further details~\cite{SM}).

In Fig.~\ref{fig:gfactor_exp}, we summarize the experimental values of the $g$-factors extracted for the neutral and charged excitons measured on the MoSe$_2$, WSe$_2$, and MoWSe$_2$ MLs with different Mo/W ratios.
The complete magneto-optical data, comprising magneto-PL spectra, Zeeman splitting analysis and degree of circular polarization, of the end compounds (MoSe$_2$ and WSe$_2$) and of the alloys (Mo$_{x}$W$_{1-x}$Se$_2$) are reported in SM~\cite{SM} Sec. S4 and S5, respectively.
Two main effects can be appreciated. First, the $g$-factors for the X lines change gradually from almost -4.5 for the Mo$_{0.85}$W$_{0.15}$Se$_2$ ML through -6 for the Mo$_{0.68}$W$_{0.32}$Se$_2$ ML, to about -7 for MLs with almost equal compositions of Mo and W atoms, up to reaching a value of the order of -10 for the ML with $\sim 23 \%$ of Mo atoms. In the latter sample, the possible influence of local strain and compositional variations are addressed in Sec. S6 in SM~\cite{SM} showing the robustness of the observed exceptionally high $g$-factor value. Second, we note that the corresponding T $g$-factors are in the range of around -3.5 to almost -5, similar to the values of around -4 found in standard S-TMD MLs~~\cite{Arora_MoTe2, Koperski2019, Zinkiewicz2021}; leading to a substantial difference in the $g$-factors values found for the X and T lines. 
This is indeed in contrast with the almost identical values reported for standard S-TMD MLs~\cite{Arora_MoTe2, Koperski2019, Zinkiewicz2021}.\\


\textit{First principles calculations---}To elucidate the origin of the exciton $g$-factor's striking behavior, we performed density functional theory (DFT) calculations using the modern methodology for $g$-factor evaluation, based on the summation-over-bands approach for the orbital angular momentum of Bloch states~\cite{Wozniak2020, Deilmann2020PRL, Forste2020NatComm, Xuan2020PRR}. 
We are aware that our DFT calculations provide the underestimated value of a band gap, but the relevant valley-Zeeman physics does not depend on it~\cite{Wozniak2020, Deilmann2020PRL, FariaJunior_2022}.
Particularly, we model the Mo$_x$W$_{1-x}$Se$_2$ systems using 2$\times$2 supercells, allowing us to probe Mo concentrations of $x=0.25, 0.50, 0.75$. 
We used experimental in-plane lattice constants of $a_{\text{MoSe}_2}=3.288\;\textrm{\AA}$~\cite{MoSe2} and $a_{\text{WSe}_2}=3.282\;\textrm{\AA}$~\cite{WSe2}. 
The lattice constants of the alloys are interpolated linearly, that is $a_0(x) = x \cdot a_{\text{MoSe}_2} + (1-x) \cdot a_{\text{WSe}_2}$. 
Additionally, we explore different values of $a_0$ to account for typical uncertainties, such as the choice of the exchange-correlation functional within DFT~\cite{Zollner2019PRB, Dirnberger2021SciAdv}, spread of the experimentally determined values~\cite{MoSe2, C4NR06874B, Wilson01051969, ALHILLI197293, Bronsema1986, Evans1981, WSe2}, and possible disorder effects due to alloying and strain~\cite{Mennel2018NatComm, Kolesnichenko2020TDM, Darlington2020NatNano, Alexeev2020ACSNano} (which can reach values as high as 2\% in TMD MLs \cite{BlundoAPR}). 
	The lattice parameters we considered here are within the 1\% range. 
	The computational details are given in the SM~\cite{SM} Sec. 7.


\begin{figure}[t]
	\centering
	\includegraphics[width=1\linewidth]{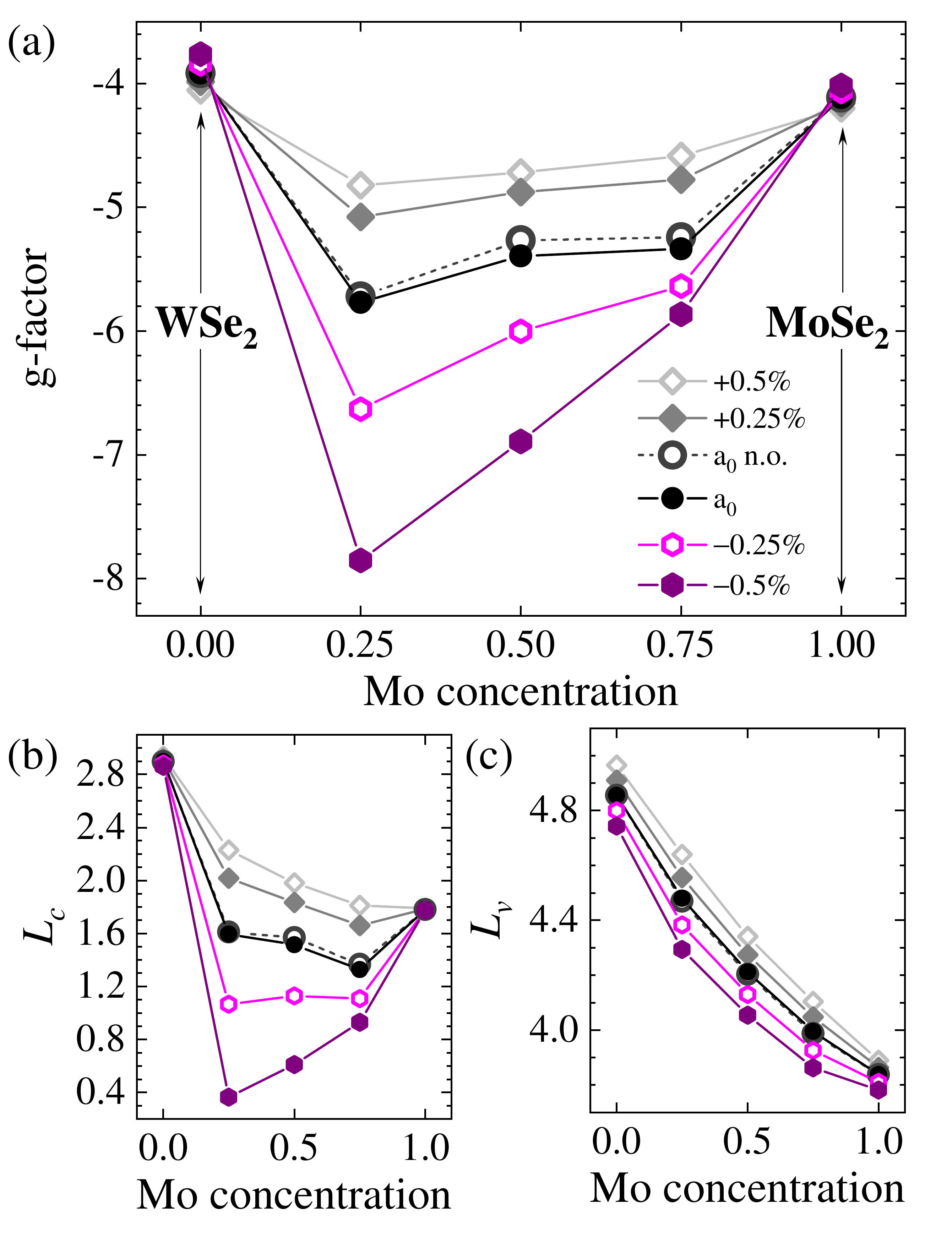}
	\caption{
		(a) Exciton $g$-factors calculated from first principles. Colored lines correspond to different lattice constants of MoSe$_2$. The non-monotonic dependence reproduces the experimental trend shown in Fig.~\ref{fig:gfactor_exp}. The dashed lines indicate the results without optimization of the atomic positions (n.o.), which has a minor impact on the final values. Calculated $g$-factors for the (b) conduction, $g_c$, and (c) valence, $g_v$, bands at the K point, allowing us to pinpoint the electronic origin of the non-monotonic dependence observed in the exciton $g$-factor.}
	\label{fig:g_calc}
\end{figure}

Let us first look at the electronic properties of the Mo$_x$W$_{1-x}$Se$_2$ alloys. 
Fig.~\ref{fig:band} shows the calculated electronic band structures near the K point for different values of the Mo concentration, with the color code indicating the degree of wavefunction localization at the W and Mo atoms. 
While the VBs gradually change their composition and spin splitting as a function of $x$, the scenario is quite different for CBs. 
Particularly, in the conduction bands, we observe 8 energy branches, 2 originally from the K point, and 6 originating from the 2 spin branches of the CBs from the Q valleys, as depicted in Sec.\ S7.1 of SM~\cite{SM}. 
The vertical black arrows in Fig.~\ref{fig:band} indicate the relevant bands involved in the (spin conserving) lowest energy direct optical transitions with circular polarization at the K point. 
We note that the folded Q bands do not participate in the optical processes but, due to their nearly-resonant energetic alignment with the K point bands, they contribute to the wavefunction hybridization by modifying the orbital composition.

The calculated $g$-factors for the lowest energy transitions are displayed in Fig.~\ref{fig:g_calc}(a).
Our calculations reveal that, for intermediate values of $x=0.25, 0.50, 0.75$, the exciton $g$-factor is significantly enhanced and reaches the most negative value at $x=0.25$. 
This non-monotonic dependence revealed by our first principles calculations is in excellent agreement with the experimental trends shown in Fig.~\ref{fig:gfactor_exp}. 
Moreover, the calculated trend is observed regardless of the choice of the lattice parameter. However, the distinct non-monotonic feature is more pronounced for the lowest value of 
$a=3.264$$ \textrm{ \AA}$ for $x$=0.25. 
In fact, the sensitivity of the $g$-factor trend with respect to the chosen in-plane lattice parameter suggests that external strain can be employed to efficiently alter the valley Zeeman features of alloyed S-TMDs, even much more strongly than what has been reported for their parent MLs~\cite{Blundo_magneto_prl, FariaJunior_2022, Covre2022Nanoscale}. 
Note that further insights into the progressive evolution around $x = 0.25$ and random alloys may be obtained in future work using larger supercells.

To understand the non-monotonic dependence of the exciton $g$-factor on $x$, let us look at the calculated $g$-factors of the conduction ($g_c$) and valence ($g_v$) bands, shown in Figs.~\ref{fig:g_calc}(b) and \ref{fig:g_calc}(c), respectively. 
We readily identify that $g_v$ decreases monotonically and remains in-between the values of $g_v$ of the parent compounds. 
In contrast,$g_c$ displays a distinct non-monotonic behavior, with calculated values reaching magnitudes smaller than the $g_c$ values of the parent compounds depending on the chosen value of the lattice parameter. 
This non-monotonic dependence of $g_c$ reflects the non-trivial changes in the orbital nature of the CBs in the K valleys (mostly $d_{z^2}$ orbitals) due to the mixing with conduction bands from Q valleys (mostly $d_{x^2-y^2}+d_{xy}$ and $p$ orbitals). 
Interestingly, for larger values of the in-plane lattice parameter (with Mo and W atoms further apart), the Q and K conduction bands decouple and, consequently, the $g_c$ acquires a monotonic dependence (see Sec.\ S7.3 of SM~\cite{SM}). 
These dramatic changes in $g_c$ revealed by our first principles calculations indeed demonstrate that engineering the wavefunction hybridization in S-TMD alloys to enhance exciton $g$-factors is much more efficient than using proximity effects in van der Waals heterostructures, as recently reported in WS$_2$/graphene systems~\cite{FariaJunior2023TDM}.

\textit{Conclusions---}In this Letter, we outlined striking valley properties of S-TMD alloys. 
In particular, we experimentally observed a high valley polarization close to unity for excitonic complexes in the Mo$_x$W$_{1-x}$Se$_2$ alloys, and a highly stoichiometry-dependent behavior of the exciton $g$-factor. 
Specifically, we found that the $g$-factors for neutral excitons exhibit a strongly non-linear dependence from about -4 in the MoSe$_2$ and WSe$_2$ MLs, up to -10 for $x \approx 0.2$.
Our first-principles calculations successfully reproduced the experimental trends and identified the origin of such strong $g$-factor modulation due to the orbital mixing in the conduction band. 
We also revealed that the $g$-factor values can be significantly altered by strain.
Our results thus show that S-TMD alloys represent a unique, yet not much explored material platform with remarkable valley-tunable properties that can be manipulated on demand.\\

\textit{Acknowledgements---}The work was supported by the National Science Centre, Poland (grant no. 2022/45/N/ST3/01887 and 2021/41/N/ST3/04516).
We acknowledge the support of the LNCMI-CNRS, a member of the European Magnetic Field Laboratory (EMFL). 
The publication was created as part of a project co-financed by the Polish Ministry of Science and Higher Education under contract no. 2025/WK/01.
This project was funded within the QuantERA II Programme that has received funding from the European Union’s Horizon 2020 research and innovation programme under Grant Agreement No 101017733, and with the funding organizations Ministero dell'Universit\'{a} e della Ricerca (MUR) and Consiglio Nazionale delle Ricerche (CNR). E.B. acknowledges support from La Sapienza through the grant Avvio alla Ricerca 2021 (grant no. AR12117A8A090764).
A.P. acknowledges support from PNRR MUR Project PE0000023-NQSTI. 
K.W. and T.T. acknowledge support from the JSPS KAKENHI (Grant Numbers 21H05233 and 23H02052), the CREST (JPMJCR24A5), JST and World Premier International Research Center Initiative (WPI), MEXT, Japan.
M.P. acknowledges support from the CENTERA2, FENG.02.01-IP.05-5 T004/23 project funded within the IRA program of the FNP Poland, co-financed by the EU FENG Programme.
Z.S. was supported by ERC-CZ program (project LL2101) from Ministry of Education Youth and Sports (MEYS) and by the project Advanced Functional Nanorobots (reg. No. CZ.02.1.01/0.0/0.0/15\_003/0000444 financed by the EFRR).
T.W. gratefully acknowledges Poland’s high-performance Infrastructure PLGrid ACC Cyfronet AGH for providing computer facilities and support within computational grant no. PLG/2025/018073.
\bibliographystyle{apsrev4-2}
\bibliography{biblio}

\begin{thebibliography}{123}%
\makeatletter
\providecommand \@ifxundefined [1]{%
 \@ifx{#1\undefined}
}%
\providecommand \@ifnum [1]{%
 \ifnum #1\expandafter \@firstoftwo
 \else \expandafter \@secondoftwo
 \fi
}%
\providecommand \@ifx [1]{%
 \ifx #1\expandafter \@firstoftwo
 \else \expandafter \@secondoftwo
 \fi
}%
\providecommand \natexlab [1]{#1}%
\providecommand \enquote  [1]{``#1''}%
\providecommand \bibnamefont  [1]{#1}%
\providecommand \bibfnamefont [1]{#1}%
\providecommand \citenamefont [1]{#1}%
\providecommand \href@noop [0]{\@secondoftwo}%
\providecommand \href [0]{\begingroup \@sanitize@url \@href}%
\providecommand \@href[1]{\@@startlink{#1}\@@href}%
\providecommand \@@href[1]{\endgroup#1\@@endlink}%
\providecommand \@sanitize@url [0]{\catcode `\\12\catcode `\$12\catcode
  `\&12\catcode `\#12\catcode `\^12\catcode `\_12\catcode `\%12\relax}%
\providecommand \@@startlink[1]{}%
\providecommand \@@endlink[0]{}%
\providecommand \url  [0]{\begingroup\@sanitize@url \@url }%
\providecommand \@url [1]{\endgroup\@href {#1}{\urlprefix }}%
\providecommand \urlprefix  [0]{URL }%
\providecommand \Eprint [0]{\href }%
\providecommand \doibase [0]{https://doi.org/}%
\providecommand \selectlanguage [0]{\@gobble}%
\providecommand \bibinfo  [0]{\@secondoftwo}%
\providecommand \bibfield  [0]{\@secondoftwo}%
\providecommand \translation [1]{[#1]}%
\providecommand \BibitemOpen [0]{}%
\providecommand \bibitemStop [0]{}%
\providecommand \bibitemNoStop [0]{.\EOS\space}%
\providecommand \EOS [0]{\spacefactor3000\relax}%
\providecommand \BibitemShut  [1]{\csname bibitem#1\endcsname}%
\let\auto@bib@innerbib\@empty
\bibitem [{\citenamefont {Mak}\ \emph {et~al.}(2010)\citenamefont {Mak},
  \citenamefont {Lee}, \citenamefont {Hone}, \citenamefont {Shan},\ and\
  \citenamefont {Heinz}}]{Mak2010}%
  \BibitemOpen
  \bibfield  {author} {\bibinfo {author} {\bibfnamefont {K.~F.}\ \bibnamefont
  {Mak}}, \bibinfo {author} {\bibfnamefont {C.}~\bibnamefont {Lee}}, \bibinfo
  {author} {\bibfnamefont {J.}~\bibnamefont {Hone}}, \bibinfo {author}
  {\bibfnamefont {J.}~\bibnamefont {Shan}},\ and\ \bibinfo {author}
  {\bibfnamefont {T.~F.}\ \bibnamefont {Heinz}},\ }\href
  {https://doi.org/10.1103/PhysRevLett.105.136805} {\bibfield  {journal}
  {\bibinfo  {journal} {Phys. Rev. Lett.}\ }\textbf {\bibinfo {volume} {105}},\
  \bibinfo {pages} {136805} (\bibinfo {year} {2010})}\BibitemShut {NoStop}%
\bibitem [{\citenamefont {Cao}\ \emph {et~al.}(2012)\citenamefont {Cao},
  \citenamefont {Wang}, \citenamefont {Han}, \citenamefont {Ye}, \citenamefont
  {Zhu}, \citenamefont {Shi}, \citenamefont {Niu}, \citenamefont {Tan},
  \citenamefont {Wang}, \citenamefont {Liu},\ and\ \citenamefont
  {Feng}}]{Cao2012}%
  \BibitemOpen
  \bibfield  {author} {\bibinfo {author} {\bibfnamefont {T.}~\bibnamefont
  {Cao}}, \bibinfo {author} {\bibfnamefont {G.}~\bibnamefont {Wang}}, \bibinfo
  {author} {\bibfnamefont {W.}~\bibnamefont {Han}}, \bibinfo {author}
  {\bibfnamefont {H.}~\bibnamefont {Ye}}, \bibinfo {author} {\bibfnamefont
  {C.}~\bibnamefont {Zhu}}, \bibinfo {author} {\bibfnamefont {J.}~\bibnamefont
  {Shi}}, \bibinfo {author} {\bibfnamefont {Q.}~\bibnamefont {Niu}}, \bibinfo
  {author} {\bibfnamefont {P.}~\bibnamefont {Tan}}, \bibinfo {author}
  {\bibfnamefont {E.}~\bibnamefont {Wang}}, \bibinfo {author} {\bibfnamefont
  {B.}~\bibnamefont {Liu}},\ and\ \bibinfo {author} {\bibfnamefont
  {J.}~\bibnamefont {Feng}},\ }\href {https://doi.org/10.1038/ncomms1882}
  {\bibfield  {journal} {\bibinfo  {journal} {Nature Communications}\ }\textbf
  {\bibinfo {volume} {3}},\ \bibinfo {pages} {887} (\bibinfo {year}
  {2012})}\BibitemShut {NoStop}%
\bibitem [{\citenamefont {Wang}\ \emph {et~al.}(2014)\citenamefont {Wang},
  \citenamefont {Bouet}, \citenamefont {Lagarde}, \citenamefont {Vidal},
  \citenamefont {Balocchi}, \citenamefont {Amand}, \citenamefont {Marie},\ and\
  \citenamefont {Urbaszek}}]{Wang2014}%
  \BibitemOpen
  \bibfield  {author} {\bibinfo {author} {\bibfnamefont {G.}~\bibnamefont
  {Wang}}, \bibinfo {author} {\bibfnamefont {L.}~\bibnamefont {Bouet}},
  \bibinfo {author} {\bibfnamefont {D.}~\bibnamefont {Lagarde}}, \bibinfo
  {author} {\bibfnamefont {M.}~\bibnamefont {Vidal}}, \bibinfo {author}
  {\bibfnamefont {A.}~\bibnamefont {Balocchi}}, \bibinfo {author}
  {\bibfnamefont {T.}~\bibnamefont {Amand}}, \bibinfo {author} {\bibfnamefont
  {X.}~\bibnamefont {Marie}},\ and\ \bibinfo {author} {\bibfnamefont
  {B.}~\bibnamefont {Urbaszek}},\ }\href
  {https://doi.org/10.1103/PhysRevB.90.075413} {\bibfield  {journal} {\bibinfo
  {journal} {Phys. Rev. B}\ }\textbf {\bibinfo {volume} {90}},\ \bibinfo
  {pages} {075413} (\bibinfo {year} {2014})}\BibitemShut {NoStop}%
\bibitem [{\citenamefont {Arora}\ \emph
  {et~al.}(2015{\natexlab{a}})\citenamefont {Arora}, \citenamefont {Koperski},
  \citenamefont {Nogajewski}, \citenamefont {Marcus}, \citenamefont
  {Faugeras},\ and\ \citenamefont {Potemski}}]{Arora2015W}%
  \BibitemOpen
  \bibfield  {author} {\bibinfo {author} {\bibfnamefont {A.}~\bibnamefont
  {Arora}}, \bibinfo {author} {\bibfnamefont {M.}~\bibnamefont {Koperski}},
  \bibinfo {author} {\bibfnamefont {K.}~\bibnamefont {Nogajewski}}, \bibinfo
  {author} {\bibfnamefont {J.}~\bibnamefont {Marcus}}, \bibinfo {author}
  {\bibfnamefont {C.}~\bibnamefont {Faugeras}},\ and\ \bibinfo {author}
  {\bibfnamefont {M.}~\bibnamefont {Potemski}},\ }\href
  {https://doi.org/10.1039/C5NR01536G} {\bibfield  {journal} {\bibinfo
  {journal} {Nanoscale}\ }\textbf {\bibinfo {volume} {7}},\ \bibinfo {pages}
  {10421} (\bibinfo {year} {2015}{\natexlab{a}})}\BibitemShut {NoStop}%
\bibitem [{\citenamefont {Arora}\ \emph
  {et~al.}(2015{\natexlab{b}})\citenamefont {Arora}, \citenamefont
  {Nogajewski}, \citenamefont {Molas}, \citenamefont {Koperski},\ and\
  \citenamefont {Potemski}}]{aroramose2}%
  \BibitemOpen
  \bibfield  {author} {\bibinfo {author} {\bibfnamefont {A.}~\bibnamefont
  {Arora}}, \bibinfo {author} {\bibfnamefont {K.}~\bibnamefont {Nogajewski}},
  \bibinfo {author} {\bibfnamefont {M.}~\bibnamefont {Molas}}, \bibinfo
  {author} {\bibfnamefont {M.}~\bibnamefont {Koperski}},\ and\ \bibinfo
  {author} {\bibfnamefont {M.}~\bibnamefont {Potemski}},\ }\href
  {https://doi.org/10.1039/C5NR06782K} {\bibfield  {journal} {\bibinfo
  {journal} {Nanoscale}\ }\textbf {\bibinfo {volume} {7}},\ \bibinfo {pages}
  {20769} (\bibinfo {year} {2015}{\natexlab{b}})}\BibitemShut {NoStop}%
\bibitem [{\citenamefont {Tonndorf}\ \emph {et~al.}(2013)\citenamefont
  {Tonndorf}, \citenamefont {Schmidt}, \citenamefont {B\"{o}ttger},
  \citenamefont {Zhang}, \citenamefont {B\"{o}rner}, \citenamefont {Liebig},
  \citenamefont {Albrecht}, \citenamefont {Kloc}, \citenamefont {Gordan},
  \citenamefont {Zahn}, \citenamefont {de~Vasconcellos},\ and\ \citenamefont
  {Bratschitsch}}]{Tonndorf2013}%
  \BibitemOpen
  \bibfield  {author} {\bibinfo {author} {\bibfnamefont {P.}~\bibnamefont
  {Tonndorf}}, \bibinfo {author} {\bibfnamefont {R.}~\bibnamefont {Schmidt}},
  \bibinfo {author} {\bibfnamefont {P.}~\bibnamefont {B\"{o}ttger}}, \bibinfo
  {author} {\bibfnamefont {X.}~\bibnamefont {Zhang}}, \bibinfo {author}
  {\bibfnamefont {J.}~\bibnamefont {B\"{o}rner}}, \bibinfo {author}
  {\bibfnamefont {A.}~\bibnamefont {Liebig}}, \bibinfo {author} {\bibfnamefont
  {M.}~\bibnamefont {Albrecht}}, \bibinfo {author} {\bibfnamefont
  {C.}~\bibnamefont {Kloc}}, \bibinfo {author} {\bibfnamefont {O.}~\bibnamefont
  {Gordan}}, \bibinfo {author} {\bibfnamefont {D.~R.~T.}\ \bibnamefont {Zahn}},
  \bibinfo {author} {\bibfnamefont {S.~M.}\ \bibnamefont {de~Vasconcellos}},\
  and\ \bibinfo {author} {\bibfnamefont {R.}~\bibnamefont {Bratschitsch}},\
  }\href {https://doi.org/10.1364/OE.21.004908} {\bibfield  {journal} {\bibinfo
   {journal} {Opt. Express}\ }\textbf {\bibinfo {volume} {21}},\ \bibinfo
  {pages} {4908} (\bibinfo {year} {2013})}\BibitemShut {NoStop}%
\bibitem [{\citenamefont {Lezama}\ \emph {et~al.}(2015)\citenamefont {Lezama},
  \citenamefont {Arora}, \citenamefont {Ubaldini}, \citenamefont {Barreteau},
  \citenamefont {Giannini}, \citenamefont {Potemski},\ and\ \citenamefont
  {Morpurgo}}]{Lezama2015}%
  \BibitemOpen
  \bibfield  {author} {\bibinfo {author} {\bibfnamefont {I.~G.}\ \bibnamefont
  {Lezama}}, \bibinfo {author} {\bibfnamefont {A.}~\bibnamefont {Arora}},
  \bibinfo {author} {\bibfnamefont {A.}~\bibnamefont {Ubaldini}}, \bibinfo
  {author} {\bibfnamefont {C.}~\bibnamefont {Barreteau}}, \bibinfo {author}
  {\bibfnamefont {E.}~\bibnamefont {Giannini}}, \bibinfo {author}
  {\bibfnamefont {M.}~\bibnamefont {Potemski}},\ and\ \bibinfo {author}
  {\bibfnamefont {A.~F.}\ \bibnamefont {Morpurgo}},\ }\href
  {https://doi.org/10.1021/nl5045007} {\bibfield  {journal} {\bibinfo
  {journal} {Nano Letters}\ }\textbf {\bibinfo {volume} {15}},\ \bibinfo
  {pages} {2336–2342} (\bibinfo {year} {2015})}\BibitemShut {NoStop}%
\bibitem [{\citenamefont {Koperski}\ \emph {et~al.}(2017)\citenamefont
  {Koperski}, \citenamefont {Molas}, \citenamefont {Arora}, \citenamefont
  {Nogajewski}, \citenamefont {Slobodeniuk}, \citenamefont {Faugeras},\ and\
  \citenamefont {Potemski}}]{Koperski2017}%
  \BibitemOpen
  \bibfield  {author} {\bibinfo {author} {\bibfnamefont {M.}~\bibnamefont
  {Koperski}}, \bibinfo {author} {\bibfnamefont {M.~R.}\ \bibnamefont {Molas}},
  \bibinfo {author} {\bibfnamefont {A.}~\bibnamefont {Arora}}, \bibinfo
  {author} {\bibfnamefont {K.}~\bibnamefont {Nogajewski}}, \bibinfo {author}
  {\bibfnamefont {A.~O.}\ \bibnamefont {Slobodeniuk}}, \bibinfo {author}
  {\bibfnamefont {C.}~\bibnamefont {Faugeras}},\ and\ \bibinfo {author}
  {\bibfnamefont {M.}~\bibnamefont {Potemski}},\ }\href
  {https://doi.org/10.1515/nanoph-2016-0165} {\bibfield  {journal} {\bibinfo
  {journal} {Nanophotonics}\ }\textbf {\bibinfo {volume} {6}},\ \bibinfo
  {pages} {1289} (\bibinfo {year} {2017})}\BibitemShut {NoStop}%
\bibitem [{\citenamefont {Molas}\ \emph
  {et~al.}(2017{\natexlab{a}})\citenamefont {Molas}, \citenamefont
  {Nogajewski}, \citenamefont {Slobodeniuk}, \citenamefont {Binder},
  \citenamefont {Bartos},\ and\ \citenamefont {Potemski}}]{molasNanoscale}%
  \BibitemOpen
  \bibfield  {author} {\bibinfo {author} {\bibfnamefont {M.~R.}\ \bibnamefont
  {Molas}}, \bibinfo {author} {\bibfnamefont {K.}~\bibnamefont {Nogajewski}},
  \bibinfo {author} {\bibfnamefont {A.~O.}\ \bibnamefont {Slobodeniuk}},
  \bibinfo {author} {\bibfnamefont {J.}~\bibnamefont {Binder}}, \bibinfo
  {author} {\bibfnamefont {M.}~\bibnamefont {Bartos}},\ and\ \bibinfo {author}
  {\bibfnamefont {M.}~\bibnamefont {Potemski}},\ }\href
  {https://doi.org/10.1039/C7NR04672C} {\bibfield  {journal} {\bibinfo
  {journal} {Nanoscale}\ }\textbf {\bibinfo {volume} {9}},\ \bibinfo {pages}
  {13128} (\bibinfo {year} {2017}{\natexlab{a}})}\BibitemShut {NoStop}%
\bibitem [{\citenamefont {Molas}\ \emph
  {et~al.}(2017{\natexlab{b}})\citenamefont {Molas}, \citenamefont {Faugeras},
  \citenamefont {Slobodeniuk}, \citenamefont {Nogajewski}, \citenamefont
  {Bartos}, \citenamefont {Basko},\ and\ \citenamefont {Potemski}}]{Molas2017}%
  \BibitemOpen
  \bibfield  {author} {\bibinfo {author} {\bibfnamefont {M.~R.}\ \bibnamefont
  {Molas}}, \bibinfo {author} {\bibfnamefont {C.}~\bibnamefont {Faugeras}},
  \bibinfo {author} {\bibfnamefont {A.~O.}\ \bibnamefont {Slobodeniuk}},
  \bibinfo {author} {\bibfnamefont {K.}~\bibnamefont {Nogajewski}}, \bibinfo
  {author} {\bibfnamefont {M.}~\bibnamefont {Bartos}}, \bibinfo {author}
  {\bibfnamefont {D.~M.}\ \bibnamefont {Basko}},\ and\ \bibinfo {author}
  {\bibfnamefont {M.}~\bibnamefont {Potemski}},\ }\href
  {https://doi.org/10.1088/2053-1583/aa5521} {\bibfield  {journal} {\bibinfo
  {journal} {2D Materials}\ }\textbf {\bibinfo {volume} {4}},\ \bibinfo {pages}
  {021003} (\bibinfo {year} {2017}{\natexlab{b}})}\BibitemShut {NoStop}%
\bibitem [{\citenamefont {Liu}\ \emph {et~al.}(2020)\citenamefont {Liu},
  \citenamefont {van Baren}, \citenamefont {Liang}, \citenamefont {Taniguchi},
  \citenamefont {Watanabe}, \citenamefont {Gabor}, \citenamefont {Chang},\ and\
  \citenamefont {Lui}}]{Liu2020}%
  \BibitemOpen
  \bibfield  {author} {\bibinfo {author} {\bibfnamefont {E.}~\bibnamefont
  {Liu}}, \bibinfo {author} {\bibfnamefont {J.}~\bibnamefont {van Baren}},
  \bibinfo {author} {\bibfnamefont {C.-T.}\ \bibnamefont {Liang}}, \bibinfo
  {author} {\bibfnamefont {T.}~\bibnamefont {Taniguchi}}, \bibinfo {author}
  {\bibfnamefont {K.}~\bibnamefont {Watanabe}}, \bibinfo {author}
  {\bibfnamefont {N.~M.}\ \bibnamefont {Gabor}}, \bibinfo {author}
  {\bibfnamefont {Y.-C.}\ \bibnamefont {Chang}},\ and\ \bibinfo {author}
  {\bibfnamefont {C.~H.}\ \bibnamefont {Lui}},\ }\href
  {https://doi.org/10.1103/PhysRevLett.124.196802} {\bibfield  {journal}
  {\bibinfo  {journal} {Phys. Rev. Lett.}\ }\textbf {\bibinfo {volume} {124}},\
  \bibinfo {pages} {196802} (\bibinfo {year} {2020})}\BibitemShut {NoStop}%
\bibitem [{\citenamefont {Grzeszczyk}\ \emph {et~al.}(2021)\citenamefont
  {Grzeszczyk}, \citenamefont {Olkowska-Pucko}, \citenamefont {Nogajewski},
  \citenamefont {Watanabe}, \citenamefont {Taniguchi}, \citenamefont
  {Kossacki}, \citenamefont {Babiński},\ and\ \citenamefont
  {Molas}}]{Grzeszczyk2021}%
  \BibitemOpen
  \bibfield  {author} {\bibinfo {author} {\bibfnamefont {M.}~\bibnamefont
  {Grzeszczyk}}, \bibinfo {author} {\bibfnamefont {K.}~\bibnamefont
  {Olkowska-Pucko}}, \bibinfo {author} {\bibfnamefont {K.}~\bibnamefont
  {Nogajewski}}, \bibinfo {author} {\bibfnamefont {K.}~\bibnamefont
  {Watanabe}}, \bibinfo {author} {\bibfnamefont {T.}~\bibnamefont {Taniguchi}},
  \bibinfo {author} {\bibfnamefont {P.}~\bibnamefont {Kossacki}}, \bibinfo
  {author} {\bibfnamefont {A.}~\bibnamefont {Babiński}},\ and\ \bibinfo
  {author} {\bibfnamefont {M.~R.}\ \bibnamefont {Molas}},\ }\href
  {https://doi.org/10.1039/D1NR03855A} {\bibfield  {journal} {\bibinfo
  {journal} {Nanoscale}\ }\textbf {\bibinfo {volume} {13}},\ \bibinfo {pages}
  {18726} (\bibinfo {year} {2021})}\BibitemShut {NoStop}%
\bibitem [{\citenamefont {Cadiz}\ \emph {et~al.}(2017)\citenamefont {Cadiz},
  \citenamefont {Courtade}, \citenamefont {Robert}, \citenamefont {Wang},
  \citenamefont {Shen}, \citenamefont {Cai}, \citenamefont {Taniguchi},
  \citenamefont {Watanabe}, \citenamefont {Carrere}, \citenamefont {Lagarde},
  \citenamefont {Manca}, \citenamefont {Amand}, \citenamefont {Renucci},
  \citenamefont {Tongay}, \citenamefont {Marie},\ and\ \citenamefont
  {Urbaszek}}]{Cadiz2017}%
  \BibitemOpen
  \bibfield  {author} {\bibinfo {author} {\bibfnamefont {F.}~\bibnamefont
  {Cadiz}}, \bibinfo {author} {\bibfnamefont {E.}~\bibnamefont {Courtade}},
  \bibinfo {author} {\bibfnamefont {C.}~\bibnamefont {Robert}}, \bibinfo
  {author} {\bibfnamefont {G.}~\bibnamefont {Wang}}, \bibinfo {author}
  {\bibfnamefont {Y.}~\bibnamefont {Shen}}, \bibinfo {author} {\bibfnamefont
  {H.}~\bibnamefont {Cai}}, \bibinfo {author} {\bibfnamefont {T.}~\bibnamefont
  {Taniguchi}}, \bibinfo {author} {\bibfnamefont {K.}~\bibnamefont {Watanabe}},
  \bibinfo {author} {\bibfnamefont {H.}~\bibnamefont {Carrere}}, \bibinfo
  {author} {\bibfnamefont {D.}~\bibnamefont {Lagarde}}, \bibinfo {author}
  {\bibfnamefont {M.}~\bibnamefont {Manca}}, \bibinfo {author} {\bibfnamefont
  {T.}~\bibnamefont {Amand}}, \bibinfo {author} {\bibfnamefont
  {P.}~\bibnamefont {Renucci}}, \bibinfo {author} {\bibfnamefont
  {S.}~\bibnamefont {Tongay}}, \bibinfo {author} {\bibfnamefont
  {X.}~\bibnamefont {Marie}},\ and\ \bibinfo {author} {\bibfnamefont
  {B.}~\bibnamefont {Urbaszek}},\ }\href
  {https://doi.org/10.1103/PhysRevX.7.021026} {\bibfield  {journal} {\bibinfo
  {journal} {Phys. Rev. X}\ }\textbf {\bibinfo {volume} {7}},\ \bibinfo {pages}
  {021026} (\bibinfo {year} {2017})}\BibitemShut {NoStop}%
\bibitem [{\citenamefont {Malic}\ \emph {et~al.}(2018)\citenamefont {Malic},
  \citenamefont {Selig}, \citenamefont {Feierabend}, \citenamefont {Brem},
  \citenamefont {Christiansen}, \citenamefont {Wendler}, \citenamefont
  {Knorr},\ and\ \citenamefont {Bergh\"auser}}]{Malic2018}%
  \BibitemOpen
  \bibfield  {author} {\bibinfo {author} {\bibfnamefont {E.}~\bibnamefont
  {Malic}}, \bibinfo {author} {\bibfnamefont {M.}~\bibnamefont {Selig}},
  \bibinfo {author} {\bibfnamefont {M.}~\bibnamefont {Feierabend}}, \bibinfo
  {author} {\bibfnamefont {S.}~\bibnamefont {Brem}}, \bibinfo {author}
  {\bibfnamefont {D.}~\bibnamefont {Christiansen}}, \bibinfo {author}
  {\bibfnamefont {F.}~\bibnamefont {Wendler}}, \bibinfo {author} {\bibfnamefont
  {A.}~\bibnamefont {Knorr}},\ and\ \bibinfo {author} {\bibfnamefont
  {G.}~\bibnamefont {Bergh\"auser}},\ }\href
  {https://doi.org/10.1103/PhysRevMaterials.2.014002} {\bibfield  {journal}
  {\bibinfo  {journal} {Phys. Rev. Mater.}\ }\textbf {\bibinfo {volume} {2}},\
  \bibinfo {pages} {014002} (\bibinfo {year} {2018})}\BibitemShut {NoStop}%
\bibitem [{\citenamefont {Taghizadeh}\ and\ \citenamefont
  {Pedersen}(2019)}]{Taghizadeh2019}%
  \BibitemOpen
  \bibfield  {author} {\bibinfo {author} {\bibfnamefont {A.}~\bibnamefont
  {Taghizadeh}}\ and\ \bibinfo {author} {\bibfnamefont {T.~G.}\ \bibnamefont
  {Pedersen}},\ }\href {https://doi.org/10.1103/PhysRevB.99.235433} {\bibfield
  {journal} {\bibinfo  {journal} {Phys. Rev. B}\ }\textbf {\bibinfo {volume}
  {99}},\ \bibinfo {pages} {235433} (\bibinfo {year} {2019})}\BibitemShut
  {NoStop}%
\bibitem [{\citenamefont {Gerber}\ \emph {et~al.}(2019)\citenamefont {Gerber},
  \citenamefont {Courtade}, \citenamefont {Shree}, \citenamefont {Robert},
  \citenamefont {Taniguchi}, \citenamefont {Watanabe}, \citenamefont
  {Balocchi}, \citenamefont {Renucci}, \citenamefont {Lagarde}, \citenamefont
  {Marie},\ and\ \citenamefont {Urbaszek}}]{Gerber2019}%
  \BibitemOpen
  \bibfield  {author} {\bibinfo {author} {\bibfnamefont {I.~C.}\ \bibnamefont
  {Gerber}}, \bibinfo {author} {\bibfnamefont {E.}~\bibnamefont {Courtade}},
  \bibinfo {author} {\bibfnamefont {S.}~\bibnamefont {Shree}}, \bibinfo
  {author} {\bibfnamefont {C.}~\bibnamefont {Robert}}, \bibinfo {author}
  {\bibfnamefont {T.}~\bibnamefont {Taniguchi}}, \bibinfo {author}
  {\bibfnamefont {K.}~\bibnamefont {Watanabe}}, \bibinfo {author}
  {\bibfnamefont {A.}~\bibnamefont {Balocchi}}, \bibinfo {author}
  {\bibfnamefont {P.}~\bibnamefont {Renucci}}, \bibinfo {author} {\bibfnamefont
  {D.}~\bibnamefont {Lagarde}}, \bibinfo {author} {\bibfnamefont
  {X.}~\bibnamefont {Marie}},\ and\ \bibinfo {author} {\bibfnamefont
  {B.}~\bibnamefont {Urbaszek}},\ }\href
  {https://doi.org/10.1103/PhysRevB.99.035443} {\bibfield  {journal} {\bibinfo
  {journal} {Phys. Rev. B}\ }\textbf {\bibinfo {volume} {99}},\ \bibinfo
  {pages} {035443} (\bibinfo {year} {2019})}\BibitemShut {NoStop}%
\bibitem [{\citenamefont {Robert}\ \emph {et~al.}(2016)\citenamefont {Robert},
  \citenamefont {Picard}, \citenamefont {Lagarde}, \citenamefont {Wang},
  \citenamefont {Echeverry}, \citenamefont {Cadiz}, \citenamefont {Renucci},
  \citenamefont {H\"ogele}, \citenamefont {Amand}, \citenamefont {Marie},
  \citenamefont {Gerber},\ and\ \citenamefont {Urbaszek}}]{Robert2016}%
  \BibitemOpen
  \bibfield  {author} {\bibinfo {author} {\bibfnamefont {C.}~\bibnamefont
  {Robert}}, \bibinfo {author} {\bibfnamefont {R.}~\bibnamefont {Picard}},
  \bibinfo {author} {\bibfnamefont {D.}~\bibnamefont {Lagarde}}, \bibinfo
  {author} {\bibfnamefont {G.}~\bibnamefont {Wang}}, \bibinfo {author}
  {\bibfnamefont {J.~P.}\ \bibnamefont {Echeverry}}, \bibinfo {author}
  {\bibfnamefont {F.}~\bibnamefont {Cadiz}}, \bibinfo {author} {\bibfnamefont
  {P.}~\bibnamefont {Renucci}}, \bibinfo {author} {\bibfnamefont
  {A.}~\bibnamefont {H\"ogele}}, \bibinfo {author} {\bibfnamefont
  {T.}~\bibnamefont {Amand}}, \bibinfo {author} {\bibfnamefont
  {X.}~\bibnamefont {Marie}}, \bibinfo {author} {\bibfnamefont {I.~C.}\
  \bibnamefont {Gerber}},\ and\ \bibinfo {author} {\bibfnamefont
  {B.}~\bibnamefont {Urbaszek}},\ }\href
  {https://doi.org/10.1103/PhysRevB.94.155425} {\bibfield  {journal} {\bibinfo
  {journal} {Phys. Rev. B}\ }\textbf {\bibinfo {volume} {94}},\ \bibinfo
  {pages} {155425} (\bibinfo {year} {2016})}\BibitemShut {NoStop}%
\bibitem [{\citenamefont {Zhang}\ \emph {et~al.}(2017)\citenamefont {Zhang},
  \citenamefont {Cao}, \citenamefont {Lu}, \citenamefont {Lin}, \citenamefont
  {Zhang}, \citenamefont {Wang}, \citenamefont {Li}, \citenamefont {Hone},
  \citenamefont {Robinson}, \citenamefont {Smirnov}, \citenamefont {Louie},\
  and\ \citenamefont {Heinz}}]{Zhang2017}%
  \BibitemOpen
  \bibfield  {author} {\bibinfo {author} {\bibfnamefont {X.-X.}\ \bibnamefont
  {Zhang}}, \bibinfo {author} {\bibfnamefont {T.}~\bibnamefont {Cao}}, \bibinfo
  {author} {\bibfnamefont {Z.}~\bibnamefont {Lu}}, \bibinfo {author}
  {\bibfnamefont {Y.-C.}\ \bibnamefont {Lin}}, \bibinfo {author} {\bibfnamefont
  {F.}~\bibnamefont {Zhang}}, \bibinfo {author} {\bibfnamefont
  {Y.}~\bibnamefont {Wang}}, \bibinfo {author} {\bibfnamefont {Z.}~\bibnamefont
  {Li}}, \bibinfo {author} {\bibfnamefont {J.~C.}\ \bibnamefont {Hone}},
  \bibinfo {author} {\bibfnamefont {J.~A.}\ \bibnamefont {Robinson}}, \bibinfo
  {author} {\bibfnamefont {D.}~\bibnamefont {Smirnov}}, \bibinfo {author}
  {\bibfnamefont {S.~G.}\ \bibnamefont {Louie}},\ and\ \bibinfo {author}
  {\bibfnamefont {T.~F.}\ \bibnamefont {Heinz}},\ }\href
  {https://doi.org/10.1038/nnano.2017.105} {\bibfield  {journal} {\bibinfo
  {journal} {Nature Nanotechnology}\ }\textbf {\bibinfo {volume} {12}},\
  \bibinfo {pages} {883} (\bibinfo {year} {2017})}\BibitemShut {NoStop}%
\bibitem [{\citenamefont {Robert}\ \emph {et~al.}(2017)\citenamefont {Robert},
  \citenamefont {Amand}, \citenamefont {Cadiz}, \citenamefont {Lagarde},
  \citenamefont {Courtade}, \citenamefont {Manca}, \citenamefont {Taniguchi},
  \citenamefont {Watanabe}, \citenamefont {Urbaszek},\ and\ \citenamefont
  {Marie}}]{Robert2017}%
  \BibitemOpen
  \bibfield  {author} {\bibinfo {author} {\bibfnamefont {C.}~\bibnamefont
  {Robert}}, \bibinfo {author} {\bibfnamefont {T.}~\bibnamefont {Amand}},
  \bibinfo {author} {\bibfnamefont {F.}~\bibnamefont {Cadiz}}, \bibinfo
  {author} {\bibfnamefont {D.}~\bibnamefont {Lagarde}}, \bibinfo {author}
  {\bibfnamefont {E.}~\bibnamefont {Courtade}}, \bibinfo {author}
  {\bibfnamefont {M.}~\bibnamefont {Manca}}, \bibinfo {author} {\bibfnamefont
  {T.}~\bibnamefont {Taniguchi}}, \bibinfo {author} {\bibfnamefont
  {K.}~\bibnamefont {Watanabe}}, \bibinfo {author} {\bibfnamefont
  {B.}~\bibnamefont {Urbaszek}},\ and\ \bibinfo {author} {\bibfnamefont
  {X.}~\bibnamefont {Marie}},\ }\href
  {https://doi.org/10.1103/PhysRevB.96.155423} {\bibfield  {journal} {\bibinfo
  {journal} {Phys. Rev. B}\ }\textbf {\bibinfo {volume} {96}},\ \bibinfo
  {pages} {155423} (\bibinfo {year} {2017})}\BibitemShut {NoStop}%
\bibitem [{\citenamefont {Molas}\ \emph
  {et~al.}(2019{\natexlab{a}})\citenamefont {Molas}, \citenamefont
  {Slobodeniuk}, \citenamefont {Kazimierczuk}, \citenamefont {Nogajewski},
  \citenamefont {Bartos}, \citenamefont {Kapu\ifmmode \acute{s}\else
  \'{s}\fi{}ci\ifmmode~\acute{n}\else \'{n}\fi{}ski}, \citenamefont
  {Oreszczuk}, \citenamefont {Watanabe}, \citenamefont {Taniguchi},
  \citenamefont {Faugeras}, \citenamefont {Kossacki}, \citenamefont {Basko},\
  and\ \citenamefont {Potemski}}]{Molas2019}%
  \BibitemOpen
  \bibfield  {author} {\bibinfo {author} {\bibfnamefont {M.~R.}\ \bibnamefont
  {Molas}}, \bibinfo {author} {\bibfnamefont {A.~O.}\ \bibnamefont
  {Slobodeniuk}}, \bibinfo {author} {\bibfnamefont {T.}~\bibnamefont
  {Kazimierczuk}}, \bibinfo {author} {\bibfnamefont {K.}~\bibnamefont
  {Nogajewski}}, \bibinfo {author} {\bibfnamefont {M.}~\bibnamefont {Bartos}},
  \bibinfo {author} {\bibfnamefont {P.}~\bibnamefont {Kapu\ifmmode
  \acute{s}\else \'{s}\fi{}ci\ifmmode~\acute{n}\else \'{n}\fi{}ski}}, \bibinfo
  {author} {\bibfnamefont {K.}~\bibnamefont {Oreszczuk}}, \bibinfo {author}
  {\bibfnamefont {K.}~\bibnamefont {Watanabe}}, \bibinfo {author}
  {\bibfnamefont {T.}~\bibnamefont {Taniguchi}}, \bibinfo {author}
  {\bibfnamefont {C.}~\bibnamefont {Faugeras}}, \bibinfo {author}
  {\bibfnamefont {P.}~\bibnamefont {Kossacki}}, \bibinfo {author}
  {\bibfnamefont {D.~M.}\ \bibnamefont {Basko}},\ and\ \bibinfo {author}
  {\bibfnamefont {M.}~\bibnamefont {Potemski}},\ }\href
  {https://doi.org/10.1103/PhysRevLett.123.096803} {\bibfield  {journal}
  {\bibinfo  {journal} {Phys. Rev. Lett.}\ }\textbf {\bibinfo {volume} {123}},\
  \bibinfo {pages} {096803} (\bibinfo {year} {2019}{\natexlab{a}})}\BibitemShut
  {NoStop}%
\bibitem [{\citenamefont {Liu}\ \emph {et~al.}(2019{\natexlab{a}})\citenamefont
  {Liu}, \citenamefont {van Baren}, \citenamefont {Lu}, \citenamefont
  {Altaiary}, \citenamefont {Taniguchi}, \citenamefont {Watanabe},
  \citenamefont {Smirnov},\ and\ \citenamefont {Lui}}]{LiuGate2019}%
  \BibitemOpen
  \bibfield  {author} {\bibinfo {author} {\bibfnamefont {E.}~\bibnamefont
  {Liu}}, \bibinfo {author} {\bibfnamefont {J.}~\bibnamefont {van Baren}},
  \bibinfo {author} {\bibfnamefont {Z.}~\bibnamefont {Lu}}, \bibinfo {author}
  {\bibfnamefont {M.~M.}\ \bibnamefont {Altaiary}}, \bibinfo {author}
  {\bibfnamefont {T.}~\bibnamefont {Taniguchi}}, \bibinfo {author}
  {\bibfnamefont {K.}~\bibnamefont {Watanabe}}, \bibinfo {author}
  {\bibfnamefont {D.}~\bibnamefont {Smirnov}},\ and\ \bibinfo {author}
  {\bibfnamefont {C.~H.}\ \bibnamefont {Lui}},\ }\href
  {https://doi.org/10.1103/PhysRevLett.123.027401} {\bibfield  {journal}
  {\bibinfo  {journal} {Phys. Rev. Lett.}\ }\textbf {\bibinfo {volume} {123}},\
  \bibinfo {pages} {027401} (\bibinfo {year} {2019}{\natexlab{a}})}\BibitemShut
  {NoStop}%
\bibitem [{\citenamefont {Lu}\ \emph {et~al.}(2019)\citenamefont {Lu},
  \citenamefont {Rhodes}, \citenamefont {Li}, \citenamefont {Tuan},
  \citenamefont {Jiang}, \citenamefont {Ludwig}, \citenamefont {Jiang},
  \citenamefont {Lian}, \citenamefont {Shi}, \citenamefont {Hone},
  \citenamefont {Dery},\ and\ \citenamefont {Smirnov}}]{Lu_2020}%
  \BibitemOpen
  \bibfield  {author} {\bibinfo {author} {\bibfnamefont {Z.}~\bibnamefont
  {Lu}}, \bibinfo {author} {\bibfnamefont {D.}~\bibnamefont {Rhodes}}, \bibinfo
  {author} {\bibfnamefont {Z.}~\bibnamefont {Li}}, \bibinfo {author}
  {\bibfnamefont {D.~V.}\ \bibnamefont {Tuan}}, \bibinfo {author}
  {\bibfnamefont {Y.}~\bibnamefont {Jiang}}, \bibinfo {author} {\bibfnamefont
  {J.}~\bibnamefont {Ludwig}}, \bibinfo {author} {\bibfnamefont
  {Z.}~\bibnamefont {Jiang}}, \bibinfo {author} {\bibfnamefont
  {Z.}~\bibnamefont {Lian}}, \bibinfo {author} {\bibfnamefont {S.-F.}\
  \bibnamefont {Shi}}, \bibinfo {author} {\bibfnamefont {J.}~\bibnamefont
  {Hone}}, \bibinfo {author} {\bibfnamefont {H.}~\bibnamefont {Dery}},\ and\
  \bibinfo {author} {\bibfnamefont {D.}~\bibnamefont {Smirnov}},\ }\href
  {https://doi.org/10.1088/2053-1583/ab5614} {\bibfield  {journal} {\bibinfo
  {journal} {2D Materials}\ }\textbf {\bibinfo {volume} {7}},\ \bibinfo {pages}
  {015017} (\bibinfo {year} {2019})}\BibitemShut {NoStop}%
\bibitem [{\citenamefont {Arora}\ \emph {et~al.}(2020)\citenamefont {Arora},
  \citenamefont {Wessling}, \citenamefont {Deilmann}, \citenamefont
  {Reichenauer}, \citenamefont {Steeger}, \citenamefont {Kossacki},
  \citenamefont {Potemski}, \citenamefont {Michaelis~de Vasconcellos},
  \citenamefont {Rohlfing},\ and\ \citenamefont {Bratschitsch}}]{Arora2020}%
  \BibitemOpen
  \bibfield  {author} {\bibinfo {author} {\bibfnamefont {A.}~\bibnamefont
  {Arora}}, \bibinfo {author} {\bibfnamefont {N.~K.}\ \bibnamefont {Wessling}},
  \bibinfo {author} {\bibfnamefont {T.}~\bibnamefont {Deilmann}}, \bibinfo
  {author} {\bibfnamefont {T.}~\bibnamefont {Reichenauer}}, \bibinfo {author}
  {\bibfnamefont {P.}~\bibnamefont {Steeger}}, \bibinfo {author} {\bibfnamefont
  {P.}~\bibnamefont {Kossacki}}, \bibinfo {author} {\bibfnamefont
  {M.}~\bibnamefont {Potemski}}, \bibinfo {author} {\bibfnamefont
  {S.}~\bibnamefont {Michaelis~de Vasconcellos}}, \bibinfo {author}
  {\bibfnamefont {M.}~\bibnamefont {Rohlfing}},\ and\ \bibinfo {author}
  {\bibfnamefont {R.}~\bibnamefont {Bratschitsch}},\ }\href
  {https://doi.org/10.1103/PhysRevB.101.241413} {\bibfield  {journal} {\bibinfo
   {journal} {Phys. Rev. B}\ }\textbf {\bibinfo {volume} {101}},\ \bibinfo
  {pages} {241413} (\bibinfo {year} {2020})}\BibitemShut {NoStop}%
\bibitem [{\citenamefont {He}\ \emph {et~al.}(2020)\citenamefont {He},
  \citenamefont {Rivera}, \citenamefont {Van~Tuan}, \citenamefont {Wilson},
  \citenamefont {Yang}, \citenamefont {Taniguchi}, \citenamefont {Watanabe},
  \citenamefont {Yan}, \citenamefont {Mandrus}, \citenamefont {Yu},
  \citenamefont {Dery}, \citenamefont {Yao},\ and\ \citenamefont
  {Xu}}]{He2020}%
  \BibitemOpen
  \bibfield  {author} {\bibinfo {author} {\bibfnamefont {M.}~\bibnamefont
  {He}}, \bibinfo {author} {\bibfnamefont {P.}~\bibnamefont {Rivera}}, \bibinfo
  {author} {\bibfnamefont {D.}~\bibnamefont {Van~Tuan}}, \bibinfo {author}
  {\bibfnamefont {N.~P.}\ \bibnamefont {Wilson}}, \bibinfo {author}
  {\bibfnamefont {M.}~\bibnamefont {Yang}}, \bibinfo {author} {\bibfnamefont
  {T.}~\bibnamefont {Taniguchi}}, \bibinfo {author} {\bibfnamefont
  {K.}~\bibnamefont {Watanabe}}, \bibinfo {author} {\bibfnamefont
  {J.}~\bibnamefont {Yan}}, \bibinfo {author} {\bibfnamefont {D.~G.}\
  \bibnamefont {Mandrus}}, \bibinfo {author} {\bibfnamefont {H.}~\bibnamefont
  {Yu}}, \bibinfo {author} {\bibfnamefont {H.}~\bibnamefont {Dery}}, \bibinfo
  {author} {\bibfnamefont {W.}~\bibnamefont {Yao}},\ and\ \bibinfo {author}
  {\bibfnamefont {X.}~\bibnamefont {Xu}},\ }\href
  {https://doi.org/10.1038/s41467-020-14472-0} {\bibfield  {journal} {\bibinfo
  {journal} {Nature Communications}\ }\textbf {\bibinfo {volume} {11}},\
  \bibinfo {pages} {618} (\bibinfo {year} {2020})}\BibitemShut {NoStop}%
\bibitem [{\citenamefont {Robert}\ \emph {et~al.}(2020)\citenamefont {Robert},
  \citenamefont {Han}, \citenamefont {Kapuscinski}, \citenamefont {Delhomme},
  \citenamefont {Faugeras}, \citenamefont {Amand}, \citenamefont {Molas},
  \citenamefont {Bartos}, \citenamefont {Watanabe}, \citenamefont {Taniguchi},
  \citenamefont {Urbaszek}, \citenamefont {Potemski},\ and\ \citenamefont
  {Marie}}]{Robert2020}%
  \BibitemOpen
  \bibfield  {author} {\bibinfo {author} {\bibfnamefont {C.}~\bibnamefont
  {Robert}}, \bibinfo {author} {\bibfnamefont {B.}~\bibnamefont {Han}},
  \bibinfo {author} {\bibfnamefont {P.}~\bibnamefont {Kapuscinski}}, \bibinfo
  {author} {\bibfnamefont {A.}~\bibnamefont {Delhomme}}, \bibinfo {author}
  {\bibfnamefont {C.}~\bibnamefont {Faugeras}}, \bibinfo {author}
  {\bibfnamefont {T.}~\bibnamefont {Amand}}, \bibinfo {author} {\bibfnamefont
  {M.~R.}\ \bibnamefont {Molas}}, \bibinfo {author} {\bibfnamefont
  {M.}~\bibnamefont {Bartos}}, \bibinfo {author} {\bibfnamefont
  {K.}~\bibnamefont {Watanabe}}, \bibinfo {author} {\bibfnamefont
  {T.}~\bibnamefont {Taniguchi}}, \bibinfo {author} {\bibfnamefont
  {B.}~\bibnamefont {Urbaszek}}, \bibinfo {author} {\bibfnamefont
  {M.}~\bibnamefont {Potemski}},\ and\ \bibinfo {author} {\bibfnamefont
  {X.}~\bibnamefont {Marie}},\ }\href
  {https://doi.org/10.1038/s41467-020-17608-4} {\bibfield  {journal} {\bibinfo
  {journal} {Nature Communications}\ }\textbf {\bibinfo {volume} {11}},\
  \bibinfo {pages} {4037} (\bibinfo {year} {2020})}\BibitemShut {NoStop}%
\bibitem [{\citenamefont {Zinkiewicz}\ \emph {et~al.}(2020)\citenamefont
  {Zinkiewicz}, \citenamefont {Slobodeniuk}, \citenamefont {Kazimierczuk},
  \citenamefont {Kapuściński}, \citenamefont {Oreszczuk}, \citenamefont
  {Grzeszczyk}, \citenamefont {Bartos}, \citenamefont {Nogajewski},
  \citenamefont {Watanabe}, \citenamefont {Taniguchi}, \citenamefont
  {Faugeras}, \citenamefont {Kossacki}, \citenamefont {Potemski}, \citenamefont
  {Babiński},\ and\ \citenamefont {Molas}}]{Zinkiewicz2020}%
  \BibitemOpen
  \bibfield  {author} {\bibinfo {author} {\bibfnamefont {M.}~\bibnamefont
  {Zinkiewicz}}, \bibinfo {author} {\bibfnamefont {A.~O.}\ \bibnamefont
  {Slobodeniuk}}, \bibinfo {author} {\bibfnamefont {T.}~\bibnamefont
  {Kazimierczuk}}, \bibinfo {author} {\bibfnamefont {P.}~\bibnamefont
  {Kapuściński}}, \bibinfo {author} {\bibfnamefont {K.}~\bibnamefont
  {Oreszczuk}}, \bibinfo {author} {\bibfnamefont {M.}~\bibnamefont
  {Grzeszczyk}}, \bibinfo {author} {\bibfnamefont {M.}~\bibnamefont {Bartos}},
  \bibinfo {author} {\bibfnamefont {K.}~\bibnamefont {Nogajewski}}, \bibinfo
  {author} {\bibfnamefont {K.}~\bibnamefont {Watanabe}}, \bibinfo {author}
  {\bibfnamefont {T.}~\bibnamefont {Taniguchi}}, \bibinfo {author}
  {\bibfnamefont {C.}~\bibnamefont {Faugeras}}, \bibinfo {author}
  {\bibfnamefont {P.}~\bibnamefont {Kossacki}}, \bibinfo {author}
  {\bibfnamefont {M.}~\bibnamefont {Potemski}}, \bibinfo {author}
  {\bibfnamefont {A.}~\bibnamefont {Babiński}},\ and\ \bibinfo {author}
  {\bibfnamefont {M.~R.}\ \bibnamefont {Molas}},\ }\href
  {https://doi.org/10.1039/D0NR04243A} {\bibfield  {journal} {\bibinfo
  {journal} {Nanoscale}\ }\textbf {\bibinfo {volume} {12}},\ \bibinfo {pages}
  {18153} (\bibinfo {year} {2020})}\BibitemShut {NoStop}%
\bibitem [{\citenamefont {Zinkiewicz}\ \emph {et~al.}(2021)\citenamefont
  {Zinkiewicz}, \citenamefont {Woźniak}, \citenamefont {Kazimierczuk},
  \citenamefont {Kapuscinski}, \citenamefont {Oreszczuk}, \citenamefont
  {Grzeszczyk}, \citenamefont {Bartoš}, \citenamefont {Nogajewski},
  \citenamefont {Watanabe}, \citenamefont {Taniguchi}, \citenamefont
  {Faugeras}, \citenamefont {Kossacki}, \citenamefont {Potemski}, \citenamefont
  {Babiński},\ and\ \citenamefont {Molas}}]{Zinkiewicz2021}%
  \BibitemOpen
  \bibfield  {author} {\bibinfo {author} {\bibfnamefont {M.}~\bibnamefont
  {Zinkiewicz}}, \bibinfo {author} {\bibfnamefont {T.}~\bibnamefont
  {Woźniak}}, \bibinfo {author} {\bibfnamefont {T.}~\bibnamefont
  {Kazimierczuk}}, \bibinfo {author} {\bibfnamefont {P.}~\bibnamefont
  {Kapuscinski}}, \bibinfo {author} {\bibfnamefont {K.}~\bibnamefont
  {Oreszczuk}}, \bibinfo {author} {\bibfnamefont {M.}~\bibnamefont
  {Grzeszczyk}}, \bibinfo {author} {\bibfnamefont {M.}~\bibnamefont {Bartoš}},
  \bibinfo {author} {\bibfnamefont {K.}~\bibnamefont {Nogajewski}}, \bibinfo
  {author} {\bibfnamefont {K.}~\bibnamefont {Watanabe}}, \bibinfo {author}
  {\bibfnamefont {T.}~\bibnamefont {Taniguchi}}, \bibinfo {author}
  {\bibfnamefont {C.}~\bibnamefont {Faugeras}}, \bibinfo {author}
  {\bibfnamefont {P.}~\bibnamefont {Kossacki}}, \bibinfo {author}
  {\bibfnamefont {M.}~\bibnamefont {Potemski}}, \bibinfo {author}
  {\bibfnamefont {A.}~\bibnamefont {Babiński}},\ and\ \bibinfo {author}
  {\bibfnamefont {M.~R.}\ \bibnamefont {Molas}},\ }\href
  {https://doi.org/10.1021/acs.nanolett.0c05021} {\bibfield  {journal}
  {\bibinfo  {journal} {Nano Letters}\ }\textbf {\bibinfo {volume} {21}},\
  \bibinfo {pages} {2519} (\bibinfo {year} {2021})}\BibitemShut {NoStop}%
\bibitem [{\citenamefont {Kapu{\'{s}}ci{\'{n}}ski}\ \emph
  {et~al.}(2021)\citenamefont {Kapu{\'{s}}ci{\'{n}}ski}, \citenamefont
  {Delhomme}, \citenamefont {Vaclavkova}, \citenamefont {Slobodeniuk},
  \citenamefont {Grzeszczyk}, \citenamefont {Bartos}, \citenamefont {Watanabe},
  \citenamefont {Taniguchi}, \citenamefont {Faugeras},\ and\ \citenamefont
  {Potemski}}]{Kapuscinski2021}%
  \BibitemOpen
  \bibfield  {author} {\bibinfo {author} {\bibfnamefont {P.}~\bibnamefont
  {Kapu{\'{s}}ci{\'{n}}ski}}, \bibinfo {author} {\bibfnamefont
  {A.}~\bibnamefont {Delhomme}}, \bibinfo {author} {\bibfnamefont
  {D.}~\bibnamefont {Vaclavkova}}, \bibinfo {author} {\bibfnamefont {A.~O.}\
  \bibnamefont {Slobodeniuk}}, \bibinfo {author} {\bibfnamefont
  {M.}~\bibnamefont {Grzeszczyk}}, \bibinfo {author} {\bibfnamefont
  {M.}~\bibnamefont {Bartos}}, \bibinfo {author} {\bibfnamefont
  {K.}~\bibnamefont {Watanabe}}, \bibinfo {author} {\bibfnamefont
  {T.}~\bibnamefont {Taniguchi}}, \bibinfo {author} {\bibfnamefont
  {C.}~\bibnamefont {Faugeras}},\ and\ \bibinfo {author} {\bibfnamefont
  {M.}~\bibnamefont {Potemski}},\ }\href
  {https://doi.org/10.1038/s42005-021-00692-3} {\bibfield  {journal} {\bibinfo
  {journal} {Communications Physics}\ }\textbf {\bibinfo {volume} {4}},\
  \bibinfo {pages} {186} (\bibinfo {year} {2021})}\BibitemShut {NoStop}%
\bibitem [{\citenamefont {Łucja Kipczak}\ \emph {et~al.}(2024)\citenamefont
  {Łucja Kipczak}, \citenamefont {Zawadzka}, \citenamefont {Jana},
  \citenamefont {Antoniazzi}, \citenamefont {Grzeszczyk}, \citenamefont
  {Zinkiewicz}, \citenamefont {Watanabe}, \citenamefont {Taniguchi},
  \citenamefont {Potemski}, \citenamefont {Faugeras}, \citenamefont
  {Babiński},\ and\ \citenamefont {Molas}}]{Kipczak2024}%
  \BibitemOpen
  \bibfield  {author} {\bibinfo {author} {\bibnamefont {Łucja Kipczak}},
  \bibinfo {author} {\bibfnamefont {N.}~\bibnamefont {Zawadzka}}, \bibinfo
  {author} {\bibfnamefont {D.}~\bibnamefont {Jana}}, \bibinfo {author}
  {\bibfnamefont {I.}~\bibnamefont {Antoniazzi}}, \bibinfo {author}
  {\bibfnamefont {M.}~\bibnamefont {Grzeszczyk}}, \bibinfo {author}
  {\bibfnamefont {M.}~\bibnamefont {Zinkiewicz}}, \bibinfo {author}
  {\bibfnamefont {K.}~\bibnamefont {Watanabe}}, \bibinfo {author}
  {\bibfnamefont {T.}~\bibnamefont {Taniguchi}}, \bibinfo {author}
  {\bibfnamefont {M.}~\bibnamefont {Potemski}}, \bibinfo {author}
  {\bibfnamefont {C.}~\bibnamefont {Faugeras}}, \bibinfo {author}
  {\bibfnamefont {A.}~\bibnamefont {Babiński}},\ and\ \bibinfo {author}
  {\bibfnamefont {M.~R.}\ \bibnamefont {Molas}},\ }\bibfield  {journal}
  {\bibinfo  {journal} {Nanophotonics}\ }\href
  {https://doi.org/doi:10.1515/nanoph-2024-0385} {doi:10.1515/nanoph-2024-0385}
  (\bibinfo {year} {2024})\BibitemShut {NoStop}%
\bibitem [{\citenamefont {Zhao}\ \emph {et~al.}(2018)\citenamefont {Zhao},
  \citenamefont {Zhang},\ and\ \citenamefont {Ouyang}}]{Zhao2018}%
  \BibitemOpen
  \bibfield  {author} {\bibinfo {author} {\bibfnamefont {Y.}~\bibnamefont
  {Zhao}}, \bibinfo {author} {\bibfnamefont {Z.}~\bibnamefont {Zhang}},\ and\
  \bibinfo {author} {\bibfnamefont {G.}~\bibnamefont {Ouyang}},\ }\href
  {https://doi.org/10.1007/s00339-018-1730-2} {\bibfield  {journal} {\bibinfo
  {journal} {Applied Physics A}\ }\textbf {\bibinfo {volume} {124}},\ \bibinfo
  {pages} {1432} (\bibinfo {year} {2018})}\BibitemShut {NoStop}%
\bibitem [{\citenamefont {Xie}(2015)}]{Xie2015}%
  \BibitemOpen
  \bibfield  {author} {\bibinfo {author} {\bibfnamefont {L.~M.}\ \bibnamefont
  {Xie}},\ }\href {https://doi.org/10.1039/C5NR05712D} {\bibfield  {journal}
  {\bibinfo  {journal} {Nanoscale}\ }\textbf {\bibinfo {volume} {7}},\ \bibinfo
  {pages} {18392} (\bibinfo {year} {2015})}\BibitemShut {NoStop}%
\bibitem [{\citenamefont {Ma}\ \emph {et~al.}(2018)\citenamefont {Ma},
  \citenamefont {Wu}, \citenamefont {Wang},\ and\ \citenamefont
  {Wang}}]{Yucheng2018}%
  \BibitemOpen
  \bibfield  {author} {\bibinfo {author} {\bibfnamefont {X.}~\bibnamefont
  {Ma}}, \bibinfo {author} {\bibfnamefont {X.}~\bibnamefont {Wu}}, \bibinfo
  {author} {\bibfnamefont {H.}~\bibnamefont {Wang}},\ and\ \bibinfo {author}
  {\bibfnamefont {Y.}~\bibnamefont {Wang}},\ }\href
  {https://doi.org/10.1039/C7TA10015A} {\bibfield  {journal} {\bibinfo
  {journal} {J. Mater. Chem. A}\ }\textbf {\bibinfo {volume} {6}},\ \bibinfo
  {pages} {2295} (\bibinfo {year} {2018})}\BibitemShut {NoStop}%
\bibitem [{\citenamefont {Jing}\ \emph {et~al.}(2020)\citenamefont {Jing},
  \citenamefont {Liu}, \citenamefont {Zhu}, \citenamefont {Ouyang},
  \citenamefont {Sun},\ and\ \citenamefont {Zhou}}]{Yumei2020}%
  \BibitemOpen
  \bibfield  {author} {\bibinfo {author} {\bibfnamefont {Y.}~\bibnamefont
  {Jing}}, \bibinfo {author} {\bibfnamefont {B.}~\bibnamefont {Liu}}, \bibinfo
  {author} {\bibfnamefont {X.}~\bibnamefont {Zhu}}, \bibinfo {author}
  {\bibfnamefont {F.}~\bibnamefont {Ouyang}}, \bibinfo {author} {\bibfnamefont
  {J.}~\bibnamefont {Sun}},\ and\ \bibinfo {author} {\bibfnamefont
  {Y.}~\bibnamefont {Zhou}},\ }\href
  {https://doi.org/doi:10.1515/nanoph-2019-0574} {\bibfield  {journal}
  {\bibinfo  {journal} {Nanophotonics}\ }\textbf {\bibinfo {volume} {9}},\
  \bibinfo {pages} {1675} (\bibinfo {year} {2020})}\BibitemShut {NoStop}%
\bibitem [{\citenamefont {Bikerouin}\ and\ \citenamefont
  {Balli}(2022)}]{BIKEROUIN2022}%
  \BibitemOpen
  \bibfield  {author} {\bibinfo {author} {\bibfnamefont {M.}~\bibnamefont
  {Bikerouin}}\ and\ \bibinfo {author} {\bibfnamefont {M.}~\bibnamefont
  {Balli}},\ }\href
  {https://doi.org/https://doi.org/10.1016/j.apsusc.2022.153835} {\bibfield
  {journal} {\bibinfo  {journal} {Applied Surface Science}\ }\textbf {\bibinfo
  {volume} {598}},\ \bibinfo {pages} {153835} (\bibinfo {year}
  {2022})}\BibitemShut {NoStop}%
\bibitem [{\citenamefont {Waheed}\ \emph {et~al.}(2022)\citenamefont {Waheed},
  \citenamefont {Ullah}, \citenamefont {{Waqas Iqbal}},\ and\ \citenamefont
  {Shin}}]{Sumaira2022}%
  \BibitemOpen
  \bibfield  {author} {\bibinfo {author} {\bibfnamefont {H.~S.}\ \bibnamefont
  {Waheed}}, \bibinfo {author} {\bibfnamefont {H.}~\bibnamefont {Ullah}},
  \bibinfo {author} {\bibfnamefont {M.}~\bibnamefont {{Waqas Iqbal}}},\ and\
  \bibinfo {author} {\bibfnamefont {Y.-H.}\ \bibnamefont {Shin}},\ }\href
  {https://doi.org/https://doi.org/10.1016/j.ijleo.2022.170071} {\bibfield
  {journal} {\bibinfo  {journal} {Optik}\ }\textbf {\bibinfo {volume} {271}},\
  \bibinfo {pages} {170071} (\bibinfo {year} {2022})}\BibitemShut {NoStop}%
\bibitem [{\citenamefont {Wu}\ \emph {et~al.}(2021)\citenamefont {Wu},
  \citenamefont {Cong}, \citenamefont {Yang}, \citenamefont {Chen},
  \citenamefont {Shao}, \citenamefont {Do}, \citenamefont {Wen}, \citenamefont
  {Feng}, \citenamefont {Zou}, \citenamefont {Zhang}, \citenamefont {Du},
  \citenamefont {Cao}, \citenamefont {Shang}, \citenamefont {Xiong},
  \citenamefont {Loh},\ and\ \citenamefont {Yu}}]{Wu2021}%
  \BibitemOpen
  \bibfield  {author} {\bibinfo {author} {\bibfnamefont {L.}~\bibnamefont
  {Wu}}, \bibinfo {author} {\bibfnamefont {C.}~\bibnamefont {Cong}}, \bibinfo
  {author} {\bibfnamefont {W.}~\bibnamefont {Yang}}, \bibinfo {author}
  {\bibfnamefont {Y.}~\bibnamefont {Chen}}, \bibinfo {author} {\bibfnamefont
  {Y.}~\bibnamefont {Shao}}, \bibinfo {author} {\bibfnamefont {T.~T.~H.}\
  \bibnamefont {Do}}, \bibinfo {author} {\bibfnamefont {W.}~\bibnamefont
  {Wen}}, \bibinfo {author} {\bibfnamefont {S.}~\bibnamefont {Feng}}, \bibinfo
  {author} {\bibfnamefont {C.}~\bibnamefont {Zou}}, \bibinfo {author}
  {\bibfnamefont {H.}~\bibnamefont {Zhang}}, \bibinfo {author} {\bibfnamefont
  {B.}~\bibnamefont {Du}}, \bibinfo {author} {\bibfnamefont {B.}~\bibnamefont
  {Cao}}, \bibinfo {author} {\bibfnamefont {J.}~\bibnamefont {Shang}}, \bibinfo
  {author} {\bibfnamefont {Q.}~\bibnamefont {Xiong}}, \bibinfo {author}
  {\bibfnamefont {K.~P.}\ \bibnamefont {Loh}},\ and\ \bibinfo {author}
  {\bibfnamefont {T.}~\bibnamefont {Yu}},\ }\href
  {https://doi.org/10.1021/acsnano.0c10478} {\bibfield  {journal} {\bibinfo
  {journal} {ACS Nano}\ }\textbf {\bibinfo {volume} {15}},\ \bibinfo {pages}
  {8397} (\bibinfo {year} {2021})}\BibitemShut {NoStop}%
\bibitem [{\citenamefont {Schaibley}\ \emph {et~al.}(2016)\citenamefont
  {Schaibley}, \citenamefont {Yu}, \citenamefont {Clark}, \citenamefont
  {Rivera}, \citenamefont {Ross}, \citenamefont {Seyler}, \citenamefont {Yao},\
  and\ \citenamefont {Xu}}]{Schaibley2016}%
  \BibitemOpen
  \bibfield  {author} {\bibinfo {author} {\bibfnamefont {J.~R.}\ \bibnamefont
  {Schaibley}}, \bibinfo {author} {\bibfnamefont {H.}~\bibnamefont {Yu}},
  \bibinfo {author} {\bibfnamefont {G.}~\bibnamefont {Clark}}, \bibinfo
  {author} {\bibfnamefont {P.}~\bibnamefont {Rivera}}, \bibinfo {author}
  {\bibfnamefont {J.~S.}\ \bibnamefont {Ross}}, \bibinfo {author}
  {\bibfnamefont {K.~L.}\ \bibnamefont {Seyler}}, \bibinfo {author}
  {\bibfnamefont {W.}~\bibnamefont {Yao}},\ and\ \bibinfo {author}
  {\bibfnamefont {X.}~\bibnamefont {Xu}},\ }\href
  {https://doi.org/10.1038/natrevmats.2016.55} {\bibfield  {journal} {\bibinfo
  {journal} {Nature Reviews Materials}\ }\textbf {\bibinfo {volume} {1}},\
  \bibinfo {pages} {16055} (\bibinfo {year} {2016})}\BibitemShut {NoStop}%
\bibitem [{\citenamefont {Mak}\ \emph {et~al.}(2018)\citenamefont {Mak},
  \citenamefont {Xiao},\ and\ \citenamefont {Shan}}]{Mak2018}%
  \BibitemOpen
  \bibfield  {author} {\bibinfo {author} {\bibfnamefont {K.~F.}\ \bibnamefont
  {Mak}}, \bibinfo {author} {\bibfnamefont {D.}~\bibnamefont {Xiao}},\ and\
  \bibinfo {author} {\bibfnamefont {J.}~\bibnamefont {Shan}},\ }\href
  {https://doi.org/10.1038/s41566-018-0204-6} {\bibfield  {journal} {\bibinfo
  {journal} {Nature Photonics}\ }\textbf {\bibinfo {volume} {12}},\ \bibinfo
  {pages} {451} (\bibinfo {year} {2018})}\BibitemShut {NoStop}%
\bibitem [{\citenamefont {Pal}\ \emph {et~al.}(2023)\citenamefont {Pal},
  \citenamefont {Zhang}, \citenamefont {Chavan}, \citenamefont {Agashiwala},
  \citenamefont {Yeh}, \citenamefont {Cao},\ and\ \citenamefont
  {Banerjee}}]{Pal2023}%
  \BibitemOpen
  \bibfield  {author} {\bibinfo {author} {\bibfnamefont {A.}~\bibnamefont
  {Pal}}, \bibinfo {author} {\bibfnamefont {S.}~\bibnamefont {Zhang}}, \bibinfo
  {author} {\bibfnamefont {T.}~\bibnamefont {Chavan}}, \bibinfo {author}
  {\bibfnamefont {K.}~\bibnamefont {Agashiwala}}, \bibinfo {author}
  {\bibfnamefont {C.-H.}\ \bibnamefont {Yeh}}, \bibinfo {author} {\bibfnamefont
  {W.}~\bibnamefont {Cao}},\ and\ \bibinfo {author} {\bibfnamefont
  {K.}~\bibnamefont {Banerjee}},\ }\href
  {https://doi.org/https://doi.org/10.1002/adma.202109894} {\bibfield
  {journal} {\bibinfo  {journal} {Advanced Materials}\ }\textbf {\bibinfo
  {volume} {35}},\ \bibinfo {pages} {2109894} (\bibinfo {year}
  {2023})}\BibitemShut {NoStop}%
\bibitem [{\citenamefont {Vitale}\ \emph {et~al.}(2018)\citenamefont {Vitale},
  \citenamefont {Nezich}, \citenamefont {Varghese}, \citenamefont {Kim},
  \citenamefont {Gedik}, \citenamefont {Jarillo-Herrero}, \citenamefont
  {Xiao},\ and\ \citenamefont {Rothschild}}]{Vitale2018}%
  \BibitemOpen
  \bibfield  {author} {\bibinfo {author} {\bibfnamefont {S.~A.}\ \bibnamefont
  {Vitale}}, \bibinfo {author} {\bibfnamefont {D.}~\bibnamefont {Nezich}},
  \bibinfo {author} {\bibfnamefont {J.~O.}\ \bibnamefont {Varghese}}, \bibinfo
  {author} {\bibfnamefont {P.}~\bibnamefont {Kim}}, \bibinfo {author}
  {\bibfnamefont {N.}~\bibnamefont {Gedik}}, \bibinfo {author} {\bibfnamefont
  {P.}~\bibnamefont {Jarillo-Herrero}}, \bibinfo {author} {\bibfnamefont
  {D.}~\bibnamefont {Xiao}},\ and\ \bibinfo {author} {\bibfnamefont
  {M.}~\bibnamefont {Rothschild}},\ }\href
  {https://doi.org/https://doi.org/10.1002/smll.201801483} {\bibfield
  {journal} {\bibinfo  {journal} {Small}\ }\textbf {\bibinfo {volume} {14}},\
  \bibinfo {pages} {1801483} (\bibinfo {year} {2018})}\BibitemShut {NoStop}%
\bibitem [{\citenamefont {Blundo}\ and\ \citenamefont
  {Polimeni}(2024)}]{alice_and_bob}%
  \BibitemOpen
  \bibfield  {author} {\bibinfo {author} {\bibfnamefont {E.}~\bibnamefont
  {Blundo}}\ and\ \bibinfo {author} {\bibfnamefont {A.}~\bibnamefont
  {Polimeni}},\ }\href
  {https://doi.org/https://doi.org/10.1021/acs.nanolett.4c02702} {\bibfield
  {journal} {\bibinfo  {journal} {Nano Lett.}\ }\textbf {\bibinfo {volume}
  {24}},\ \bibinfo {pages} {9777} (\bibinfo {year} {2024})}\BibitemShut
  {NoStop}%
\bibitem [{\citenamefont {Kormányos}\ \emph {et~al.}(2015)\citenamefont
  {Kormányos}, \citenamefont {Burkard}, \citenamefont {Gmitra}, \citenamefont
  {Fabian}, \citenamefont {Zólyomi}, \citenamefont {Drummond},\ and\
  \citenamefont {Fal’ko}}]{Kormányos_2015}%
  \BibitemOpen
  \bibfield  {author} {\bibinfo {author} {\bibfnamefont {A.}~\bibnamefont
  {Kormányos}}, \bibinfo {author} {\bibfnamefont {G.}~\bibnamefont {Burkard}},
  \bibinfo {author} {\bibfnamefont {M.}~\bibnamefont {Gmitra}}, \bibinfo
  {author} {\bibfnamefont {J.}~\bibnamefont {Fabian}}, \bibinfo {author}
  {\bibfnamefont {V.}~\bibnamefont {Zólyomi}}, \bibinfo {author}
  {\bibfnamefont {N.~D.}\ \bibnamefont {Drummond}},\ and\ \bibinfo {author}
  {\bibfnamefont {V.}~\bibnamefont {Fal’ko}},\ }\href
  {https://doi.org/10.1088/2053-1583/2/2/022001} {\bibfield  {journal}
  {\bibinfo  {journal} {2D Materials}\ }\textbf {\bibinfo {volume} {2}},\
  \bibinfo {pages} {022001} (\bibinfo {year} {2015})}\BibitemShut {NoStop}%
\bibitem [{\citenamefont {Wang}\ \emph {et~al.}(2018)\citenamefont {Wang},
  \citenamefont {Chernikov}, \citenamefont {Glazov}, \citenamefont {Heinz},
  \citenamefont {Marie}, \citenamefont {Amand},\ and\ \citenamefont
  {Urbaszek}}]{Wang2018}%
  \BibitemOpen
  \bibfield  {author} {\bibinfo {author} {\bibfnamefont {G.}~\bibnamefont
  {Wang}}, \bibinfo {author} {\bibfnamefont {A.}~\bibnamefont {Chernikov}},
  \bibinfo {author} {\bibfnamefont {M.~M.}\ \bibnamefont {Glazov}}, \bibinfo
  {author} {\bibfnamefont {T.~F.}\ \bibnamefont {Heinz}}, \bibinfo {author}
  {\bibfnamefont {X.}~\bibnamefont {Marie}}, \bibinfo {author} {\bibfnamefont
  {T.}~\bibnamefont {Amand}},\ and\ \bibinfo {author} {\bibfnamefont
  {B.}~\bibnamefont {Urbaszek}},\ }\href
  {https://doi.org/10.1103/RevModPhys.90.021001} {\bibfield  {journal}
  {\bibinfo  {journal} {Rev. Mod. Phys.}\ }\textbf {\bibinfo {volume} {90}},\
  \bibinfo {pages} {021001} (\bibinfo {year} {2018})}\BibitemShut {NoStop}%
\bibitem [{\citenamefont {Koperski}\ \emph {et~al.}(2019)\citenamefont
  {Koperski}, \citenamefont {Molas}, \citenamefont {Arora}, \citenamefont
  {Nogajewski}, \citenamefont {Bartos}, \citenamefont {Wyzula}, \citenamefont
  {Vaclavkova}, \citenamefont {Kossacki},\ and\ \citenamefont
  {Potemski}}]{Koperski2019}%
  \BibitemOpen
  \bibfield  {author} {\bibinfo {author} {\bibfnamefont {M.}~\bibnamefont
  {Koperski}}, \bibinfo {author} {\bibfnamefont {M.~R.}\ \bibnamefont {Molas}},
  \bibinfo {author} {\bibfnamefont {A.}~\bibnamefont {Arora}}, \bibinfo
  {author} {\bibfnamefont {K.}~\bibnamefont {Nogajewski}}, \bibinfo {author}
  {\bibfnamefont {M.}~\bibnamefont {Bartos}}, \bibinfo {author} {\bibfnamefont
  {J.}~\bibnamefont {Wyzula}}, \bibinfo {author} {\bibfnamefont
  {D.}~\bibnamefont {Vaclavkova}}, \bibinfo {author} {\bibfnamefont
  {P.}~\bibnamefont {Kossacki}},\ and\ \bibinfo {author} {\bibfnamefont
  {M.}~\bibnamefont {Potemski}},\ }\href
  {https://doi.org/10.1088/2053-1583/aae14b} {\bibfield  {journal} {\bibinfo
  {journal} {2D Materials}\ }\textbf {\bibinfo {volume} {6}},\ \bibinfo {pages}
  {015001} (\bibinfo {year} {2019})}\BibitemShut {NoStop}%
\bibitem [{\citenamefont {Wo\ifmmode~\acute{z}\else \'{z}\fi{}niak}\ \emph
  {et~al.}(2020)\citenamefont {Wo\ifmmode~\acute{z}\else \'{z}\fi{}niak},
  \citenamefont {{Faria Junior}}, \citenamefont {Seifert}, \citenamefont
  {Chaves},\ and\ \citenamefont {Kunstmann}}]{Wozniak2020}%
  \BibitemOpen
  \bibfield  {author} {\bibinfo {author} {\bibfnamefont {T.}~\bibnamefont
  {Wo\ifmmode~\acute{z}\else \'{z}\fi{}niak}}, \bibinfo {author} {\bibfnamefont
  {P.~E.}\ \bibnamefont {{Faria Junior}}}, \bibinfo {author} {\bibfnamefont
  {G.}~\bibnamefont {Seifert}}, \bibinfo {author} {\bibfnamefont
  {A.}~\bibnamefont {Chaves}},\ and\ \bibinfo {author} {\bibfnamefont
  {J.}~\bibnamefont {Kunstmann}},\ }\href
  {https://doi.org/10.1103/PhysRevB.101.235408} {\bibfield  {journal} {\bibinfo
   {journal} {Phys. Rev. B}\ }\textbf {\bibinfo {volume} {101}},\ \bibinfo
  {pages} {235408} (\bibinfo {year} {2020})}\BibitemShut {NoStop}%
\bibitem [{\citenamefont {Ivchenko}(2005)}]{Ivchenko2005}%
  \BibitemOpen
  \bibfield  {author} {\bibinfo {author} {\bibfnamefont {E.~L.}\ \bibnamefont
  {Ivchenko}},\ }\href@noop {} {\emph {\bibinfo {title} {Optical Spectroscopy
  of Semiconductor Nanostructures}}}\ (\bibinfo  {publisher} {Alpha Science
  International},\ \bibinfo {address} {Harrow, UK},\ \bibinfo {year}
  {2005})\BibitemShut {NoStop}%
\bibitem [{\citenamefont {Stier}\ \emph {et~al.}(2018)\citenamefont {Stier},
  \citenamefont {Wilson}, \citenamefont {Velizhanin}, \citenamefont {Kono},
  \citenamefont {Xu},\ and\ \citenamefont {Crooker}}]{Stier2018}%
  \BibitemOpen
  \bibfield  {author} {\bibinfo {author} {\bibfnamefont {A.~V.}\ \bibnamefont
  {Stier}}, \bibinfo {author} {\bibfnamefont {N.~P.}\ \bibnamefont {Wilson}},
  \bibinfo {author} {\bibfnamefont {K.~A.}\ \bibnamefont {Velizhanin}},
  \bibinfo {author} {\bibfnamefont {J.}~\bibnamefont {Kono}}, \bibinfo {author}
  {\bibfnamefont {X.}~\bibnamefont {Xu}},\ and\ \bibinfo {author}
  {\bibfnamefont {S.~A.}\ \bibnamefont {Crooker}},\ }\href
  {https://doi.org/10.1103/PhysRevLett.120.057405} {\bibfield  {journal}
  {\bibinfo  {journal} {Phys. Rev. Lett.}\ }\textbf {\bibinfo {volume} {120}},\
  \bibinfo {pages} {057405} (\bibinfo {year} {2018})}\BibitemShut {NoStop}%
\bibitem [{\citenamefont {Goryca}\ \emph {et~al.}(2019)\citenamefont {Goryca},
  \citenamefont {Li}, \citenamefont {Stier}, \citenamefont {Taniguchi},
  \citenamefont {Watanabe}, \citenamefont {Courtade}, \citenamefont {Shree},
  \citenamefont {Robert}, \citenamefont {Urbaszek}, \citenamefont {Marie},\
  and\ \citenamefont {Crooker}}]{Goryca2019}%
  \BibitemOpen
  \bibfield  {author} {\bibinfo {author} {\bibfnamefont {M.}~\bibnamefont
  {Goryca}}, \bibinfo {author} {\bibfnamefont {J.}~\bibnamefont {Li}}, \bibinfo
  {author} {\bibfnamefont {A.~V.}\ \bibnamefont {Stier}}, \bibinfo {author}
  {\bibfnamefont {T.}~\bibnamefont {Taniguchi}}, \bibinfo {author}
  {\bibfnamefont {K.}~\bibnamefont {Watanabe}}, \bibinfo {author}
  {\bibfnamefont {E.}~\bibnamefont {Courtade}}, \bibinfo {author}
  {\bibfnamefont {S.}~\bibnamefont {Shree}}, \bibinfo {author} {\bibfnamefont
  {C.}~\bibnamefont {Robert}}, \bibinfo {author} {\bibfnamefont
  {B.}~\bibnamefont {Urbaszek}}, \bibinfo {author} {\bibfnamefont
  {X.}~\bibnamefont {Marie}},\ and\ \bibinfo {author} {\bibfnamefont {S.~A.}\
  \bibnamefont {Crooker}},\ }\href {https://doi.org/10.1038/s41467-019-12180-y}
  {\bibfield  {journal} {\bibinfo  {journal} {Nature Communications}\ }\textbf
  {\bibinfo {volume} {10}},\ \bibinfo {pages} {4172} (\bibinfo {year}
  {2019})}\BibitemShut {NoStop}%
\bibitem [{\citenamefont {F{\"o}rste}\ \emph
  {et~al.}(2020{\natexlab{a}})\citenamefont {F{\"o}rste}, \citenamefont
  {Tepliakov}, \citenamefont {Kruchinin}, \citenamefont {Lindlau},
  \citenamefont {Funk}, \citenamefont {F{\"o}rg}, \citenamefont {Watanabe},
  \citenamefont {Taniguchi}, \citenamefont {Baimuratov},\ and\ \citenamefont
  {H{\"o}gele}}]{Förste2020}%
  \BibitemOpen
  \bibfield  {author} {\bibinfo {author} {\bibfnamefont {J.}~\bibnamefont
  {F{\"o}rste}}, \bibinfo {author} {\bibfnamefont {N.~V.}\ \bibnamefont
  {Tepliakov}}, \bibinfo {author} {\bibfnamefont {S.~Y.}\ \bibnamefont
  {Kruchinin}}, \bibinfo {author} {\bibfnamefont {J.}~\bibnamefont {Lindlau}},
  \bibinfo {author} {\bibfnamefont {V.}~\bibnamefont {Funk}}, \bibinfo {author}
  {\bibfnamefont {M.}~\bibnamefont {F{\"o}rg}}, \bibinfo {author}
  {\bibfnamefont {K.}~\bibnamefont {Watanabe}}, \bibinfo {author}
  {\bibfnamefont {T.}~\bibnamefont {Taniguchi}}, \bibinfo {author}
  {\bibfnamefont {A.~S.}\ \bibnamefont {Baimuratov}},\ and\ \bibinfo {author}
  {\bibfnamefont {A.}~\bibnamefont {H{\"o}gele}},\ }\href
  {https://doi.org/10.1038/s41467-020-18019-1} {\bibfield  {journal} {\bibinfo
  {journal} {Nature Communications}\ }\textbf {\bibinfo {volume} {11}},\
  \bibinfo {pages} {4539} (\bibinfo {year} {2020}{\natexlab{a}})}\BibitemShut
  {NoStop}%
\bibitem [{\citenamefont {Jadczak}\ \emph {et~al.}(2021)\citenamefont
  {Jadczak}, \citenamefont {Kutrowska-Girzycka}, \citenamefont {Bieniek},
  \citenamefont {Kazimierczuk}, \citenamefont {Kossacki}, \citenamefont
  {Schindler}, \citenamefont {Debus}, \citenamefont {Watanabe}, \citenamefont
  {Taniguchi}, \citenamefont {Ho}, \citenamefont {Wójs}, \citenamefont
  {Hawrylak},\ and\ \citenamefont {Bryja}}]{Jadczak_2021}%
  \BibitemOpen
  \bibfield  {author} {\bibinfo {author} {\bibfnamefont {J.}~\bibnamefont
  {Jadczak}}, \bibinfo {author} {\bibfnamefont {J.}~\bibnamefont
  {Kutrowska-Girzycka}}, \bibinfo {author} {\bibfnamefont {M.}~\bibnamefont
  {Bieniek}}, \bibinfo {author} {\bibfnamefont {T.}~\bibnamefont
  {Kazimierczuk}}, \bibinfo {author} {\bibfnamefont {P.}~\bibnamefont
  {Kossacki}}, \bibinfo {author} {\bibfnamefont {J.~J.}\ \bibnamefont
  {Schindler}}, \bibinfo {author} {\bibfnamefont {J.}~\bibnamefont {Debus}},
  \bibinfo {author} {\bibfnamefont {K.}~\bibnamefont {Watanabe}}, \bibinfo
  {author} {\bibfnamefont {T.}~\bibnamefont {Taniguchi}}, \bibinfo {author}
  {\bibfnamefont {C.~H.}\ \bibnamefont {Ho}}, \bibinfo {author} {\bibfnamefont
  {A.}~\bibnamefont {Wójs}}, \bibinfo {author} {\bibfnamefont
  {P.}~\bibnamefont {Hawrylak}},\ and\ \bibinfo {author} {\bibfnamefont
  {L.}~\bibnamefont {Bryja}},\ }\href
  {https://doi.org/10.1088/1361-6528/abd507} {\bibfield  {journal} {\bibinfo
  {journal} {Nanotechnology}\ }\textbf {\bibinfo {volume} {32}},\ \bibinfo
  {pages} {145717} (\bibinfo {year} {2021})}\BibitemShut {NoStop}%
\bibitem [{\citenamefont {Komsa}\ and\ \citenamefont
  {Krasheninnikov}(2012)}]{Komsa2012}%
  \BibitemOpen
  \bibfield  {author} {\bibinfo {author} {\bibfnamefont {H.-P.}\ \bibnamefont
  {Komsa}}\ and\ \bibinfo {author} {\bibfnamefont {A.~V.}\ \bibnamefont
  {Krasheninnikov}},\ }\href {https://doi.org/10.1103/PhysRevB.86.241201}
  {\bibfield  {journal} {\bibinfo  {journal} {Phys. Rev. B}\ }\textbf {\bibinfo
  {volume} {86}},\ \bibinfo {pages} {241201} (\bibinfo {year}
  {2012})}\BibitemShut {NoStop}%
\bibitem [{\citenamefont {Lin}\ \emph {et~al.}(2014)\citenamefont {Lin},
  \citenamefont {Ling}, \citenamefont {Yu}, \citenamefont {Huang},
  \citenamefont {Hsu}, \citenamefont {Lee}, \citenamefont {Kong}, \citenamefont
  {Dresselhaus},\ and\ \citenamefont {Palacios}}]{Lin2014}%
  \BibitemOpen
  \bibfield  {author} {\bibinfo {author} {\bibfnamefont {Y.}~\bibnamefont
  {Lin}}, \bibinfo {author} {\bibfnamefont {X.}~\bibnamefont {Ling}}, \bibinfo
  {author} {\bibfnamefont {L.}~\bibnamefont {Yu}}, \bibinfo {author}
  {\bibfnamefont {S.}~\bibnamefont {Huang}}, \bibinfo {author} {\bibfnamefont
  {A.~L.}\ \bibnamefont {Hsu}}, \bibinfo {author} {\bibfnamefont {Y.-H.}\
  \bibnamefont {Lee}}, \bibinfo {author} {\bibfnamefont {J.}~\bibnamefont
  {Kong}}, \bibinfo {author} {\bibfnamefont {M.~S.}\ \bibnamefont
  {Dresselhaus}},\ and\ \bibinfo {author} {\bibfnamefont {T.}~\bibnamefont
  {Palacios}},\ }\href {https://doi.org/10.1021/nl501988y} {\bibfield
  {journal} {\bibinfo  {journal} {Nano Letters}\ }\textbf {\bibinfo {volume}
  {14}},\ \bibinfo {pages} {5569} (\bibinfo {year} {2014})}\BibitemShut
  {NoStop}%
\bibitem [{\citenamefont {Latini}\ \emph {et~al.}(2015)\citenamefont {Latini},
  \citenamefont {Olsen},\ and\ \citenamefont {Thygesen}}]{Latini2015}%
  \BibitemOpen
  \bibfield  {author} {\bibinfo {author} {\bibfnamefont {S.}~\bibnamefont
  {Latini}}, \bibinfo {author} {\bibfnamefont {T.}~\bibnamefont {Olsen}},\ and\
  \bibinfo {author} {\bibfnamefont {K.~S.}\ \bibnamefont {Thygesen}},\ }\href
  {https://doi.org/10.1103/PhysRevB.92.245123} {\bibfield  {journal} {\bibinfo
  {journal} {Phys. Rev. B}\ }\textbf {\bibinfo {volume} {92}},\ \bibinfo
  {pages} {245123} (\bibinfo {year} {2015})}\BibitemShut {NoStop}%
\bibitem [{\citenamefont {Andersen}\ \emph {et~al.}(2015)\citenamefont
  {Andersen}, \citenamefont {Latini},\ and\ \citenamefont
  {Thygesen}}]{Andersen2015}%
  \BibitemOpen
  \bibfield  {author} {\bibinfo {author} {\bibfnamefont {K.}~\bibnamefont
  {Andersen}}, \bibinfo {author} {\bibfnamefont {S.}~\bibnamefont {Latini}},\
  and\ \bibinfo {author} {\bibfnamefont {K.~S.}\ \bibnamefont {Thygesen}},\
  }\href {https://doi.org/10.1021/acs.nanolett.5b01251} {\bibfield  {journal}
  {\bibinfo  {journal} {Nano Letters}\ }\textbf {\bibinfo {volume} {15}},\
  \bibinfo {pages} {4616} (\bibinfo {year} {2015})}\BibitemShut {NoStop}%
\bibitem [{\citenamefont {Stier}\ \emph {et~al.}(2016)\citenamefont {Stier},
  \citenamefont {Wilson}, \citenamefont {Clark}, \citenamefont {Xu},\ and\
  \citenamefont {Crooker}}]{Stier2016}%
  \BibitemOpen
  \bibfield  {author} {\bibinfo {author} {\bibfnamefont {A.~V.}\ \bibnamefont
  {Stier}}, \bibinfo {author} {\bibfnamefont {N.~P.}\ \bibnamefont {Wilson}},
  \bibinfo {author} {\bibfnamefont {G.}~\bibnamefont {Clark}}, \bibinfo
  {author} {\bibfnamefont {X.}~\bibnamefont {Xu}},\ and\ \bibinfo {author}
  {\bibfnamefont {S.~A.}\ \bibnamefont {Crooker}},\ }\href
  {https://doi.org/10.1021/acs.nanolett.6b03276} {\bibfield  {journal}
  {\bibinfo  {journal} {Nano Letters}\ }\textbf {\bibinfo {volume} {16}},\
  \bibinfo {pages} {7054} (\bibinfo {year} {2016})}\BibitemShut {NoStop}%
\bibitem [{\citenamefont {Junior}\ \emph {et~al.}(2023)\citenamefont {Junior},
  \citenamefont {Naimer}, \citenamefont {McCreary}, \citenamefont {Jonker},
  \citenamefont {Finley}, \citenamefont {Crooker}, \citenamefont {Fabian},\
  and\ \citenamefont {Stier}}]{Faria2023}%
  \BibitemOpen
  \bibfield  {author} {\bibinfo {author} {\bibfnamefont {P.~E.~F.}\
  \bibnamefont {Junior}}, \bibinfo {author} {\bibfnamefont {T.}~\bibnamefont
  {Naimer}}, \bibinfo {author} {\bibfnamefont {K.~M.}\ \bibnamefont
  {McCreary}}, \bibinfo {author} {\bibfnamefont {B.~T.}\ \bibnamefont
  {Jonker}}, \bibinfo {author} {\bibfnamefont {J.~J.}\ \bibnamefont {Finley}},
  \bibinfo {author} {\bibfnamefont {S.~A.}\ \bibnamefont {Crooker}}, \bibinfo
  {author} {\bibfnamefont {J.}~\bibnamefont {Fabian}},\ and\ \bibinfo {author}
  {\bibfnamefont {A.~V.}\ \bibnamefont {Stier}},\ }\href
  {https://doi.org/10.1088/2053-1583/acd5df} {\bibfield  {journal} {\bibinfo
  {journal} {2D Materials}\ }\textbf {\bibinfo {volume} {10}},\ \bibinfo
  {pages} {034002} (\bibinfo {year} {2023})}\BibitemShut {NoStop}%
\bibitem [{\citenamefont {Pucko}\ \emph {et~al.}(2022)\citenamefont {Pucko},
  \citenamefont {Blundo}, \citenamefont {Zawadzka}, \citenamefont {Cianci},
  \citenamefont {Vaclavkova}, \citenamefont {Kapuściński}, \citenamefont
  {Jana}, \citenamefont {Pettinari}, \citenamefont {Felici}, \citenamefont
  {Nogajewski}, \citenamefont {Bartoš}, \citenamefont {Watanabe},
  \citenamefont {Taniguchi}, \citenamefont {Faugeras}, \citenamefont
  {Potemski}, \citenamefont {Babiński}, \citenamefont {Polimeni},\ and\
  \citenamefont {Molas}}]{Pucko_2023}%
  \BibitemOpen
  \bibfield  {author} {\bibinfo {author} {\bibfnamefont {K.~O.}\ \bibnamefont
  {Pucko}}, \bibinfo {author} {\bibfnamefont {E.}~\bibnamefont {Blundo}},
  \bibinfo {author} {\bibfnamefont {N.}~\bibnamefont {Zawadzka}}, \bibinfo
  {author} {\bibfnamefont {S.}~\bibnamefont {Cianci}}, \bibinfo {author}
  {\bibfnamefont {D.}~\bibnamefont {Vaclavkova}}, \bibinfo {author}
  {\bibfnamefont {P.}~\bibnamefont {Kapuściński}}, \bibinfo {author}
  {\bibfnamefont {D.}~\bibnamefont {Jana}}, \bibinfo {author} {\bibfnamefont
  {G.}~\bibnamefont {Pettinari}}, \bibinfo {author} {\bibfnamefont
  {M.}~\bibnamefont {Felici}}, \bibinfo {author} {\bibfnamefont
  {K.}~\bibnamefont {Nogajewski}}, \bibinfo {author} {\bibfnamefont
  {M.}~\bibnamefont {Bartoš}}, \bibinfo {author} {\bibfnamefont
  {K.}~\bibnamefont {Watanabe}}, \bibinfo {author} {\bibfnamefont
  {T.}~\bibnamefont {Taniguchi}}, \bibinfo {author} {\bibfnamefont
  {C.}~\bibnamefont {Faugeras}}, \bibinfo {author} {\bibfnamefont
  {M.}~\bibnamefont {Potemski}}, \bibinfo {author} {\bibfnamefont
  {A.}~\bibnamefont {Babiński}}, \bibinfo {author} {\bibfnamefont
  {A.}~\bibnamefont {Polimeni}},\ and\ \bibinfo {author} {\bibfnamefont
  {M.~R.}\ \bibnamefont {Molas}},\ }\href
  {https://doi.org/10.1088/2053-1583/aca915} {\bibfield  {journal} {\bibinfo
  {journal} {2D Materials}\ }\textbf {\bibinfo {volume} {10}},\ \bibinfo
  {pages} {015018} (\bibinfo {year} {2022})}\BibitemShut {NoStop}%
\bibitem [{\citenamefont {Seyler}\ \emph {et~al.}(2019)\citenamefont {Seyler},
  \citenamefont {Rivera}, \citenamefont {Yu}, \citenamefont {Wilson},
  \citenamefont {Ray}, \citenamefont {Mandrus}, \citenamefont {Yan},
  \citenamefont {Yao},\ and\ \citenamefont {Xu}}]{Seyler2019}%
  \BibitemOpen
  \bibfield  {author} {\bibinfo {author} {\bibfnamefont {K.~L.}\ \bibnamefont
  {Seyler}}, \bibinfo {author} {\bibfnamefont {P.}~\bibnamefont {Rivera}},
  \bibinfo {author} {\bibfnamefont {H.}~\bibnamefont {Yu}}, \bibinfo {author}
  {\bibfnamefont {N.~P.}\ \bibnamefont {Wilson}}, \bibinfo {author}
  {\bibfnamefont {E.~L.}\ \bibnamefont {Ray}}, \bibinfo {author} {\bibfnamefont
  {D.~G.}\ \bibnamefont {Mandrus}}, \bibinfo {author} {\bibfnamefont
  {J.}~\bibnamefont {Yan}}, \bibinfo {author} {\bibfnamefont {W.}~\bibnamefont
  {Yao}},\ and\ \bibinfo {author} {\bibfnamefont {X.}~\bibnamefont {Xu}},\
  }\href {https://doi.org/10.1038/s41586-019-0957-1} {\bibfield  {journal}
  {\bibinfo  {journal} {Nature}\ }\textbf {\bibinfo {volume} {567}},\ \bibinfo
  {pages} {66} (\bibinfo {year} {2019})}\BibitemShut {NoStop}%
\bibitem [{\citenamefont {Liu}\ \emph {et~al.}(2021)\citenamefont {Liu},
  \citenamefont {Barr{\'e}}, \citenamefont {van Baren}, \citenamefont {Wilson},
  \citenamefont {Taniguchi}, \citenamefont {Watanabe}, \citenamefont {Cui},
  \citenamefont {Gabor}, \citenamefont {Heinz}, \citenamefont {Chang},\ and\
  \citenamefont {Lui}}]{Liu2021}%
  \BibitemOpen
  \bibfield  {author} {\bibinfo {author} {\bibfnamefont {E.}~\bibnamefont
  {Liu}}, \bibinfo {author} {\bibfnamefont {E.}~\bibnamefont {Barr{\'e}}},
  \bibinfo {author} {\bibfnamefont {J.}~\bibnamefont {van Baren}}, \bibinfo
  {author} {\bibfnamefont {M.}~\bibnamefont {Wilson}}, \bibinfo {author}
  {\bibfnamefont {T.}~\bibnamefont {Taniguchi}}, \bibinfo {author}
  {\bibfnamefont {K.}~\bibnamefont {Watanabe}}, \bibinfo {author}
  {\bibfnamefont {Y.-T.}\ \bibnamefont {Cui}}, \bibinfo {author} {\bibfnamefont
  {N.~M.}\ \bibnamefont {Gabor}}, \bibinfo {author} {\bibfnamefont {T.~F.}\
  \bibnamefont {Heinz}}, \bibinfo {author} {\bibfnamefont {Y.-C.}\ \bibnamefont
  {Chang}},\ and\ \bibinfo {author} {\bibfnamefont {C.~H.}\ \bibnamefont
  {Lui}},\ }\href {https://doi.org/10.1038/s41586-021-03541-z} {\bibfield
  {journal} {\bibinfo  {journal} {Nature}\ }\textbf {\bibinfo {volume} {594}},\
  \bibinfo {pages} {46} (\bibinfo {year} {2021})}\BibitemShut {NoStop}%
\bibitem [{\citenamefont {Jiang}\ \emph {et~al.}(2021)\citenamefont {Jiang},
  \citenamefont {Chen}, \citenamefont {Zheng}, \citenamefont {Zheng},\ and\
  \citenamefont {Pan}}]{Jiang2021}%
  \BibitemOpen
  \bibfield  {author} {\bibinfo {author} {\bibfnamefont {Y.}~\bibnamefont
  {Jiang}}, \bibinfo {author} {\bibfnamefont {S.}~\bibnamefont {Chen}},
  \bibinfo {author} {\bibfnamefont {W.}~\bibnamefont {Zheng}}, \bibinfo
  {author} {\bibfnamefont {B.}~\bibnamefont {Zheng}},\ and\ \bibinfo {author}
  {\bibfnamefont {A.}~\bibnamefont {Pan}},\ }\href
  {https://doi.org/10.1038/s41377-021-00500-1} {\bibfield  {journal} {\bibinfo
  {journal} {Light: Science {\&} Applications}\ }\textbf {\bibinfo {volume}
  {10}},\ \bibinfo {pages} {72} (\bibinfo {year} {2021})}\BibitemShut {NoStop}%
\bibitem [{\citenamefont {Brotons-Gisbert}\ \emph {et~al.}(2021)\citenamefont
  {Brotons-Gisbert}, \citenamefont {Baek}, \citenamefont {Campbell},
  \citenamefont {Watanabe}, \citenamefont {Taniguchi},\ and\ \citenamefont
  {Gerardot}}]{Brotons2021}%
  \BibitemOpen
  \bibfield  {author} {\bibinfo {author} {\bibfnamefont {M.}~\bibnamefont
  {Brotons-Gisbert}}, \bibinfo {author} {\bibfnamefont {H.}~\bibnamefont
  {Baek}}, \bibinfo {author} {\bibfnamefont {A.}~\bibnamefont {Campbell}},
  \bibinfo {author} {\bibfnamefont {K.}~\bibnamefont {Watanabe}}, \bibinfo
  {author} {\bibfnamefont {T.}~\bibnamefont {Taniguchi}},\ and\ \bibinfo
  {author} {\bibfnamefont {B.~D.}\ \bibnamefont {Gerardot}},\ }\href
  {https://doi.org/10.1103/PhysRevX.11.031033} {\bibfield  {journal} {\bibinfo
  {journal} {Phys. Rev. X}\ }\textbf {\bibinfo {volume} {11}},\ \bibinfo
  {pages} {031033} (\bibinfo {year} {2021})}\BibitemShut {NoStop}%
\bibitem [{\citenamefont {Blundo}\ \emph {et~al.}(2024)\citenamefont {Blundo},
  \citenamefont {Tuzi}, \citenamefont {Cianci}, \citenamefont {Cuccu},
  \citenamefont {Olkowska-Pucko}, \citenamefont {Kipczak}, \citenamefont
  {Contestabile}, \citenamefont {Miriametro}, \citenamefont {Felici},
  \citenamefont {Pettinari}, \citenamefont {Taniguchi}, \citenamefont
  {Watanabe}, \citenamefont {Babi{\'{n}}ski}, \citenamefont {Molas},\ and\
  \citenamefont {Polimeni}}]{Blundo2024}%
  \BibitemOpen
  \bibfield  {author} {\bibinfo {author} {\bibfnamefont {E.}~\bibnamefont
  {Blundo}}, \bibinfo {author} {\bibfnamefont {F.}~\bibnamefont {Tuzi}},
  \bibinfo {author} {\bibfnamefont {S.}~\bibnamefont {Cianci}}, \bibinfo
  {author} {\bibfnamefont {M.}~\bibnamefont {Cuccu}}, \bibinfo {author}
  {\bibfnamefont {K.}~\bibnamefont {Olkowska-Pucko}}, \bibinfo {author}
  {\bibfnamefont {{\L}.}~\bibnamefont {Kipczak}}, \bibinfo {author}
  {\bibfnamefont {G.}~\bibnamefont {Contestabile}}, \bibinfo {author}
  {\bibfnamefont {A.}~\bibnamefont {Miriametro}}, \bibinfo {author}
  {\bibfnamefont {M.}~\bibnamefont {Felici}}, \bibinfo {author} {\bibfnamefont
  {G.}~\bibnamefont {Pettinari}}, \bibinfo {author} {\bibfnamefont
  {T.}~\bibnamefont {Taniguchi}}, \bibinfo {author} {\bibfnamefont
  {K.}~\bibnamefont {Watanabe}}, \bibinfo {author} {\bibfnamefont
  {A.}~\bibnamefont {Babi{\'{n}}ski}}, \bibinfo {author} {\bibfnamefont
  {M.~R.}\ \bibnamefont {Molas}},\ and\ \bibinfo {author} {\bibfnamefont
  {A.}~\bibnamefont {Polimeni}},\ }\href
  {https://doi.org/10.1038/s41467-024-44739-9} {\bibfield  {journal} {\bibinfo
  {journal} {Nature Communications}\ }\textbf {\bibinfo {volume} {15}},\
  \bibinfo {pages} {1057} (\bibinfo {year} {2024})}\BibitemShut {NoStop}%
\bibitem [{\citenamefont {Prado}\ \emph {et~al.}(2004)\citenamefont {Prado},
  \citenamefont {López-Richard}, \citenamefont {Alcalde}, \citenamefont
  {Marques},\ and\ \citenamefont {Hai}}]{SJPrado_2004}%
  \BibitemOpen
  \bibfield  {author} {\bibinfo {author} {\bibfnamefont {S.~J.}\ \bibnamefont
  {Prado}}, \bibinfo {author} {\bibfnamefont {V.}~\bibnamefont
  {López-Richard}}, \bibinfo {author} {\bibfnamefont {A.~M.}\ \bibnamefont
  {Alcalde}}, \bibinfo {author} {\bibfnamefont {G.~E.}\ \bibnamefont
  {Marques}},\ and\ \bibinfo {author} {\bibfnamefont {G.~Q.}\ \bibnamefont
  {Hai}},\ }\href {https://doi.org/10.1088/0953-8984/16/39/027} {\bibfield
  {journal} {\bibinfo  {journal} {Journal of Physics: Condensed Matter}\
  }\textbf {\bibinfo {volume} {16}},\ \bibinfo {pages} {6949} (\bibinfo {year}
  {2004})}\BibitemShut {NoStop}%
\bibitem [{SM()}]{SM}%
  \BibitemOpen
  \href@noop {} {}\bibinfo {note} {See Supplemental Material at URL for the
  Methods section containing details of the sample preparation, experimental
  setups, and theoretical calculations; the elemental characterization of the
  alloys; temperature and power evolutions of the PL spectra measured on
  MoWSe$_2$ MLs and on the parent compounds MoSe$_2$ and WSe$_2$ MLs;
  magnetic-field evolutions of the PL spectra measured on the MoWSe$_2$ MLs;
  computational details on band folding and mixing in the first principles
  calculations}\BibitemShut {NoStop}%
\bibitem [{\citenamefont {Molas}\ \emph
  {et~al.}(2019{\natexlab{b}})\citenamefont {Molas}, \citenamefont
  {Slobodeniuk}, \citenamefont {Nogajewski}, \citenamefont {Bartos},
  \citenamefont {Bala}, \citenamefont {Babi\ifmmode~\acute{n}\else
  \'{n}\fi{}ski}, \citenamefont {Watanabe}, \citenamefont {Taniguchi},
  \citenamefont {Faugeras},\ and\ \citenamefont {Potemski}}]{Molas2019Energy}%
  \BibitemOpen
  \bibfield  {author} {\bibinfo {author} {\bibfnamefont {M.~R.}\ \bibnamefont
  {Molas}}, \bibinfo {author} {\bibfnamefont {A.~O.}\ \bibnamefont
  {Slobodeniuk}}, \bibinfo {author} {\bibfnamefont {K.}~\bibnamefont
  {Nogajewski}}, \bibinfo {author} {\bibfnamefont {M.}~\bibnamefont {Bartos}},
  \bibinfo {author} {\bibfnamefont {L.}~\bibnamefont {Bala}}, \bibinfo {author}
  {\bibfnamefont {A.}~\bibnamefont {Babi\ifmmode~\acute{n}\else
  \'{n}\fi{}ski}}, \bibinfo {author} {\bibfnamefont {K.}~\bibnamefont
  {Watanabe}}, \bibinfo {author} {\bibfnamefont {T.}~\bibnamefont {Taniguchi}},
  \bibinfo {author} {\bibfnamefont {C.}~\bibnamefont {Faugeras}},\ and\
  \bibinfo {author} {\bibfnamefont {M.}~\bibnamefont {Potemski}},\ }\href
  {https://doi.org/10.1103/PhysRevLett.123.136801} {\bibfield  {journal}
  {\bibinfo  {journal} {Phys. Rev. Lett.}\ }\textbf {\bibinfo {volume} {123}},\
  \bibinfo {pages} {136801} (\bibinfo {year} {2019}{\natexlab{b}})}\BibitemShut
  {NoStop}%
\bibitem [{\citenamefont {Oreszczuk}\ \emph {et~al.}(2023)\citenamefont
  {Oreszczuk}, \citenamefont {Rodek}, \citenamefont {Goryca}, \citenamefont
  {Kazimierczuk}, \citenamefont {Raczyński}, \citenamefont {Howarth},
  \citenamefont {Taniguchi}, \citenamefont {Watanabe}, \citenamefont
  {Potemski},\ and\ \citenamefont {Kossacki}}]{Oreszczuk2023}%
  \BibitemOpen
  \bibfield  {author} {\bibinfo {author} {\bibfnamefont {K.}~\bibnamefont
  {Oreszczuk}}, \bibinfo {author} {\bibfnamefont {A.}~\bibnamefont {Rodek}},
  \bibinfo {author} {\bibfnamefont {M.}~\bibnamefont {Goryca}}, \bibinfo
  {author} {\bibfnamefont {T.}~\bibnamefont {Kazimierczuk}}, \bibinfo {author}
  {\bibfnamefont {M.}~\bibnamefont {Raczyński}}, \bibinfo {author}
  {\bibfnamefont {J.}~\bibnamefont {Howarth}}, \bibinfo {author} {\bibfnamefont
  {T.}~\bibnamefont {Taniguchi}}, \bibinfo {author} {\bibfnamefont
  {K.}~\bibnamefont {Watanabe}}, \bibinfo {author} {\bibfnamefont
  {M.}~\bibnamefont {Potemski}},\ and\ \bibinfo {author} {\bibfnamefont
  {P.}~\bibnamefont {Kossacki}},\ }\href
  {https://doi.org/10.1088/2053-1583/acefe3} {\bibfield  {journal} {\bibinfo
  {journal} {2D Materials}\ }\textbf {\bibinfo {volume} {10}},\ \bibinfo
  {pages} {045019} (\bibinfo {year} {2023})}\BibitemShut {NoStop}%
\bibitem [{\citenamefont {Courtade}\ \emph {et~al.}(2017)\citenamefont
  {Courtade}, \citenamefont {Semina}, \citenamefont {Manca}, \citenamefont
  {Glazov}, \citenamefont {Robert}, \citenamefont {Cadiz}, \citenamefont
  {Wang}, \citenamefont {Taniguchi}, \citenamefont {Watanabe}, \citenamefont
  {Pierre}, \citenamefont {Escoffier}, \citenamefont {Ivchenko}, \citenamefont
  {Renucci}, \citenamefont {Marie}, \citenamefont {Amand},\ and\ \citenamefont
  {Urbaszek}}]{Courtade2017}%
  \BibitemOpen
  \bibfield  {author} {\bibinfo {author} {\bibfnamefont {E.}~\bibnamefont
  {Courtade}}, \bibinfo {author} {\bibfnamefont {M.}~\bibnamefont {Semina}},
  \bibinfo {author} {\bibfnamefont {M.}~\bibnamefont {Manca}}, \bibinfo
  {author} {\bibfnamefont {M.~M.}\ \bibnamefont {Glazov}}, \bibinfo {author}
  {\bibfnamefont {C.}~\bibnamefont {Robert}}, \bibinfo {author} {\bibfnamefont
  {F.}~\bibnamefont {Cadiz}}, \bibinfo {author} {\bibfnamefont
  {G.}~\bibnamefont {Wang}}, \bibinfo {author} {\bibfnamefont {T.}~\bibnamefont
  {Taniguchi}}, \bibinfo {author} {\bibfnamefont {K.}~\bibnamefont {Watanabe}},
  \bibinfo {author} {\bibfnamefont {M.}~\bibnamefont {Pierre}}, \bibinfo
  {author} {\bibfnamefont {W.}~\bibnamefont {Escoffier}}, \bibinfo {author}
  {\bibfnamefont {E.~L.}\ \bibnamefont {Ivchenko}}, \bibinfo {author}
  {\bibfnamefont {P.}~\bibnamefont {Renucci}}, \bibinfo {author} {\bibfnamefont
  {X.}~\bibnamefont {Marie}}, \bibinfo {author} {\bibfnamefont
  {T.}~\bibnamefont {Amand}},\ and\ \bibinfo {author} {\bibfnamefont
  {B.}~\bibnamefont {Urbaszek}},\ }\href
  {https://doi.org/10.1103/PhysRevB.96.085302} {\bibfield  {journal} {\bibinfo
  {journal} {Phys. Rev. B}\ }\textbf {\bibinfo {volume} {96}},\ \bibinfo
  {pages} {085302} (\bibinfo {year} {2017})}\BibitemShut {NoStop}%
\bibitem [{\citenamefont {Li}\ \emph {et~al.}(2018)\citenamefont {Li},
  \citenamefont {Wang}, \citenamefont {Lu}, \citenamefont {Jin}, \citenamefont
  {Chen}, \citenamefont {Meng}, \citenamefont {Lian}, \citenamefont
  {Taniguchi}, \citenamefont {Watanabe}, \citenamefont {Zhang}, \citenamefont
  {Smirnov},\ and\ \citenamefont {Shi}}]{Li2018}%
  \BibitemOpen
  \bibfield  {author} {\bibinfo {author} {\bibfnamefont {Z.}~\bibnamefont
  {Li}}, \bibinfo {author} {\bibfnamefont {T.}~\bibnamefont {Wang}}, \bibinfo
  {author} {\bibfnamefont {Z.}~\bibnamefont {Lu}}, \bibinfo {author}
  {\bibfnamefont {C.}~\bibnamefont {Jin}}, \bibinfo {author} {\bibfnamefont
  {Y.}~\bibnamefont {Chen}}, \bibinfo {author} {\bibfnamefont {Y.}~\bibnamefont
  {Meng}}, \bibinfo {author} {\bibfnamefont {Z.}~\bibnamefont {Lian}}, \bibinfo
  {author} {\bibfnamefont {T.}~\bibnamefont {Taniguchi}}, \bibinfo {author}
  {\bibfnamefont {K.}~\bibnamefont {Watanabe}}, \bibinfo {author}
  {\bibfnamefont {S.}~\bibnamefont {Zhang}}, \bibinfo {author} {\bibfnamefont
  {D.}~\bibnamefont {Smirnov}},\ and\ \bibinfo {author} {\bibfnamefont {S.-F.}\
  \bibnamefont {Shi}},\ }\href {https://doi.org/10.1038/s41467-018-05863-5}
  {\bibfield  {journal} {\bibinfo  {journal} {Nature Communications}\ }\textbf
  {\bibinfo {volume} {9}},\ \bibinfo {pages} {3719} (\bibinfo {year}
  {2018})}\BibitemShut {NoStop}%
\bibitem [{\citenamefont {Chen}\ \emph {et~al.}(2018)\citenamefont {Chen},
  \citenamefont {Goldstein}, \citenamefont {Taniguchi}, \citenamefont
  {Watanabe},\ and\ \citenamefont {Yan}}]{Chen2018}%
  \BibitemOpen
  \bibfield  {author} {\bibinfo {author} {\bibfnamefont {S.-Y.}\ \bibnamefont
  {Chen}}, \bibinfo {author} {\bibfnamefont {T.}~\bibnamefont {Goldstein}},
  \bibinfo {author} {\bibfnamefont {T.}~\bibnamefont {Taniguchi}}, \bibinfo
  {author} {\bibfnamefont {K.}~\bibnamefont {Watanabe}},\ and\ \bibinfo
  {author} {\bibfnamefont {J.}~\bibnamefont {Yan}},\ }\href
  {https://doi.org/10.1038/s41467-018-05558-x} {\bibfield  {journal} {\bibinfo
  {journal} {Nature Communications}\ }\textbf {\bibinfo {volume} {9}},\
  \bibinfo {pages} {3717} (\bibinfo {year} {2018})}\BibitemShut {NoStop}%
\bibitem [{\citenamefont {Barbone}\ \emph {et~al.}(2018)\citenamefont
  {Barbone}, \citenamefont {Montblanch}, \citenamefont {Kara}, \citenamefont
  {Palacios-Berraquero}, \citenamefont {Cadore}, \citenamefont {De~Fazio},
  \citenamefont {Pingault}, \citenamefont {Mostaani}, \citenamefont {Li},
  \citenamefont {Chen}, \citenamefont {Watanabe}, \citenamefont {Taniguchi},
  \citenamefont {Tongay}, \citenamefont {Wang}, \citenamefont {Ferrari},\ and\
  \citenamefont {Atat{\"u}re}}]{Barbone2018}%
  \BibitemOpen
  \bibfield  {author} {\bibinfo {author} {\bibfnamefont {M.}~\bibnamefont
  {Barbone}}, \bibinfo {author} {\bibfnamefont {A.~R.-P.}\ \bibnamefont
  {Montblanch}}, \bibinfo {author} {\bibfnamefont {D.~M.}\ \bibnamefont
  {Kara}}, \bibinfo {author} {\bibfnamefont {C.}~\bibnamefont
  {Palacios-Berraquero}}, \bibinfo {author} {\bibfnamefont {A.~R.}\
  \bibnamefont {Cadore}}, \bibinfo {author} {\bibfnamefont {D.}~\bibnamefont
  {De~Fazio}}, \bibinfo {author} {\bibfnamefont {B.}~\bibnamefont {Pingault}},
  \bibinfo {author} {\bibfnamefont {E.}~\bibnamefont {Mostaani}}, \bibinfo
  {author} {\bibfnamefont {H.}~\bibnamefont {Li}}, \bibinfo {author}
  {\bibfnamefont {B.}~\bibnamefont {Chen}}, \bibinfo {author} {\bibfnamefont
  {K.}~\bibnamefont {Watanabe}}, \bibinfo {author} {\bibfnamefont
  {T.}~\bibnamefont {Taniguchi}}, \bibinfo {author} {\bibfnamefont
  {S.}~\bibnamefont {Tongay}}, \bibinfo {author} {\bibfnamefont
  {G.}~\bibnamefont {Wang}}, \bibinfo {author} {\bibfnamefont {A.~C.}\
  \bibnamefont {Ferrari}},\ and\ \bibinfo {author} {\bibfnamefont
  {M.}~\bibnamefont {Atat{\"u}re}},\ }\href
  {https://doi.org/10.1038/s41467-018-05632-4} {\bibfield  {journal} {\bibinfo
  {journal} {Nature Communications}\ }\textbf {\bibinfo {volume} {9}},\
  \bibinfo {pages} {3721} (\bibinfo {year} {2018})}\BibitemShut {NoStop}%
\bibitem [{\citenamefont {Paur}\ \emph {et~al.}(2019)\citenamefont {Paur},
  \citenamefont {Molina-Mendoza}, \citenamefont {Bratschitsch}, \citenamefont
  {Watanabe}, \citenamefont {Taniguchi},\ and\ \citenamefont
  {Mueller}}]{Paur2019}%
  \BibitemOpen
  \bibfield  {author} {\bibinfo {author} {\bibfnamefont {M.}~\bibnamefont
  {Paur}}, \bibinfo {author} {\bibfnamefont {A.~J.}\ \bibnamefont
  {Molina-Mendoza}}, \bibinfo {author} {\bibfnamefont {R.}~\bibnamefont
  {Bratschitsch}}, \bibinfo {author} {\bibfnamefont {K.}~\bibnamefont
  {Watanabe}}, \bibinfo {author} {\bibfnamefont {T.}~\bibnamefont
  {Taniguchi}},\ and\ \bibinfo {author} {\bibfnamefont {T.}~\bibnamefont
  {Mueller}},\ }\href {https://doi.org/10.1038/s41467-019-09781-y} {\bibfield
  {journal} {\bibinfo  {journal} {Nature Communications}\ }\textbf {\bibinfo
  {volume} {10}},\ \bibinfo {pages} {1709} (\bibinfo {year}
  {2019})}\BibitemShut {NoStop}%
\bibitem [{\citenamefont {Li}\ \emph {et~al.}(2019{\natexlab{a}})\citenamefont
  {Li}, \citenamefont {Wang}, \citenamefont {Lu}, \citenamefont {Khatoniar},
  \citenamefont {Lian}, \citenamefont {Meng}, \citenamefont {Blei},
  \citenamefont {Taniguchi}, \citenamefont {Watanabe}, \citenamefont {McGill},
  \citenamefont {Tongay}, \citenamefont {Menon}, \citenamefont {Smirnov},\ and\
  \citenamefont {Shi}}]{Li2019}%
  \BibitemOpen
  \bibfield  {author} {\bibinfo {author} {\bibfnamefont {Z.}~\bibnamefont
  {Li}}, \bibinfo {author} {\bibfnamefont {T.}~\bibnamefont {Wang}}, \bibinfo
  {author} {\bibfnamefont {Z.}~\bibnamefont {Lu}}, \bibinfo {author}
  {\bibfnamefont {M.}~\bibnamefont {Khatoniar}}, \bibinfo {author}
  {\bibfnamefont {Z.}~\bibnamefont {Lian}}, \bibinfo {author} {\bibfnamefont
  {Y.}~\bibnamefont {Meng}}, \bibinfo {author} {\bibfnamefont {M.}~\bibnamefont
  {Blei}}, \bibinfo {author} {\bibfnamefont {T.}~\bibnamefont {Taniguchi}},
  \bibinfo {author} {\bibfnamefont {K.}~\bibnamefont {Watanabe}}, \bibinfo
  {author} {\bibfnamefont {S.~A.}\ \bibnamefont {McGill}}, \bibinfo {author}
  {\bibfnamefont {S.}~\bibnamefont {Tongay}}, \bibinfo {author} {\bibfnamefont
  {V.~M.}\ \bibnamefont {Menon}}, \bibinfo {author} {\bibfnamefont
  {D.}~\bibnamefont {Smirnov}},\ and\ \bibinfo {author} {\bibfnamefont {S.-F.}\
  \bibnamefont {Shi}},\ }\href {https://doi.org/10.1021/acs.nanolett.9b02132}
  {\bibfield  {journal} {\bibinfo  {journal} {Nano Letters}\ }\textbf {\bibinfo
  {volume} {19}},\ \bibinfo {pages} {6886} (\bibinfo {year}
  {2019}{\natexlab{a}})}\BibitemShut {NoStop}%
\bibitem [{\citenamefont {Li}\ \emph {et~al.}(2019{\natexlab{b}})\citenamefont
  {Li}, \citenamefont {Wang}, \citenamefont {Jin}, \citenamefont {Lu},
  \citenamefont {Lian}, \citenamefont {Meng}, \citenamefont {Blei},
  \citenamefont {Gao}, \citenamefont {Taniguchi}, \citenamefont {Watanabe},
  \citenamefont {Ren}, \citenamefont {Tongay}, \citenamefont {Yang},
  \citenamefont {Smirnov}, \citenamefont {Cao},\ and\ \citenamefont
  {Shi}}]{Lireplica2019}%
  \BibitemOpen
  \bibfield  {author} {\bibinfo {author} {\bibfnamefont {Z.}~\bibnamefont
  {Li}}, \bibinfo {author} {\bibfnamefont {T.}~\bibnamefont {Wang}}, \bibinfo
  {author} {\bibfnamefont {C.}~\bibnamefont {Jin}}, \bibinfo {author}
  {\bibfnamefont {Z.}~\bibnamefont {Lu}}, \bibinfo {author} {\bibfnamefont
  {Z.}~\bibnamefont {Lian}}, \bibinfo {author} {\bibfnamefont {Y.}~\bibnamefont
  {Meng}}, \bibinfo {author} {\bibfnamefont {M.}~\bibnamefont {Blei}}, \bibinfo
  {author} {\bibfnamefont {S.}~\bibnamefont {Gao}}, \bibinfo {author}
  {\bibfnamefont {T.}~\bibnamefont {Taniguchi}}, \bibinfo {author}
  {\bibfnamefont {K.}~\bibnamefont {Watanabe}}, \bibinfo {author}
  {\bibfnamefont {T.}~\bibnamefont {Ren}}, \bibinfo {author} {\bibfnamefont
  {S.}~\bibnamefont {Tongay}}, \bibinfo {author} {\bibfnamefont
  {L.}~\bibnamefont {Yang}}, \bibinfo {author} {\bibfnamefont {D.}~\bibnamefont
  {Smirnov}}, \bibinfo {author} {\bibfnamefont {T.}~\bibnamefont {Cao}},\ and\
  \bibinfo {author} {\bibfnamefont {S.-F.}\ \bibnamefont {Shi}},\ }\href
  {https://doi.org/10.1038/s41467-019-10477-6} {\bibfield  {journal} {\bibinfo
  {journal} {Nature Communications}\ }\textbf {\bibinfo {volume} {10}},\
  \bibinfo {pages} {2469} (\bibinfo {year} {2019}{\natexlab{b}})}\BibitemShut
  {NoStop}%
\bibitem [{\citenamefont {Li}\ \emph {et~al.}(2019{\natexlab{c}})\citenamefont
  {Li}, \citenamefont {Wang}, \citenamefont {Jin}, \citenamefont {Lu},
  \citenamefont {Lian}, \citenamefont {Meng}, \citenamefont {Blei},
  \citenamefont {Gao}, \citenamefont {Taniguchi}, \citenamefont {Watanabe},
  \citenamefont {Ren}, \citenamefont {Cao}, \citenamefont {Tongay},
  \citenamefont {Smirnov}, \citenamefont {Zhang},\ and\ \citenamefont
  {Shi}}]{Li2019momentum}%
  \BibitemOpen
  \bibfield  {author} {\bibinfo {author} {\bibfnamefont {Z.}~\bibnamefont
  {Li}}, \bibinfo {author} {\bibfnamefont {T.}~\bibnamefont {Wang}}, \bibinfo
  {author} {\bibfnamefont {C.}~\bibnamefont {Jin}}, \bibinfo {author}
  {\bibfnamefont {Z.}~\bibnamefont {Lu}}, \bibinfo {author} {\bibfnamefont
  {Z.}~\bibnamefont {Lian}}, \bibinfo {author} {\bibfnamefont {Y.}~\bibnamefont
  {Meng}}, \bibinfo {author} {\bibfnamefont {M.}~\bibnamefont {Blei}}, \bibinfo
  {author} {\bibfnamefont {M.}~\bibnamefont {Gao}}, \bibinfo {author}
  {\bibfnamefont {T.}~\bibnamefont {Taniguchi}}, \bibinfo {author}
  {\bibfnamefont {K.}~\bibnamefont {Watanabe}}, \bibinfo {author}
  {\bibfnamefont {T.}~\bibnamefont {Ren}}, \bibinfo {author} {\bibfnamefont
  {T.}~\bibnamefont {Cao}}, \bibinfo {author} {\bibfnamefont {S.}~\bibnamefont
  {Tongay}}, \bibinfo {author} {\bibfnamefont {D.}~\bibnamefont {Smirnov}},
  \bibinfo {author} {\bibfnamefont {L.}~\bibnamefont {Zhang}},\ and\ \bibinfo
  {author} {\bibfnamefont {S.-F.}\ \bibnamefont {Shi}},\ }\href
  {https://doi.org/10.1021/acsnano.9b06682} {\bibfield  {journal} {\bibinfo
  {journal} {ACS Nano}\ }\textbf {\bibinfo {volume} {13}},\ \bibinfo {pages}
  {14107} (\bibinfo {year} {2019}{\natexlab{c}})}\BibitemShut {NoStop}%
\bibitem [{\citenamefont {Liu}\ \emph {et~al.}(2019{\natexlab{b}})\citenamefont
  {Liu}, \citenamefont {van Baren}, \citenamefont {Taniguchi}, \citenamefont
  {Watanabe}, \citenamefont {Chang},\ and\ \citenamefont {Lui}}]{LiuValley}%
  \BibitemOpen
  \bibfield  {author} {\bibinfo {author} {\bibfnamefont {E.}~\bibnamefont
  {Liu}}, \bibinfo {author} {\bibfnamefont {J.}~\bibnamefont {van Baren}},
  \bibinfo {author} {\bibfnamefont {T.}~\bibnamefont {Taniguchi}}, \bibinfo
  {author} {\bibfnamefont {K.}~\bibnamefont {Watanabe}}, \bibinfo {author}
  {\bibfnamefont {Y.-C.}\ \bibnamefont {Chang}},\ and\ \bibinfo {author}
  {\bibfnamefont {C.~H.}\ \bibnamefont {Lui}},\ }\href
  {https://doi.org/10.1103/PhysRevResearch.1.032007} {\bibfield  {journal}
  {\bibinfo  {journal} {Phys. Rev. Research}\ }\textbf {\bibinfo {volume}
  {1}},\ \bibinfo {pages} {032007} (\bibinfo {year}
  {2019}{\natexlab{b}})}\BibitemShut {NoStop}%
\bibitem [{\citenamefont {Robert}\ \emph
  {et~al.}(2021{\natexlab{a}})\citenamefont {Robert}, \citenamefont {Park},
  \citenamefont {Cadiz}, \citenamefont {Lombez}, \citenamefont {Ren},
  \citenamefont {Tornatzky}, \citenamefont {Rowe}, \citenamefont {Paget},
  \citenamefont {Sirotti}, \citenamefont {Yang}, \citenamefont {Van~Tuan},
  \citenamefont {Taniguchi}, \citenamefont {Urbaszek}, \citenamefont
  {Watanabe}, \citenamefont {Amand}, \citenamefont {Dery},\ and\ \citenamefont
  {Marie}}]{Robert2021}%
  \BibitemOpen
  \bibfield  {author} {\bibinfo {author} {\bibfnamefont {C.}~\bibnamefont
  {Robert}}, \bibinfo {author} {\bibfnamefont {S.}~\bibnamefont {Park}},
  \bibinfo {author} {\bibfnamefont {F.}~\bibnamefont {Cadiz}}, \bibinfo
  {author} {\bibfnamefont {L.}~\bibnamefont {Lombez}}, \bibinfo {author}
  {\bibfnamefont {L.}~\bibnamefont {Ren}}, \bibinfo {author} {\bibfnamefont
  {H.}~\bibnamefont {Tornatzky}}, \bibinfo {author} {\bibfnamefont
  {A.}~\bibnamefont {Rowe}}, \bibinfo {author} {\bibfnamefont {D.}~\bibnamefont
  {Paget}}, \bibinfo {author} {\bibfnamefont {F.}~\bibnamefont {Sirotti}},
  \bibinfo {author} {\bibfnamefont {M.}~\bibnamefont {Yang}}, \bibinfo {author}
  {\bibfnamefont {D.}~\bibnamefont {Van~Tuan}}, \bibinfo {author}
  {\bibfnamefont {T.}~\bibnamefont {Taniguchi}}, \bibinfo {author}
  {\bibfnamefont {B.}~\bibnamefont {Urbaszek}}, \bibinfo {author}
  {\bibfnamefont {K.}~\bibnamefont {Watanabe}}, \bibinfo {author}
  {\bibfnamefont {T.}~\bibnamefont {Amand}}, \bibinfo {author} {\bibfnamefont
  {H.}~\bibnamefont {Dery}},\ and\ \bibinfo {author} {\bibfnamefont
  {X.}~\bibnamefont {Marie}},\ }\href
  {https://doi.org/10.1038/s41467-021-25747-5} {\bibfield  {journal} {\bibinfo
  {journal} {Nature Communications}\ }\textbf {\bibinfo {volume} {12}},\
  \bibinfo {pages} {5455} (\bibinfo {year} {2021}{\natexlab{a}})}\BibitemShut
  {NoStop}%
\bibitem [{\citenamefont {Robert}\ \emph
  {et~al.}(2021{\natexlab{b}})\citenamefont {Robert}, \citenamefont {Dery},
  \citenamefont {Ren}, \citenamefont {Van~Tuan}, \citenamefont {Courtade},
  \citenamefont {Yang}, \citenamefont {Urbaszek}, \citenamefont {Lagarde},
  \citenamefont {Watanabe}, \citenamefont {Taniguchi}, \citenamefont {Amand},\
  and\ \citenamefont {Marie}}]{Robert2021PRL}%
  \BibitemOpen
  \bibfield  {author} {\bibinfo {author} {\bibfnamefont {C.}~\bibnamefont
  {Robert}}, \bibinfo {author} {\bibfnamefont {H.}~\bibnamefont {Dery}},
  \bibinfo {author} {\bibfnamefont {L.}~\bibnamefont {Ren}}, \bibinfo {author}
  {\bibfnamefont {D.}~\bibnamefont {Van~Tuan}}, \bibinfo {author}
  {\bibfnamefont {E.}~\bibnamefont {Courtade}}, \bibinfo {author}
  {\bibfnamefont {M.}~\bibnamefont {Yang}}, \bibinfo {author} {\bibfnamefont
  {B.}~\bibnamefont {Urbaszek}}, \bibinfo {author} {\bibfnamefont
  {D.}~\bibnamefont {Lagarde}}, \bibinfo {author} {\bibfnamefont
  {K.}~\bibnamefont {Watanabe}}, \bibinfo {author} {\bibfnamefont
  {T.}~\bibnamefont {Taniguchi}}, \bibinfo {author} {\bibfnamefont
  {T.}~\bibnamefont {Amand}},\ and\ \bibinfo {author} {\bibfnamefont
  {X.}~\bibnamefont {Marie}},\ }\href
  {https://doi.org/10.1103/PhysRevLett.126.067403} {\bibfield  {journal}
  {\bibinfo  {journal} {Phys. Rev. Lett.}\ }\textbf {\bibinfo {volume} {126}},\
  \bibinfo {pages} {067403} (\bibinfo {year} {2021}{\natexlab{b}})}\BibitemShut
  {NoStop}%
\bibitem [{\citenamefont {Zinkiewicz}\ \emph {et~al.}(2022)\citenamefont
  {Zinkiewicz}, \citenamefont {Grzeszczyk}, \citenamefont {Kipczak},
  \citenamefont {Kazimierczuk}, \citenamefont {Watanabe}, \citenamefont
  {Taniguchi}, \citenamefont {Kossacki}, \citenamefont {Babinski},\ and\
  \citenamefont {Molas}}]{Zinkiewicz2022}%
  \BibitemOpen
  \bibfield  {author} {\bibinfo {author} {\bibfnamefont {M.}~\bibnamefont
  {Zinkiewicz}}, \bibinfo {author} {\bibfnamefont {M.}~\bibnamefont
  {Grzeszczyk}}, \bibinfo {author} {\bibfnamefont {L.}~\bibnamefont {Kipczak}},
  \bibinfo {author} {\bibfnamefont {T.}~\bibnamefont {Kazimierczuk}}, \bibinfo
  {author} {\bibfnamefont {K.}~\bibnamefont {Watanabe}}, \bibinfo {author}
  {\bibfnamefont {T.}~\bibnamefont {Taniguchi}}, \bibinfo {author}
  {\bibfnamefont {P.}~\bibnamefont {Kossacki}}, \bibinfo {author}
  {\bibfnamefont {A.}~\bibnamefont {Babinski}},\ and\ \bibinfo {author}
  {\bibfnamefont {M.~R.}\ \bibnamefont {Molas}},\ }\href
  {https://doi.org/10.1063/5.0085950} {\bibfield  {journal} {\bibinfo
  {journal} {Applied Physics Letters}\ }\textbf {\bibinfo {volume} {120}},\
  \bibinfo {pages} {163101} (\bibinfo {year} {2022})}\BibitemShut {NoStop}%
\bibitem [{\citenamefont {Zhang}\ \emph {et~al.}(2014)\citenamefont {Zhang},
  \citenamefont {Wu}, \citenamefont {Zhu}, \citenamefont {Dumcenco},
  \citenamefont {Hong}, \citenamefont {Mao}, \citenamefont {Deng},
  \citenamefont {Chen}, \citenamefont {Yang}, \citenamefont {Jin},
  \citenamefont {Chaki}, \citenamefont {Huang}, \citenamefont {Zhang},\ and\
  \citenamefont {Xie}}]{zhang2014}%
  \BibitemOpen
  \bibfield  {author} {\bibinfo {author} {\bibfnamefont {M.}~\bibnamefont
  {Zhang}}, \bibinfo {author} {\bibfnamefont {J.}~\bibnamefont {Wu}}, \bibinfo
  {author} {\bibfnamefont {Y.}~\bibnamefont {Zhu}}, \bibinfo {author}
  {\bibfnamefont {D.~O.}\ \bibnamefont {Dumcenco}}, \bibinfo {author}
  {\bibfnamefont {J.}~\bibnamefont {Hong}}, \bibinfo {author} {\bibfnamefont
  {N.}~\bibnamefont {Mao}}, \bibinfo {author} {\bibfnamefont {S.}~\bibnamefont
  {Deng}}, \bibinfo {author} {\bibfnamefont {Y.}~\bibnamefont {Chen}}, \bibinfo
  {author} {\bibfnamefont {Y.}~\bibnamefont {Yang}}, \bibinfo {author}
  {\bibfnamefont {C.}~\bibnamefont {Jin}}, \bibinfo {author} {\bibfnamefont
  {S.~H.}\ \bibnamefont {Chaki}}, \bibinfo {author} {\bibfnamefont {Y.-S.}\
  \bibnamefont {Huang}}, \bibinfo {author} {\bibfnamefont {J.}~\bibnamefont
  {Zhang}},\ and\ \bibinfo {author} {\bibfnamefont {L.}~\bibnamefont {Xie}},\
  }\href {https://doi.org/10.1021/nn5020566} {\bibfield  {journal} {\bibinfo
  {journal} {ACS Nano}\ }\textbf {\bibinfo {volume} {8}},\ \bibinfo {pages}
  {7130} (\bibinfo {year} {2014})}\BibitemShut {NoStop}%
\bibitem [{\citenamefont {Wang}\ \emph {et~al.}(2015)\citenamefont {Wang},
  \citenamefont {Robert}, \citenamefont {Suslu}, \citenamefont {Chen},
  \citenamefont {Yang}, \citenamefont {Alamdari}, \citenamefont {Gerber},
  \citenamefont {Amand}, \citenamefont {Marie}, \citenamefont {Tongay},\ and\
  \citenamefont {Urbaszek}}]{Wang2015}%
  \BibitemOpen
  \bibfield  {author} {\bibinfo {author} {\bibfnamefont {G.}~\bibnamefont
  {Wang}}, \bibinfo {author} {\bibfnamefont {C.}~\bibnamefont {Robert}},
  \bibinfo {author} {\bibfnamefont {A.}~\bibnamefont {Suslu}}, \bibinfo
  {author} {\bibfnamefont {B.}~\bibnamefont {Chen}}, \bibinfo {author}
  {\bibfnamefont {S.}~\bibnamefont {Yang}}, \bibinfo {author} {\bibfnamefont
  {S.}~\bibnamefont {Alamdari}}, \bibinfo {author} {\bibfnamefont {I.~C.}\
  \bibnamefont {Gerber}}, \bibinfo {author} {\bibfnamefont {T.}~\bibnamefont
  {Amand}}, \bibinfo {author} {\bibfnamefont {X.}~\bibnamefont {Marie}},
  \bibinfo {author} {\bibfnamefont {S.}~\bibnamefont {Tongay}},\ and\ \bibinfo
  {author} {\bibfnamefont {B.}~\bibnamefont {Urbaszek}},\ }\href
  {https://doi.org/10.1038/ncomms10110} {\bibfield  {journal} {\bibinfo
  {journal} {Nature Communications}\ }\textbf {\bibinfo {volume} {6}},\
  \bibinfo {pages} {10110} (\bibinfo {year} {2015})}\BibitemShut {NoStop}%
\bibitem [{\citenamefont {Catanzaro}\ \emph {et~al.}(2024)\citenamefont
  {Catanzaro}, \citenamefont {Genco}, \citenamefont {Louca}, \citenamefont
  {Ruiz-Tijerina}, \citenamefont {Gillard}, \citenamefont {Sortino},
  \citenamefont {Kozikov}, \citenamefont {Alexeev}, \citenamefont {Pisoni},
  \citenamefont {Hague}, \citenamefont {Watanabe}, \citenamefont {Taniguchi},
  \citenamefont {Ensslin}, \citenamefont {Novoselov}, \citenamefont {Fal'ko},\
  and\ \citenamefont {Tartakovskii}}]{Catanzaro2024}%
  \BibitemOpen
  \bibfield  {author} {\bibinfo {author} {\bibfnamefont {A.}~\bibnamefont
  {Catanzaro}}, \bibinfo {author} {\bibfnamefont {A.}~\bibnamefont {Genco}},
  \bibinfo {author} {\bibfnamefont {C.}~\bibnamefont {Louca}}, \bibinfo
  {author} {\bibfnamefont {D.~A.}\ \bibnamefont {Ruiz-Tijerina}}, \bibinfo
  {author} {\bibfnamefont {D.~J.}\ \bibnamefont {Gillard}}, \bibinfo {author}
  {\bibfnamefont {L.}~\bibnamefont {Sortino}}, \bibinfo {author} {\bibfnamefont
  {A.}~\bibnamefont {Kozikov}}, \bibinfo {author} {\bibfnamefont {E.~M.}\
  \bibnamefont {Alexeev}}, \bibinfo {author} {\bibfnamefont {R.}~\bibnamefont
  {Pisoni}}, \bibinfo {author} {\bibfnamefont {L.}~\bibnamefont {Hague}},
  \bibinfo {author} {\bibfnamefont {K.}~\bibnamefont {Watanabe}}, \bibinfo
  {author} {\bibfnamefont {T.}~\bibnamefont {Taniguchi}}, \bibinfo {author}
  {\bibfnamefont {K.}~\bibnamefont {Ensslin}}, \bibinfo {author} {\bibfnamefont
  {K.~S.}\ \bibnamefont {Novoselov}}, \bibinfo {author} {\bibfnamefont
  {V.}~\bibnamefont {Fal'ko}},\ and\ \bibinfo {author} {\bibfnamefont {A.~I.}\
  \bibnamefont {Tartakovskii}},\ }\href
  {https://doi.org/https://doi.org/10.1002/adma.202309644} {\bibfield
  {journal} {\bibinfo  {journal} {Advanced Materials}\ }\textbf {\bibinfo
  {volume} {36}},\ \bibinfo {pages} {2309644} (\bibinfo {year}
  {2024})}\BibitemShut {NoStop}%
\bibitem [{\citenamefont {de~Brito}\ \emph {et~al.}(2025)\citenamefont
  {de~Brito}, \citenamefont {Pinto}, \citenamefont {Plutnar}, \citenamefont
  {Sofer}, \citenamefont {Schleder}, \citenamefont {Capaz}, \citenamefont
  {Barcelos},\ and\ \citenamefont {Neves}}]{Brito_2025}%
  \BibitemOpen
  \bibfield  {author} {\bibinfo {author} {\bibfnamefont {A.~C.~F.}\
  \bibnamefont {de~Brito}}, \bibinfo {author} {\bibfnamefont {A.~A.}\
  \bibnamefont {Pinto}}, \bibinfo {author} {\bibfnamefont {J.}~\bibnamefont
  {Plutnar}}, \bibinfo {author} {\bibfnamefont {Z.}~\bibnamefont {Sofer}},
  \bibinfo {author} {\bibfnamefont {G.~R.}\ \bibnamefont {Schleder}}, \bibinfo
  {author} {\bibfnamefont {R.~B.}\ \bibnamefont {Capaz}}, \bibinfo {author}
  {\bibfnamefont {I.~D.}\ \bibnamefont {Barcelos}},\ and\ \bibinfo {author}
  {\bibfnamefont {B.~R.~A.}\ \bibnamefont {Neves}},\ }\href
  {https://doi.org/10.1088/1361-6528/ae00cd} {\bibfield  {journal} {\bibinfo
  {journal} {Nanotechnology}\ }\textbf {\bibinfo {volume} {36}},\ \bibinfo
  {pages} {375701} (\bibinfo {year} {2025})}\BibitemShut {NoStop}%
\bibitem [{\citenamefont {Blundo}\ \emph {et~al.}(2020)\citenamefont {Blundo},
  \citenamefont {Felici}, \citenamefont {Yildirim}, \citenamefont {Pettinari},
  \citenamefont {Tedeschi}, \citenamefont {Miriametro}, \citenamefont {Liu},
  \citenamefont {Ma}, \citenamefont {Lu},\ and\ \citenamefont
  {Polimeni}}]{Blundo2020}%
  \BibitemOpen
  \bibfield  {author} {\bibinfo {author} {\bibfnamefont {E.}~\bibnamefont
  {Blundo}}, \bibinfo {author} {\bibfnamefont {M.}~\bibnamefont {Felici}},
  \bibinfo {author} {\bibfnamefont {T.}~\bibnamefont {Yildirim}}, \bibinfo
  {author} {\bibfnamefont {G.}~\bibnamefont {Pettinari}}, \bibinfo {author}
  {\bibfnamefont {D.}~\bibnamefont {Tedeschi}}, \bibinfo {author}
  {\bibfnamefont {A.}~\bibnamefont {Miriametro}}, \bibinfo {author}
  {\bibfnamefont {B.}~\bibnamefont {Liu}}, \bibinfo {author} {\bibfnamefont
  {W.}~\bibnamefont {Ma}}, \bibinfo {author} {\bibfnamefont {Y.}~\bibnamefont
  {Lu}},\ and\ \bibinfo {author} {\bibfnamefont {A.}~\bibnamefont {Polimeni}},\
  }\href {https://doi.org/10.1103/PhysRevResearch.2.012024} {\bibfield
  {journal} {\bibinfo  {journal} {Phys. Rev. Res.}\ }\textbf {\bibinfo {volume}
  {2}},\ \bibinfo {pages} {012024} (\bibinfo {year} {2020})}\BibitemShut
  {NoStop}%
\bibitem [{\citenamefont {Cianci}\ \emph {et~al.}(2020)\citenamefont {Cianci},
  \citenamefont {Blundo}, \citenamefont {Tuzi}, \citenamefont {Cecchetti},
  \citenamefont {Pettinari}, \citenamefont {Felici},\ and\ \citenamefont
  {Polimeni}}]{Cianci2024}%
  \BibitemOpen
  \bibfield  {author} {\bibinfo {author} {\bibfnamefont {S.}~\bibnamefont
  {Cianci}}, \bibinfo {author} {\bibfnamefont {E.}~\bibnamefont {Blundo}},
  \bibinfo {author} {\bibfnamefont {F.}~\bibnamefont {Tuzi}}, \bibinfo {author}
  {\bibfnamefont {D.}~\bibnamefont {Cecchetti}}, \bibinfo {author}
  {\bibfnamefont {G.}~\bibnamefont {Pettinari}}, \bibinfo {author}
  {\bibfnamefont {M.}~\bibnamefont {Felici}},\ and\ \bibinfo {author}
  {\bibfnamefont {A.}~\bibnamefont {Polimeni}},\ }\href@noop {} {\bibfield
  {journal} {\bibinfo  {journal} {J. Appl. Phys.}\ }\textbf {\bibinfo {volume}
  {2}},\ \bibinfo {pages} {012024} (\bibinfo {year} {2020})}\BibitemShut
  {NoStop}%
\bibitem [{\citenamefont {Arora}\ \emph {et~al.}(2016)\citenamefont {Arora},
  \citenamefont {Schmidt}, \citenamefont {Schneider}, \citenamefont {Molas},
  \citenamefont {Breslavetz}, \citenamefont {Potemski},\ and\ \citenamefont
  {Bratschitsch}}]{Arora_MoTe2}%
  \BibitemOpen
  \bibfield  {author} {\bibinfo {author} {\bibfnamefont {A.}~\bibnamefont
  {Arora}}, \bibinfo {author} {\bibfnamefont {R.}~\bibnamefont {Schmidt}},
  \bibinfo {author} {\bibfnamefont {R.}~\bibnamefont {Schneider}}, \bibinfo
  {author} {\bibfnamefont {M.~R.}\ \bibnamefont {Molas}}, \bibinfo {author}
  {\bibfnamefont {I.}~\bibnamefont {Breslavetz}}, \bibinfo {author}
  {\bibfnamefont {M.}~\bibnamefont {Potemski}},\ and\ \bibinfo {author}
  {\bibfnamefont {R.}~\bibnamefont {Bratschitsch}},\ }\href
  {https://doi.org/10.1021/acs.nanolett.6b00748} {\bibfield  {journal}
  {\bibinfo  {journal} {Nano Letters}\ }\textbf {\bibinfo {volume} {16}},\
  \bibinfo {pages} {3624} (\bibinfo {year} {2016})}\BibitemShut {NoStop}%
\bibitem [{\citenamefont {Deilmann}\ \emph {et~al.}(2020)\citenamefont
  {Deilmann}, \citenamefont {Kr\"uger},\ and\ \citenamefont
  {Rohlfing}}]{Deilmann2020PRL}%
  \BibitemOpen
  \bibfield  {author} {\bibinfo {author} {\bibfnamefont {T.}~\bibnamefont
  {Deilmann}}, \bibinfo {author} {\bibfnamefont {P.}~\bibnamefont {Kr\"uger}},\
  and\ \bibinfo {author} {\bibfnamefont {M.}~\bibnamefont {Rohlfing}},\ }\href
  {https://doi.org/10.1103/PhysRevLett.124.226402} {\bibfield  {journal}
  {\bibinfo  {journal} {Phys. Rev. Lett.}\ }\textbf {\bibinfo {volume} {124}},\
  \bibinfo {pages} {226402} (\bibinfo {year} {2020})}\BibitemShut {NoStop}%
\bibitem [{\citenamefont {F{\"o}rste}\ \emph
  {et~al.}(2020{\natexlab{b}})\citenamefont {F{\"o}rste}, \citenamefont
  {Tepliakov}, \citenamefont {Kruchinin}, \citenamefont {Lindlau},
  \citenamefont {Funk}, \citenamefont {F{\"o}rg}, \citenamefont {Watanabe},
  \citenamefont {Taniguchi}, \citenamefont {Baimuratov},\ and\ \citenamefont
  {H{\"o}gele}}]{Forste2020NatComm}%
  \BibitemOpen
  \bibfield  {author} {\bibinfo {author} {\bibfnamefont {J.}~\bibnamefont
  {F{\"o}rste}}, \bibinfo {author} {\bibfnamefont {N.~V.}\ \bibnamefont
  {Tepliakov}}, \bibinfo {author} {\bibfnamefont {S.~Y.}\ \bibnamefont
  {Kruchinin}}, \bibinfo {author} {\bibfnamefont {J.}~\bibnamefont {Lindlau}},
  \bibinfo {author} {\bibfnamefont {V.}~\bibnamefont {Funk}}, \bibinfo {author}
  {\bibfnamefont {M.}~\bibnamefont {F{\"o}rg}}, \bibinfo {author}
  {\bibfnamefont {K.}~\bibnamefont {Watanabe}}, \bibinfo {author}
  {\bibfnamefont {T.}~\bibnamefont {Taniguchi}}, \bibinfo {author}
  {\bibfnamefont {A.~S.}\ \bibnamefont {Baimuratov}},\ and\ \bibinfo {author}
  {\bibfnamefont {A.}~\bibnamefont {H{\"o}gele}},\ }\href@noop {} {\bibfield
  {journal} {\bibinfo  {journal} {Nature Commun.}\ }\textbf {\bibinfo {volume}
  {11}},\ \bibinfo {pages} {4539} (\bibinfo {year}
  {2020}{\natexlab{b}})}\BibitemShut {NoStop}%
\bibitem [{\citenamefont {Xuan}\ and\ \citenamefont
  {Quek}(2020)}]{Xuan2020PRR}%
  \BibitemOpen
  \bibfield  {author} {\bibinfo {author} {\bibfnamefont {F.}~\bibnamefont
  {Xuan}}\ and\ \bibinfo {author} {\bibfnamefont {S.~Y.}\ \bibnamefont
  {Quek}},\ }\href {https://doi.org/10.1103/PhysRevResearch.2.033256}
  {\bibfield  {journal} {\bibinfo  {journal} {Phys. Rev. Research}\ }\textbf
  {\bibinfo {volume} {2}},\ \bibinfo {pages} {033256} (\bibinfo {year}
  {2020})}\BibitemShut {NoStop}%
\bibitem [{\citenamefont {{Faria Junior}}\ \emph {et~al.}(2022)\citenamefont
  {{Faria Junior}}, \citenamefont {Zollner}, \citenamefont {Woźniak},
  \citenamefont {Kurpas}, \citenamefont {Gmitra},\ and\ \citenamefont
  {Fabian}}]{FariaJunior_2022}%
  \BibitemOpen
  \bibfield  {author} {\bibinfo {author} {\bibfnamefont {P.~E.}\ \bibnamefont
  {{Faria Junior}}}, \bibinfo {author} {\bibfnamefont {K.}~\bibnamefont
  {Zollner}}, \bibinfo {author} {\bibfnamefont {T.}~\bibnamefont {Woźniak}},
  \bibinfo {author} {\bibfnamefont {M.}~\bibnamefont {Kurpas}}, \bibinfo
  {author} {\bibfnamefont {M.}~\bibnamefont {Gmitra}},\ and\ \bibinfo {author}
  {\bibfnamefont {J.}~\bibnamefont {Fabian}},\ }\href
  {https://doi.org/10.1088/1367-2630/ac7e21} {\bibfield  {journal} {\bibinfo
  {journal} {New Journal of Physics}\ }\textbf {\bibinfo {volume} {24}},\
  \bibinfo {pages} {083004} (\bibinfo {year} {2022})}\BibitemShut {NoStop}%
\bibitem [{\citenamefont {James}\ and\ \citenamefont {Lavik}(1963)}]{MoSe2}%
  \BibitemOpen
  \bibfield  {author} {\bibinfo {author} {\bibfnamefont {P.~B.}\ \bibnamefont
  {James}}\ and\ \bibinfo {author} {\bibfnamefont {M.~T.}\ \bibnamefont
  {Lavik}},\ }\href {https://doi.org/https://doi.org/10.1107/S0365110X6300311X}
  {\bibfield  {journal} {\bibinfo  {journal} {Acta Crystallographica}\ }\textbf
  {\bibinfo {volume} {16}},\ \bibinfo {pages} {1183} (\bibinfo {year}
  {1963})}\BibitemShut {NoStop}%
\bibitem [{\citenamefont {Schutte}\ \emph {et~al.}(1987)\citenamefont
  {Schutte}, \citenamefont {{De Boer}},\ and\ \citenamefont {Jellinek}}]{WSe2}%
  \BibitemOpen
  \bibfield  {author} {\bibinfo {author} {\bibfnamefont {W.}~\bibnamefont
  {Schutte}}, \bibinfo {author} {\bibfnamefont {J.}~\bibnamefont {{De Boer}}},\
  and\ \bibinfo {author} {\bibfnamefont {F.}~\bibnamefont {Jellinek}},\ }\href
  {https://doi.org/https://doi.org/10.1016/0022-4596(87)90057-0} {\bibfield
  {journal} {\bibinfo  {journal} {Journal of Solid State Chemistry}\ }\textbf
  {\bibinfo {volume} {70}},\ \bibinfo {pages} {207} (\bibinfo {year}
  {1987})}\BibitemShut {NoStop}%
\bibitem [{\citenamefont {Zollner}\ \emph {et~al.}(2019)\citenamefont
  {Zollner}, \citenamefont {Faria~Junior},\ and\ \citenamefont
  {Fabian}}]{Zollner2019PRB}%
  \BibitemOpen
  \bibfield  {author} {\bibinfo {author} {\bibfnamefont {K.}~\bibnamefont
  {Zollner}}, \bibinfo {author} {\bibfnamefont {P.~E.}\ \bibnamefont
  {Faria~Junior}},\ and\ \bibinfo {author} {\bibfnamefont {J.}~\bibnamefont
  {Fabian}},\ }\href {https://doi.org/10.1103/PhysRevB.100.195126} {\bibfield
  {journal} {\bibinfo  {journal} {Phys. Rev. B}\ }\textbf {\bibinfo {volume}
  {100}},\ \bibinfo {pages} {195126} (\bibinfo {year} {2019})}\BibitemShut
  {NoStop}%
\bibitem [{\citenamefont {Dirnberger}\ \emph {et~al.}(2021)\citenamefont
  {Dirnberger}, \citenamefont {Ziegler}, \citenamefont {{Faria~Junior}},
  \citenamefont {Bushati}, \citenamefont {Taniguchi}, \citenamefont {Watanabe},
  \citenamefont {Fabian}, \citenamefont {Bougeard}, \citenamefont {Chernikov},\
  and\ \citenamefont {Menon}}]{Dirnberger2021SciAdv}%
  \BibitemOpen
  \bibfield  {author} {\bibinfo {author} {\bibfnamefont {F.}~\bibnamefont
  {Dirnberger}}, \bibinfo {author} {\bibfnamefont {J.~D.}\ \bibnamefont
  {Ziegler}}, \bibinfo {author} {\bibfnamefont {P.~E.}\ \bibnamefont
  {{Faria~Junior}}}, \bibinfo {author} {\bibfnamefont {R.}~\bibnamefont
  {Bushati}}, \bibinfo {author} {\bibfnamefont {T.}~\bibnamefont {Taniguchi}},
  \bibinfo {author} {\bibfnamefont {K.}~\bibnamefont {Watanabe}}, \bibinfo
  {author} {\bibfnamefont {J.}~\bibnamefont {Fabian}}, \bibinfo {author}
  {\bibfnamefont {D.}~\bibnamefont {Bougeard}}, \bibinfo {author}
  {\bibfnamefont {A.}~\bibnamefont {Chernikov}},\ and\ \bibinfo {author}
  {\bibfnamefont {V.~M.}\ \bibnamefont {Menon}},\ }\href
  {https://doi.org/10.1126/sciadv.abj3066} {\bibfield  {journal} {\bibinfo
  {journal} {Science Advances}\ }\textbf {\bibinfo {volume} {7}},\ \bibinfo
  {pages} {eabj3066} (\bibinfo {year} {2021})}\BibitemShut {NoStop}%
\bibitem [{\citenamefont {Xenogiannopoulou}\ \emph {et~al.}(2015)\citenamefont
  {Xenogiannopoulou}, \citenamefont {Tsipas}, \citenamefont {Aretouli},
  \citenamefont {Tsoutsou}, \citenamefont {Giamini}, \citenamefont {Bazioti},
  \citenamefont {Dimitrakopulos}, \citenamefont {Komninou}, \citenamefont
  {Brems}, \citenamefont {Huyghebaert}, \citenamefont {Radu},\ and\
  \citenamefont {Dimoulas}}]{C4NR06874B}%
  \BibitemOpen
  \bibfield  {author} {\bibinfo {author} {\bibfnamefont {E.}~\bibnamefont
  {Xenogiannopoulou}}, \bibinfo {author} {\bibfnamefont {P.}~\bibnamefont
  {Tsipas}}, \bibinfo {author} {\bibfnamefont {K.~E.}\ \bibnamefont
  {Aretouli}}, \bibinfo {author} {\bibfnamefont {D.}~\bibnamefont {Tsoutsou}},
  \bibinfo {author} {\bibfnamefont {S.~A.}\ \bibnamefont {Giamini}}, \bibinfo
  {author} {\bibfnamefont {C.}~\bibnamefont {Bazioti}}, \bibinfo {author}
  {\bibfnamefont {G.~P.}\ \bibnamefont {Dimitrakopulos}}, \bibinfo {author}
  {\bibfnamefont {P.}~\bibnamefont {Komninou}}, \bibinfo {author}
  {\bibfnamefont {S.}~\bibnamefont {Brems}}, \bibinfo {author} {\bibfnamefont
  {C.}~\bibnamefont {Huyghebaert}}, \bibinfo {author} {\bibfnamefont {I.~P.}\
  \bibnamefont {Radu}},\ and\ \bibinfo {author} {\bibfnamefont
  {A.}~\bibnamefont {Dimoulas}},\ }\href {https://doi.org/10.1039/C4NR06874B}
  {\bibfield  {journal} {\bibinfo  {journal} {Nanoscale}\ }\textbf {\bibinfo
  {volume} {7}},\ \bibinfo {pages} {7896} (\bibinfo {year} {2015})}\BibitemShut
  {NoStop}%
\bibitem [{\citenamefont {Wilson}\ and\ \citenamefont
  {Yoffe}(1969)}]{Wilson01051969}%
  \BibitemOpen
  \bibfield  {author} {\bibinfo {author} {\bibfnamefont {J.}~\bibnamefont
  {Wilson}}\ and\ \bibinfo {author} {\bibfnamefont {A.}~\bibnamefont {Yoffe}},\
  }\href {https://doi.org/10.1080/00018736900101307} {\bibfield  {journal}
  {\bibinfo  {journal} {Advances in Physics}\ }\textbf {\bibinfo {volume}
  {18}},\ \bibinfo {pages} {193} (\bibinfo {year} {1969})}\BibitemShut
  {NoStop}%
\bibitem [{\citenamefont {Al-Hilli}\ and\ \citenamefont
  {Evans}(1972)}]{ALHILLI197293}%
  \BibitemOpen
  \bibfield  {author} {\bibinfo {author} {\bibfnamefont {A.}~\bibnamefont
  {Al-Hilli}}\ and\ \bibinfo {author} {\bibfnamefont {B.}~\bibnamefont
  {Evans}},\ }\href
  {https://doi.org/https://doi.org/10.1016/0022-0248(72)90129-7} {\bibfield
  {journal} {\bibinfo  {journal} {Journal of Crystal Growth}\ }\textbf
  {\bibinfo {volume} {15}},\ \bibinfo {pages} {93} (\bibinfo {year}
  {1972})}\BibitemShut {NoStop}%
\bibitem [{\citenamefont {Bronsema}\ \emph {et~al.}(1986)\citenamefont
  {Bronsema}, \citenamefont {De~Boer},\ and\ \citenamefont
  {Jellinek}}]{Bronsema1986}%
  \BibitemOpen
  \bibfield  {author} {\bibinfo {author} {\bibfnamefont {K.~D.}\ \bibnamefont
  {Bronsema}}, \bibinfo {author} {\bibfnamefont {J.~L.}\ \bibnamefont
  {De~Boer}},\ and\ \bibinfo {author} {\bibfnamefont {F.}~\bibnamefont
  {Jellinek}},\ }\href
  {https://doi.org/https://doi.org/10.1002/zaac.19865400904} {\bibfield
  {journal} {\bibinfo  {journal} {Zeitschrift für anorganische und allgemeine
  Chemie}\ }\textbf {\bibinfo {volume} {540}},\ \bibinfo {pages} {15} (\bibinfo
  {year} {1986})}\BibitemShut {NoStop}%
\bibitem [{\citenamefont {Evans}\ and\ \citenamefont
  {Hazelwood}(1971)}]{Evans1981}%
  \BibitemOpen
  \bibfield  {author} {\bibinfo {author} {\bibfnamefont {B.~L.}\ \bibnamefont
  {Evans}}\ and\ \bibinfo {author} {\bibfnamefont {R.~A.}\ \bibnamefont
  {Hazelwood}},\ }\href
  {https://doi.org/https://doi.org/10.1002/pssa.2210040119} {\bibfield
  {journal} {\bibinfo  {journal} {physica status solidi (a)}\ }\textbf
  {\bibinfo {volume} {4}},\ \bibinfo {pages} {181} (\bibinfo {year}
  {1971})}\BibitemShut {NoStop}%
\bibitem [{\citenamefont {Mennel}\ \emph {et~al.}(2018)\citenamefont {Mennel},
  \citenamefont {Furchi}, \citenamefont {Wachter}, \citenamefont {Paur},
  \citenamefont {Polyushkin},\ and\ \citenamefont
  {Mueller}}]{Mennel2018NatComm}%
  \BibitemOpen
  \bibfield  {author} {\bibinfo {author} {\bibfnamefont {L.}~\bibnamefont
  {Mennel}}, \bibinfo {author} {\bibfnamefont {M.~M.}\ \bibnamefont {Furchi}},
  \bibinfo {author} {\bibfnamefont {S.}~\bibnamefont {Wachter}}, \bibinfo
  {author} {\bibfnamefont {M.}~\bibnamefont {Paur}}, \bibinfo {author}
  {\bibfnamefont {D.~K.}\ \bibnamefont {Polyushkin}},\ and\ \bibinfo {author}
  {\bibfnamefont {T.}~\bibnamefont {Mueller}},\ }\href
  {https://doi.org/10.1038/s41467-018-02830-y} {\bibfield  {journal} {\bibinfo
  {journal} {Nature Commun.}\ }\textbf {\bibinfo {volume} {9}},\ \bibinfo
  {pages} {516} (\bibinfo {year} {2018})}\BibitemShut {NoStop}%
\bibitem [{\citenamefont {Kolesnichenko}\ \emph {et~al.}(2020)\citenamefont
  {Kolesnichenko}, \citenamefont {Zhang}, \citenamefont {Yun}, \citenamefont
  {Zheng}, \citenamefont {Fuhrer},\ and\ \citenamefont
  {Davis}}]{Kolesnichenko2020TDM}%
  \BibitemOpen
  \bibfield  {author} {\bibinfo {author} {\bibfnamefont {P.~V.}\ \bibnamefont
  {Kolesnichenko}}, \bibinfo {author} {\bibfnamefont {Q.}~\bibnamefont
  {Zhang}}, \bibinfo {author} {\bibfnamefont {T.}~\bibnamefont {Yun}}, \bibinfo
  {author} {\bibfnamefont {C.}~\bibnamefont {Zheng}}, \bibinfo {author}
  {\bibfnamefont {M.~S.}\ \bibnamefont {Fuhrer}},\ and\ \bibinfo {author}
  {\bibfnamefont {J.~A.}\ \bibnamefont {Davis}},\ }\href
  {https://doi.org/10.1088/2053-1583/ab626a} {\bibfield  {journal} {\bibinfo
  {journal} {2D Materials}\ }\textbf {\bibinfo {volume} {7}},\ \bibinfo {pages}
  {025008} (\bibinfo {year} {2020})}\BibitemShut {NoStop}%
\bibitem [{\citenamefont {Darlington}\ \emph {et~al.}(2020)\citenamefont
  {Darlington}, \citenamefont {Carmesin}, \citenamefont {Florian},
  \citenamefont {Yanev}, \citenamefont {Ajayi}, \citenamefont {Ardelean},
  \citenamefont {Rhodes}, \citenamefont {Ghiotto}, \citenamefont {Krayev},
  \citenamefont {Watanabe}, \citenamefont {Taniguchi}, \citenamefont {Kysar},
  \citenamefont {Pasupathy}, \citenamefont {Hone}, \citenamefont {Jahnke},
  \citenamefont {Borys},\ and\ \citenamefont {Schuck}}]{Darlington2020NatNano}%
  \BibitemOpen
  \bibfield  {author} {\bibinfo {author} {\bibfnamefont {T.~P.}\ \bibnamefont
  {Darlington}}, \bibinfo {author} {\bibfnamefont {C.}~\bibnamefont
  {Carmesin}}, \bibinfo {author} {\bibfnamefont {M.}~\bibnamefont {Florian}},
  \bibinfo {author} {\bibfnamefont {E.}~\bibnamefont {Yanev}}, \bibinfo
  {author} {\bibfnamefont {O.}~\bibnamefont {Ajayi}}, \bibinfo {author}
  {\bibfnamefont {J.}~\bibnamefont {Ardelean}}, \bibinfo {author}
  {\bibfnamefont {D.~A.}\ \bibnamefont {Rhodes}}, \bibinfo {author}
  {\bibfnamefont {A.}~\bibnamefont {Ghiotto}}, \bibinfo {author} {\bibfnamefont
  {A.}~\bibnamefont {Krayev}}, \bibinfo {author} {\bibfnamefont
  {K.}~\bibnamefont {Watanabe}}, \bibinfo {author} {\bibfnamefont
  {T.}~\bibnamefont {Taniguchi}}, \bibinfo {author} {\bibfnamefont {J.~W.}\
  \bibnamefont {Kysar}}, \bibinfo {author} {\bibfnamefont {A.~N.}\ \bibnamefont
  {Pasupathy}}, \bibinfo {author} {\bibfnamefont {J.~C.}\ \bibnamefont {Hone}},
  \bibinfo {author} {\bibfnamefont {F.}~\bibnamefont {Jahnke}}, \bibinfo
  {author} {\bibfnamefont {N.~J.}\ \bibnamefont {Borys}},\ and\ \bibinfo
  {author} {\bibfnamefont {P.~J.}\ \bibnamefont {Schuck}},\ }\href
  {https://doi.org/10.1038/s41565-020-0730-5} {\bibfield  {journal} {\bibinfo
  {journal} {Nature Nanotechnology}\ }\textbf {\bibinfo {volume} {15}},\
  \bibinfo {pages} {854} (\bibinfo {year} {2020})}\BibitemShut {NoStop}%
\bibitem [{\citenamefont {Alexeev}\ \emph {et~al.}(2020)\citenamefont
  {Alexeev}, \citenamefont {Mullin}, \citenamefont {Ares}, \citenamefont
  {Nevison-Andrews}, \citenamefont {Skrypka}, \citenamefont {Godde},
  \citenamefont {Kozikov}, \citenamefont {Hague}, \citenamefont {Wang},
  \citenamefont {Novoselov}, \citenamefont {Fumagalli}, \citenamefont {Hobbs},\
  and\ \citenamefont {Tartakovskii}}]{Alexeev2020ACSNano}%
  \BibitemOpen
  \bibfield  {author} {\bibinfo {author} {\bibfnamefont {E.~M.}\ \bibnamefont
  {Alexeev}}, \bibinfo {author} {\bibfnamefont {N.}~\bibnamefont {Mullin}},
  \bibinfo {author} {\bibfnamefont {P.}~\bibnamefont {Ares}}, \bibinfo {author}
  {\bibfnamefont {H.}~\bibnamefont {Nevison-Andrews}}, \bibinfo {author}
  {\bibfnamefont {O.}~\bibnamefont {Skrypka}}, \bibinfo {author} {\bibfnamefont
  {T.}~\bibnamefont {Godde}}, \bibinfo {author} {\bibfnamefont
  {A.}~\bibnamefont {Kozikov}}, \bibinfo {author} {\bibfnamefont
  {L.}~\bibnamefont {Hague}}, \bibinfo {author} {\bibfnamefont
  {Y.}~\bibnamefont {Wang}}, \bibinfo {author} {\bibfnamefont {K.~S.}\
  \bibnamefont {Novoselov}}, \bibinfo {author} {\bibfnamefont {L.}~\bibnamefont
  {Fumagalli}}, \bibinfo {author} {\bibfnamefont {J.~K.}\ \bibnamefont
  {Hobbs}},\ and\ \bibinfo {author} {\bibfnamefont {A.~I.}\ \bibnamefont
  {Tartakovskii}},\ }\href {https://doi.org/10.1021/acsnano.0c01146} {\bibfield
   {journal} {\bibinfo  {journal} {{ACS} Nano}\ }\textbf {\bibinfo {volume}
  {14}},\ \bibinfo {pages} {11110} (\bibinfo {year} {2020})}\BibitemShut
  {NoStop}%
\bibitem [{\citenamefont {Blundo}\ \emph {et~al.}(2021)\citenamefont {Blundo},
  \citenamefont {Cappelluti}, \citenamefont {Felici}, \citenamefont
  {Pettinari},\ and\ \citenamefont {Polimeni}}]{BlundoAPR}%
  \BibitemOpen
  \bibfield  {author} {\bibinfo {author} {\bibfnamefont {E.}~\bibnamefont
  {Blundo}}, \bibinfo {author} {\bibfnamefont {E.}~\bibnamefont {Cappelluti}},
  \bibinfo {author} {\bibfnamefont {M.}~\bibnamefont {Felici}}, \bibinfo
  {author} {\bibfnamefont {G.}~\bibnamefont {Pettinari}},\ and\ \bibinfo
  {author} {\bibfnamefont {A.}~\bibnamefont {Polimeni}},\ }\href
  {https://doi.org/10.1063/5.0037852} {\bibfield  {journal} {\bibinfo
  {journal} {Appl. Phys. Rev.}\ }\textbf {\bibinfo {volume} {8}},\ \bibinfo
  {pages} {021318} (\bibinfo {year} {2021})}\BibitemShut {NoStop}%
\bibitem [{\citenamefont {Blundo}\ \emph {et~al.}(2022)\citenamefont {Blundo},
  \citenamefont {{Faria Junior}}, \citenamefont {Surrente}, \citenamefont
  {Pettinari}, \citenamefont {Prosnikov}, \citenamefont {Olkowska-Pucko},
  \citenamefont {Zollner}, \citenamefont {Wo\'{z}niak}, \citenamefont {Chaves},
  \citenamefont {Kazimierczuk}, \citenamefont {Felici}, \citenamefont
  {Babi\'{n}ski}, \citenamefont {Molas}, \citenamefont {Christianen},
  \citenamefont {Fabian},\ and\ \citenamefont {Polimeni}}]{Blundo_magneto_prl}%
  \BibitemOpen
  \bibfield  {author} {\bibinfo {author} {\bibfnamefont {E.}~\bibnamefont
  {Blundo}}, \bibinfo {author} {\bibfnamefont {P.~E.}\ \bibnamefont {{Faria
  Junior}}}, \bibinfo {author} {\bibfnamefont {A.}~\bibnamefont {Surrente}},
  \bibinfo {author} {\bibfnamefont {G.}~\bibnamefont {Pettinari}}, \bibinfo
  {author} {\bibfnamefont {M.~A.}\ \bibnamefont {Prosnikov}}, \bibinfo {author}
  {\bibfnamefont {K.}~\bibnamefont {Olkowska-Pucko}}, \bibinfo {author}
  {\bibfnamefont {K.}~\bibnamefont {Zollner}}, \bibinfo {author} {\bibfnamefont
  {T.}~\bibnamefont {Wo\'{z}niak}}, \bibinfo {author} {\bibfnamefont
  {A.}~\bibnamefont {Chaves}}, \bibinfo {author} {\bibfnamefont
  {T.}~\bibnamefont {Kazimierczuk}}, \bibinfo {author} {\bibfnamefont
  {M.}~\bibnamefont {Felici}}, \bibinfo {author} {\bibfnamefont
  {A.}~\bibnamefont {Babi\'{n}ski}}, \bibinfo {author} {\bibfnamefont {M.~R.}\
  \bibnamefont {Molas}}, \bibinfo {author} {\bibfnamefont {P.~C.~M.}\
  \bibnamefont {Christianen}}, \bibinfo {author} {\bibfnamefont
  {J.}~\bibnamefont {Fabian}},\ and\ \bibinfo {author} {\bibfnamefont
  {A.}~\bibnamefont {Polimeni}},\ }\href
  {https://doi.org/10.1103/PhysRevLett.129.067402} {\bibfield  {journal}
  {\bibinfo  {journal} {Phys. Rev. Lett.}\ }\textbf {\bibinfo {volume} {129}},\
  \bibinfo {pages} {067402} (\bibinfo {year} {2022})}\BibitemShut {NoStop}%
\bibitem [{\citenamefont {Covre}\ \emph {et~al.}(2022)\citenamefont {Covre},
  \citenamefont {{Faria~Junior}}, \citenamefont {Gordo}, \citenamefont
  {de~Brito}, \citenamefont {Zhumagulov}, \citenamefont {Teodoro},
  \citenamefont {Couto}, \citenamefont {Misoguti}, \citenamefont {Pratavieira},
  \citenamefont {Andrade}, \citenamefont {Christianen}, \citenamefont {Fabian},
  \citenamefont {Withers},\ and\ \citenamefont
  {Galvão~Gobato}}]{Covre2022Nanoscale}%
  \BibitemOpen
  \bibfield  {author} {\bibinfo {author} {\bibfnamefont {F.~S.}\ \bibnamefont
  {Covre}}, \bibinfo {author} {\bibfnamefont {P.~E.}\ \bibnamefont
  {{Faria~Junior}}}, \bibinfo {author} {\bibfnamefont {V.~O.}\ \bibnamefont
  {Gordo}}, \bibinfo {author} {\bibfnamefont {C.~S.}\ \bibnamefont {de~Brito}},
  \bibinfo {author} {\bibfnamefont {Y.~V.}\ \bibnamefont {Zhumagulov}},
  \bibinfo {author} {\bibfnamefont {M.~D.}\ \bibnamefont {Teodoro}}, \bibinfo
  {author} {\bibfnamefont {O.~D.~D.}\ \bibnamefont {Couto}}, \bibinfo {author}
  {\bibfnamefont {L.}~\bibnamefont {Misoguti}}, \bibinfo {author}
  {\bibfnamefont {S.}~\bibnamefont {Pratavieira}}, \bibinfo {author}
  {\bibfnamefont {M.~B.}\ \bibnamefont {Andrade}}, \bibinfo {author}
  {\bibfnamefont {P.~C.~M.}\ \bibnamefont {Christianen}}, \bibinfo {author}
  {\bibfnamefont {J.}~\bibnamefont {Fabian}}, \bibinfo {author} {\bibfnamefont
  {F.}~\bibnamefont {Withers}},\ and\ \bibinfo {author} {\bibfnamefont
  {Y.}~\bibnamefont {Galvão~Gobato}},\ }\href
  {https://doi.org/10.1039/D2NR00315E} {\bibfield  {journal} {\bibinfo
  {journal} {Nanoscale}\ }\textbf {\bibinfo {volume} {14}},\ \bibinfo {pages}
  {5758} (\bibinfo {year} {2022})}\BibitemShut {NoStop}%
\bibitem [{\citenamefont {{Faria~Junior}}\ \emph {et~al.}(2023)\citenamefont
  {{Faria~Junior}}, \citenamefont {Naimer}, \citenamefont {McCreary},
  \citenamefont {Jonker}, \citenamefont {Finley}, \citenamefont {Crooker},
  \citenamefont {Fabian},\ and\ \citenamefont {Stier}}]{FariaJunior2023TDM}%
  \BibitemOpen
  \bibfield  {author} {\bibinfo {author} {\bibfnamefont {P.~E.}\ \bibnamefont
  {{Faria~Junior}}}, \bibinfo {author} {\bibfnamefont {T.}~\bibnamefont
  {Naimer}}, \bibinfo {author} {\bibfnamefont {K.~M.}\ \bibnamefont
  {McCreary}}, \bibinfo {author} {\bibfnamefont {B.~T.}\ \bibnamefont
  {Jonker}}, \bibinfo {author} {\bibfnamefont {J.~J.}\ \bibnamefont {Finley}},
  \bibinfo {author} {\bibfnamefont {S.~A.}\ \bibnamefont {Crooker}}, \bibinfo
  {author} {\bibfnamefont {J.}~\bibnamefont {Fabian}},\ and\ \bibinfo {author}
  {\bibfnamefont {A.~V.}\ \bibnamefont {Stier}},\ }\href
  {https://doi.org/10.1088/2053-1583/acd5df} {\bibfield  {journal} {\bibinfo
  {journal} {2D Materials}\ }\textbf {\bibinfo {volume} {10}},\ \bibinfo
  {pages} {034002} (\bibinfo {year} {2023})}\BibitemShut {NoStop}%
\bibitem [{\citenamefont {Kresse}\ and\ \citenamefont
  {Furthm{\"{u}}ller}(1996)}]{VASP}%
  \BibitemOpen
  \bibfield  {author} {\bibinfo {author} {\bibfnamefont {G.}~\bibnamefont
  {Kresse}}\ and\ \bibinfo {author} {\bibfnamefont {J.}~\bibnamefont
  {Furthm{\"{u}}ller}},\ }\href {https://doi.org/10.1103/PhysRevB.54.11169}
  {\bibfield  {journal} {\bibinfo  {journal} {Physical Review B}\ }\textbf
  {\bibinfo {volume} {54}},\ \bibinfo {pages} {11169} (\bibinfo {year}
  {1996})}\BibitemShut {NoStop}%
\bibitem [{\citenamefont {Kresse}\ and\ \citenamefont {Joubert}(1999)}]{PAW}%
  \BibitemOpen
  \bibfield  {author} {\bibinfo {author} {\bibfnamefont {G.}~\bibnamefont
  {Kresse}}\ and\ \bibinfo {author} {\bibfnamefont {D.}~\bibnamefont
  {Joubert}},\ }\href {https://doi.org/10.1103/PhysRevB.59.1758} {\bibfield
  {journal} {\bibinfo  {journal} {Physical Review B}\ }\textbf {\bibinfo
  {volume} {59}},\ \bibinfo {pages} {1758} (\bibinfo {year}
  {1999})}\BibitemShut {NoStop}%
\bibitem [{\citenamefont {Perdew}\ \emph {et~al.}(1996)\citenamefont {Perdew},
  \citenamefont {Burke},\ and\ \citenamefont {Ernzerhof}}]{PBE}%
  \BibitemOpen
  \bibfield  {author} {\bibinfo {author} {\bibfnamefont {J.~P.}\ \bibnamefont
  {Perdew}}, \bibinfo {author} {\bibfnamefont {K.}~\bibnamefont {Burke}},\ and\
  \bibinfo {author} {\bibfnamefont {M.}~\bibnamefont {Ernzerhof}},\ }\href
  {https://doi.org/10.1103/PhysRevLett.77.3865} {\bibfield  {journal} {\bibinfo
   {journal} {Physical Review Letters}\ }\textbf {\bibinfo {volume} {77}},\
  \bibinfo {pages} {3865} (\bibinfo {year} {1996})}\BibitemShut {NoStop}%
\bibitem [{\citenamefont {Grimme}\ \emph {et~al.}(2010)\citenamefont {Grimme},
  \citenamefont {Antony}, \citenamefont {Ehrlich},\ and\ \citenamefont
  {Krieg}}]{D3}%
  \BibitemOpen
  \bibfield  {author} {\bibinfo {author} {\bibfnamefont {S.}~\bibnamefont
  {Grimme}}, \bibinfo {author} {\bibfnamefont {J.}~\bibnamefont {Antony}},
  \bibinfo {author} {\bibfnamefont {S.}~\bibnamefont {Ehrlich}},\ and\ \bibinfo
  {author} {\bibfnamefont {H.}~\bibnamefont {Krieg}},\ }\href
  {https://doi.org/10.1063/1.3382344} {\bibfield  {journal} {\bibinfo
  {journal} {The Journal of Chemical Physics}\ }\textbf {\bibinfo {volume}
  {132}},\ \bibinfo {pages} {154104} (\bibinfo {year} {2010})}\BibitemShut
  {NoStop}%
\bibitem [{\citenamefont {Gajdo{\v{s}}}\ \emph {et~al.}(2006)\citenamefont
  {Gajdo{\v{s}}} \emph {et~al.}}]{Gajdos}%
  \BibitemOpen
  \bibfield  {author} {\bibinfo {author} {\bibfnamefont {M.}~\bibnamefont
  {Gajdo{\v{s}}}} \emph {et~al.},\ }\href
  {https://doi.org/10.1103/PhysRevB.73.045112} {\bibfield  {journal} {\bibinfo
  {journal} {Phys. Rev. B}\ }\textbf {\bibinfo {volume} {73}},\ \bibinfo
  {pages} {045112} (\bibinfo {year} {2006})}\BibitemShut {NoStop}%
\bibitem [{\citenamefont {Jadczak}\ \emph {et~al.}(2017)\citenamefont
  {Jadczak}, \citenamefont {Kutrowska-Girzycka}, \citenamefont
  {Kapu{\'{s}}ci{\'{n}}ski}, \citenamefont {Huang}, \citenamefont
  {W{\'{o}}js},\ and\ \citenamefont {Bryja}}]{Jadczak2017}%
  \BibitemOpen
  \bibfield  {author} {\bibinfo {author} {\bibfnamefont {J.}~\bibnamefont
  {Jadczak}}, \bibinfo {author} {\bibfnamefont {J.}~\bibnamefont
  {Kutrowska-Girzycka}}, \bibinfo {author} {\bibfnamefont {P.}~\bibnamefont
  {Kapu{\'{s}}ci{\'{n}}ski}}, \bibinfo {author} {\bibfnamefont {Y.~S.}\
  \bibnamefont {Huang}}, \bibinfo {author} {\bibfnamefont {A.}~\bibnamefont
  {W{\'{o}}js}},\ and\ \bibinfo {author} {\bibfnamefont {L.}~\bibnamefont
  {Bryja}},\ }\href {https://doi.org/10.1088/1361-6528/aa87d0} {\bibfield
  {journal} {\bibinfo  {journal} {Nanotechnology}\ }\textbf {\bibinfo {volume}
  {28}},\ \bibinfo {pages} {395702} (\bibinfo {year} {2017})}\BibitemShut
  {NoStop}%
\bibitem [{\citenamefont {Klingshirn}(2012)}]{Klingshirn2012}%
  \BibitemOpen
  \bibfield  {author} {\bibinfo {author} {\bibfnamefont {C.~F.}\ \bibnamefont
  {Klingshirn}},\ }\href@noop {} {\emph {\bibinfo {title} {Semiconductor
  Optics}}}\ (\bibinfo  {publisher} {Springer},\ \bibinfo {year}
  {2012})\BibitemShut {NoStop}%
\bibitem [{\citenamefont {Zhang}\ \emph {et~al.}(2015)\citenamefont {Zhang},
  \citenamefont {You}, \citenamefont {Zhao},\ and\ \citenamefont
  {Heinz}}]{Zhang2015}%
  \BibitemOpen
  \bibfield  {author} {\bibinfo {author} {\bibfnamefont {X.-X.}\ \bibnamefont
  {Zhang}}, \bibinfo {author} {\bibfnamefont {Y.}~\bibnamefont {You}}, \bibinfo
  {author} {\bibfnamefont {S.~Y.~F.}\ \bibnamefont {Zhao}},\ and\ \bibinfo
  {author} {\bibfnamefont {T.~F.}\ \bibnamefont {Heinz}},\ }\href
  {https://doi.org/10.1103/PhysRevLett.115.257403} {\bibfield  {journal}
  {\bibinfo  {journal} {Phys. Rev. Lett.}\ }\textbf {\bibinfo {volume} {115}},\
  \bibinfo {pages} {257403} (\bibinfo {year} {2015})}\BibitemShut {NoStop}%
\bibitem [{\citenamefont {Mitioglu}\ \emph {et~al.}(2015)\citenamefont
  {Mitioglu}, \citenamefont {Plochocka}, \citenamefont {Granados~del Aguila},
  \citenamefont {Christianen}, \citenamefont {Deligeorgis}, \citenamefont
  {Anghel}, \citenamefont {Kulyuk},\ and\ \citenamefont
  {Maude}}]{Mitioglu2105}%
  \BibitemOpen
  \bibfield  {author} {\bibinfo {author} {\bibfnamefont {A.~A.}\ \bibnamefont
  {Mitioglu}}, \bibinfo {author} {\bibfnamefont {P.}~\bibnamefont {Plochocka}},
  \bibinfo {author} {\bibfnamefont {A.}~\bibnamefont {Granados~del Aguila}},
  \bibinfo {author} {\bibfnamefont {P.~C.~M.}\ \bibnamefont {Christianen}},
  \bibinfo {author} {\bibfnamefont {G.}~\bibnamefont {Deligeorgis}}, \bibinfo
  {author} {\bibfnamefont {S.}~\bibnamefont {Anghel}}, \bibinfo {author}
  {\bibfnamefont {L.}~\bibnamefont {Kulyuk}},\ and\ \bibinfo {author}
  {\bibfnamefont {D.~K.}\ \bibnamefont {Maude}},\ }\href
  {https://doi.org/10.1021/acs.nanolett.5b00626} {\bibfield  {journal}
  {\bibinfo  {journal} {Nano Letters}\ }\textbf {\bibinfo {volume} {15}},\
  \bibinfo {pages} {4387} (\bibinfo {year} {2015})}\BibitemShut {NoStop}%
\bibitem [{\citenamefont {Plechinger}\ \emph {et~al.}(2016)\citenamefont
  {Plechinger}, \citenamefont {Nagler}, \citenamefont {Arora}, \citenamefont
  {Granados~del Águila}, \citenamefont {Ballottin}, \citenamefont {Frank},
  \citenamefont {Steinleitner}, \citenamefont {Gmitra}, \citenamefont {Fabian},
  \citenamefont {Christianen}, \citenamefont {Bratschitsch}, \citenamefont
  {Sch{\"u}ller},\ and\ \citenamefont {Korn}}]{Plechinger2016}%
  \BibitemOpen
  \bibfield  {author} {\bibinfo {author} {\bibfnamefont {G.}~\bibnamefont
  {Plechinger}}, \bibinfo {author} {\bibfnamefont {P.}~\bibnamefont {Nagler}},
  \bibinfo {author} {\bibfnamefont {A.}~\bibnamefont {Arora}}, \bibinfo
  {author} {\bibfnamefont {A.}~\bibnamefont {Granados~del Águila}}, \bibinfo
  {author} {\bibfnamefont {M.~V.}\ \bibnamefont {Ballottin}}, \bibinfo {author}
  {\bibfnamefont {T.}~\bibnamefont {Frank}}, \bibinfo {author} {\bibfnamefont
  {P.}~\bibnamefont {Steinleitner}}, \bibinfo {author} {\bibfnamefont
  {M.}~\bibnamefont {Gmitra}}, \bibinfo {author} {\bibfnamefont
  {J.}~\bibnamefont {Fabian}}, \bibinfo {author} {\bibfnamefont {P.~C.~M.}\
  \bibnamefont {Christianen}}, \bibinfo {author} {\bibfnamefont
  {R.}~\bibnamefont {Bratschitsch}}, \bibinfo {author} {\bibfnamefont
  {C.}~\bibnamefont {Sch{\"u}ller}},\ and\ \bibinfo {author} {\bibfnamefont
  {T.}~\bibnamefont {Korn}},\ }\href
  {https://doi.org/10.1021/acs.nanolett.6b04171} {\bibfield  {journal}
  {\bibinfo  {journal} {Nano Letters}\ }\textbf {\bibinfo {volume} {16}},\
  \bibinfo {pages} {7899} (\bibinfo {year} {2016})}\BibitemShut {NoStop}%
\bibitem [{\citenamefont {Nagler}\ \emph {et~al.}(2018)\citenamefont {Nagler},
  \citenamefont {Ballottin}, \citenamefont {Mitioglu}, \citenamefont {Durnev},
  \citenamefont {Taniguchi}, \citenamefont {Watanabe}, \citenamefont
  {Chernikov}, \citenamefont {Sch\"uller}, \citenamefont {Glazov},
  \citenamefont {Christianen},\ and\ \citenamefont {Korn}}]{Nagler2018}%
  \BibitemOpen
  \bibfield  {author} {\bibinfo {author} {\bibfnamefont {P.}~\bibnamefont
  {Nagler}}, \bibinfo {author} {\bibfnamefont {M.~V.}\ \bibnamefont
  {Ballottin}}, \bibinfo {author} {\bibfnamefont {A.~A.}\ \bibnamefont
  {Mitioglu}}, \bibinfo {author} {\bibfnamefont {M.~V.}\ \bibnamefont
  {Durnev}}, \bibinfo {author} {\bibfnamefont {T.}~\bibnamefont {Taniguchi}},
  \bibinfo {author} {\bibfnamefont {K.}~\bibnamefont {Watanabe}}, \bibinfo
  {author} {\bibfnamefont {A.}~\bibnamefont {Chernikov}}, \bibinfo {author}
  {\bibfnamefont {C.}~\bibnamefont {Sch\"uller}}, \bibinfo {author}
  {\bibfnamefont {M.~M.}\ \bibnamefont {Glazov}}, \bibinfo {author}
  {\bibfnamefont {P.~C.~M.}\ \bibnamefont {Christianen}},\ and\ \bibinfo
  {author} {\bibfnamefont {T.}~\bibnamefont {Korn}},\ }\href
  {https://doi.org/10.1103/PhysRevLett.121.057402} {\bibfield  {journal}
  {\bibinfo  {journal} {Phys. Rev. Lett.}\ }\textbf {\bibinfo {volume} {121}},\
  \bibinfo {pages} {057402} (\bibinfo {year} {2018})}\BibitemShut {NoStop}%
\bibitem [{\citenamefont {Prando}\ \emph {et~al.}(2021)\citenamefont {Prando},
  \citenamefont {Severijnen}, \citenamefont {Barcelos}, \citenamefont
  {Zeitler}, \citenamefont {Christianen}, \citenamefont {Withers},\ and\
  \citenamefont {Galv\~ao Gobato}}]{Prando2021}%
  \BibitemOpen
  \bibfield  {author} {\bibinfo {author} {\bibfnamefont {G.~A.}\ \bibnamefont
  {Prando}}, \bibinfo {author} {\bibfnamefont {M.~E.}\ \bibnamefont
  {Severijnen}}, \bibinfo {author} {\bibfnamefont {I.~D.}\ \bibnamefont
  {Barcelos}}, \bibinfo {author} {\bibfnamefont {U.}~\bibnamefont {Zeitler}},
  \bibinfo {author} {\bibfnamefont {P.~C.~M.}\ \bibnamefont {Christianen}},
  \bibinfo {author} {\bibfnamefont {F.}~\bibnamefont {Withers}},\ and\ \bibinfo
  {author} {\bibfnamefont {Y.}~\bibnamefont {Galv\~ao Gobato}},\ }\href
  {https://doi.org/10.1103/PhysRevApplied.16.064055} {\bibfield  {journal}
  {\bibinfo  {journal} {Phys. Rev. Appl.}\ }\textbf {\bibinfo {volume} {16}},\
  \bibinfo {pages} {064055} (\bibinfo {year} {2021})}\BibitemShut {NoStop}%
\bibitem [{\citenamefont {Łucja Kipczak}\ \emph {et~al.}(2023)\citenamefont
  {Łucja Kipczak}, \citenamefont {Slobodeniuk}, \citenamefont {Woźniak},
  \citenamefont {Bhatnagar}, \citenamefont {Zawadzka}, \citenamefont
  {Olkowska-Pucko}, \citenamefont {Grzeszczyk}, \citenamefont {Watanabe},
  \citenamefont {Taniguchi}, \citenamefont {Babiński},\ and\ \citenamefont
  {Molas}}]{Kipczak_2023}%
  \BibitemOpen
  \bibfield  {author} {\bibinfo {author} {\bibnamefont {Łucja Kipczak}},
  \bibinfo {author} {\bibfnamefont {A.~O.}\ \bibnamefont {Slobodeniuk}},
  \bibinfo {author} {\bibfnamefont {T.}~\bibnamefont {Woźniak}}, \bibinfo
  {author} {\bibfnamefont {M.}~\bibnamefont {Bhatnagar}}, \bibinfo {author}
  {\bibfnamefont {N.}~\bibnamefont {Zawadzka}}, \bibinfo {author}
  {\bibfnamefont {K.}~\bibnamefont {Olkowska-Pucko}}, \bibinfo {author}
  {\bibfnamefont {M.}~\bibnamefont {Grzeszczyk}}, \bibinfo {author}
  {\bibfnamefont {K.}~\bibnamefont {Watanabe}}, \bibinfo {author}
  {\bibfnamefont {T.}~\bibnamefont {Taniguchi}}, \bibinfo {author}
  {\bibfnamefont {A.}~\bibnamefont {Babiński}},\ and\ \bibinfo {author}
  {\bibfnamefont {M.~R.}\ \bibnamefont {Molas}},\ }\href
  {https://doi.org/10.1088/2053-1583/acbc8b} {\bibfield  {journal} {\bibinfo
  {journal} {2D Materials}\ }\textbf {\bibinfo {volume} {10}},\ \bibinfo
  {pages} {025014} (\bibinfo {year} {2023})}\BibitemShut {NoStop}%
\bibitem [{\citenamefont {Wang}\ \emph {et~al.}(2017)\citenamefont {Wang},
  \citenamefont {Robert}, \citenamefont {Glazov}, \citenamefont {Cadiz},
  \citenamefont {Courtade}, \citenamefont {Amand}, \citenamefont {Lagarde},
  \citenamefont {Taniguchi}, \citenamefont {Watanabe}, \citenamefont
  {Urbaszek},\ and\ \citenamefont {Marie}}]{Wang2017}%
  \BibitemOpen
  \bibfield  {author} {\bibinfo {author} {\bibfnamefont {G.}~\bibnamefont
  {Wang}}, \bibinfo {author} {\bibfnamefont {C.}~\bibnamefont {Robert}},
  \bibinfo {author} {\bibfnamefont {M.~M.}\ \bibnamefont {Glazov}}, \bibinfo
  {author} {\bibfnamefont {F.}~\bibnamefont {Cadiz}}, \bibinfo {author}
  {\bibfnamefont {E.}~\bibnamefont {Courtade}}, \bibinfo {author}
  {\bibfnamefont {T.}~\bibnamefont {Amand}}, \bibinfo {author} {\bibfnamefont
  {D.}~\bibnamefont {Lagarde}}, \bibinfo {author} {\bibfnamefont
  {T.}~\bibnamefont {Taniguchi}}, \bibinfo {author} {\bibfnamefont
  {K.}~\bibnamefont {Watanabe}}, \bibinfo {author} {\bibfnamefont
  {B.}~\bibnamefont {Urbaszek}},\ and\ \bibinfo {author} {\bibfnamefont
  {X.}~\bibnamefont {Marie}},\ }\href
  {https://doi.org/10.1103/PhysRevLett.119.047401} {\bibfield  {journal}
  {\bibinfo  {journal} {Phys. Rev. Lett.}\ }\textbf {\bibinfo {volume} {119}},\
  \bibinfo {pages} {047401} (\bibinfo {year} {2017})}\BibitemShut {NoStop}%
\bibitem [{\citenamefont {Smole\ifmmode~\acute{n}\else \'{n}\fi{}ski}\ \emph
  {et~al.}(2016)\citenamefont {Smole\ifmmode~\acute{n}\else \'{n}\fi{}ski},
  \citenamefont {Goryca}, \citenamefont {Koperski}, \citenamefont {Faugeras},
  \citenamefont {Kazimierczuk}, \citenamefont {Bogucki}, \citenamefont
  {Nogajewski}, \citenamefont {Kossacki},\ and\ \citenamefont
  {Potemski}}]{Smolenski2016}%
  \BibitemOpen
  \bibfield  {author} {\bibinfo {author} {\bibfnamefont {T.}~\bibnamefont
  {Smole\ifmmode~\acute{n}\else \'{n}\fi{}ski}}, \bibinfo {author}
  {\bibfnamefont {M.}~\bibnamefont {Goryca}}, \bibinfo {author} {\bibfnamefont
  {M.}~\bibnamefont {Koperski}}, \bibinfo {author} {\bibfnamefont
  {C.}~\bibnamefont {Faugeras}}, \bibinfo {author} {\bibfnamefont
  {T.}~\bibnamefont {Kazimierczuk}}, \bibinfo {author} {\bibfnamefont
  {A.}~\bibnamefont {Bogucki}}, \bibinfo {author} {\bibfnamefont
  {K.}~\bibnamefont {Nogajewski}}, \bibinfo {author} {\bibfnamefont
  {P.}~\bibnamefont {Kossacki}},\ and\ \bibinfo {author} {\bibfnamefont
  {M.}~\bibnamefont {Potemski}},\ }\href
  {https://doi.org/10.1103/PhysRevX.6.021024} {\bibfield  {journal} {\bibinfo
  {journal} {Phys. Rev. X}\ }\textbf {\bibinfo {volume} {6}},\ \bibinfo {pages}
  {021024} (\bibinfo {year} {2016})}\BibitemShut {NoStop}%
\bibitem [{\citenamefont {Liu}\ \emph {et~al.}(2019{\natexlab{c}})\citenamefont
  {Liu}, \citenamefont {Gao}, \citenamefont {Zhang}, \citenamefont {He},
  \citenamefont {Yu},\ and\ \citenamefont {Liu}}]{Liu2019valley}%
  \BibitemOpen
  \bibfield  {author} {\bibinfo {author} {\bibfnamefont {Y.}~\bibnamefont
  {Liu}}, \bibinfo {author} {\bibfnamefont {Y.}~\bibnamefont {Gao}}, \bibinfo
  {author} {\bibfnamefont {S.}~\bibnamefont {Zhang}}, \bibinfo {author}
  {\bibfnamefont {J.}~\bibnamefont {He}}, \bibinfo {author} {\bibfnamefont
  {J.}~\bibnamefont {Yu}},\ and\ \bibinfo {author} {\bibfnamefont
  {Z.}~\bibnamefont {Liu}},\ }\href {https://doi.org/10.1007/s12274-019-2497-2}
  {\bibfield  {journal} {\bibinfo  {journal} {Nano Research}\ }\textbf
  {\bibinfo {volume} {12}},\ \bibinfo {pages} {2695} (\bibinfo {year}
  {2019}{\natexlab{c}})}\BibitemShut {NoStop}%
\bibitem [{\citenamefont {Ye}\ \emph {et~al.}(2018)\citenamefont {Ye},
  \citenamefont {Waldecker}, \citenamefont {Ma}, \citenamefont {Rhodes},
  \citenamefont {Antony}, \citenamefont {Kim}, \citenamefont {Zhang},
  \citenamefont {Deng}, \citenamefont {Jiang}, \citenamefont {Lu},
  \citenamefont {Smirnov}, \citenamefont {Watanabe}, \citenamefont {Taniguchi},
  \citenamefont {Hone},\ and\ \citenamefont {Heinz}}]{Ye2018}%
  \BibitemOpen
  \bibfield  {author} {\bibinfo {author} {\bibfnamefont {Z.}~\bibnamefont
  {Ye}}, \bibinfo {author} {\bibfnamefont {L.}~\bibnamefont {Waldecker}},
  \bibinfo {author} {\bibfnamefont {E.~Y.}\ \bibnamefont {Ma}}, \bibinfo
  {author} {\bibfnamefont {D.}~\bibnamefont {Rhodes}}, \bibinfo {author}
  {\bibfnamefont {A.}~\bibnamefont {Antony}}, \bibinfo {author} {\bibfnamefont
  {B.}~\bibnamefont {Kim}}, \bibinfo {author} {\bibfnamefont {X.-X.}\
  \bibnamefont {Zhang}}, \bibinfo {author} {\bibfnamefont {M.}~\bibnamefont
  {Deng}}, \bibinfo {author} {\bibfnamefont {Y.}~\bibnamefont {Jiang}},
  \bibinfo {author} {\bibfnamefont {Z.}~\bibnamefont {Lu}}, \bibinfo {author}
  {\bibfnamefont {D.}~\bibnamefont {Smirnov}}, \bibinfo {author} {\bibfnamefont
  {K.}~\bibnamefont {Watanabe}}, \bibinfo {author} {\bibfnamefont
  {T.}~\bibnamefont {Taniguchi}}, \bibinfo {author} {\bibfnamefont
  {J.}~\bibnamefont {Hone}},\ and\ \bibinfo {author} {\bibfnamefont {T.~F.}\
  \bibnamefont {Heinz}},\ }\href {https://doi.org/10.1038/s41467-018-05917-8}
  {\bibfield  {journal} {\bibinfo  {journal} {Nature Communications}\ }\textbf
  {\bibinfo {volume} {9}},\ \bibinfo {pages} {3718} (\bibinfo {year}
  {2018})}\BibitemShut {NoStop}%
\end{thebibliography}%

\newpage
\onecolumngrid

\renewcommand{\thefigure}{S\arabic{section}.\arabic{figure}}
\renewcommand{\thesection}{S\arabic{section}}

\begin{center}
	{\large{ {\bf Supplemental Material: \\ Extremely high excitonic $g$-factors in 2D crystals \\ by alloy-induced admixing of band states}}}
	\vskip0.5\baselineskip{Katarzyna Olkowska-Pucko,{$^{1}$} Tomasz Woźniak,{$^{1}$} Elena Blundo,{$^{2}$} Natalia Zawadzka,{$^{1}$} Łucja Kipczak,{$^{1}$} Paulo E. Faria Junior,{$^{3,4}$} Jan Szpakowski,{$^{1}$} Grzegorz Krasucki,{$^{1}$} Salvatore Cianci,{$^{2}$} Diana Vaclavkova,{$^{5}$}
	Dipankar Jana,{$^{5}$} Piotr Kapuściński,{$^{5}$} Amit~Pawbake,{$^{5}$} Shalini Badola,{$^{5}$} Magdalena Grzeszczyk,{$^{1,6}$} Daniele Cecchetti,{$^{7}$} Giorgio Pettinari,{$^{7}$} Igor Antoniazzi,{$^{7}$} Zden\v{e}k Sofer,{$^{8}$} Iva Plutnarová,{$^{8}$} Kenji Watanabe,{$^{9}$} Takashi Taniguchi,{$^{10}$} Clement Faugeras,{$^{5}$} Marek Potemski,{$^{1,5,11}$} Adam Babiński,{$^{1}$} Antonio Polimeni,{$^{2}$} and Maciej R. Molas{$^{1}$}}

	\vskip0.5\baselineskip{\em$^{1}$ University of Warsaw, Faculty of Physics, 02-093 Warsaw, Poland \\$^{2}$ Physics Department, Sapienza University of Rome, 00185 Rome, Italy \\$^{3}$ Department of Physics, University of Central Florida, Orlando, Florida 32816, USA \\$^{4}$ Department of Electrical and Computer Engineering, University of Central Florida, Orlando, Florida 32816, USA \\$^{5}$ Laboratoire National des Champs Magnétiques Intenses, CNRS-UGA-UPS-INSA-EMFL, 38042 Grenoble, France \\$^{6}$ Institute for Functional Intelligent Materials, National University of Singapore, Singapore 117544, Singapore \\$^{7}$ Institute for Photonics and Nanotechnologies, National Research Council (CNR-IFN), 00133 Rome, Italy \\$^{8}$ Department of Inorganic Chemistry, University of Chemistry and Technology Prague, Technická 5, 166 28 Prague 6, Czech Republic \\$^{9}$ Research Center for Electronic and Optical Materials, National Institute for Materials Science, 1-1 Namiki, Tsukuba 305-0044, Japan \\$^{10}$ Research Center for Materials Nanoarchitectonics, National Institute for Materials Science, 1-1 Namiki, Tsukuba 305-0044, Japan \\$^{11}$ CENTERA, CEZAMAT, Warsaw University of Technology, Warsaw, Poland}
	\end{center}

This Supplemental Material provides: \ref{Methods} - Details of the sample preparation, experimental setups, and first principles calculations. \ref{S1} -  Compositional characterization of the alloys. \ref{S2} - Temperature and power evolutions of the PL spectra measured on the Mo$_{x}$W$_{1-x}$Se$_2$ MLs. \ref{S3} - degree of circular polarization.  \ref{S4} - Magnetic-field evolutions of the PL spectra of the MoSe$_2$ and WSe$_2$ MLs. \ref{S5} - Magnetic-field evolutions of the PL spectra measured on the Mo$_{x}$W$_{1-x}$Se$_2$ MLs. \ref{strain} - Local strain and compositional variations in ML of M\lowercase{o}WS\lowercase{e}$_2$ encapsulated in hBN.
\ref{theory} - Details on the band folding/unfolding and mixing from first principles calculations.

\renewcommand{\thesection}{Methods}
\section{\label{Methods}}

\noindent
\textbf{Fabrication of hBN encapsulated samples}\\

MLs of MoSe$_2$, WSe$_2$, and Mo$_{x}$W$_{1-x}$Se$_2$ were mechanically exfoliated from bulk crystals. The MLs were exfoliated with a scotch tape onto polydimethylsiloxane (PDMS). 
Relatively thick hBN flakes were exfoliated mechanically onto SiO$_2$/Si substrates (90 or 300 nm thick SiO$_2$ on top of a 500 $\mu$m thick Si). 
The MoSe$_2$, WSe$_2$, and Mo$_{x}$W$_{1-x}$Se$_2$ MLs were then deterministically transferred from the PDMS to the hBN flake. 
Thin hBN flakes were exfoliated onto PDMS and deposited on the MLs to encapsulate them. 
After each deposition step, the sample was annealed in high vacuum (10$^{-6}$ mbar) at about 120$^\circ$C for several hours to induce coalescence of the contaminants, thus improving the adhesion of the sample.\\

\noindent
\textbf{Crystal growth}\\

Four of the alloyed crystals (those characterized by a Mo percentage of 23 \%, 46 \%, 68 \% and 85 \%) were purchased from HQ Graphene. The alloy crystal Mo$_{0.49}$W$_{0.51}$Se$_2$ was instead grown by us by the chemical vapor transport method (CVT) in quartz ampoule from elements.

The stochiometric amount of molybdenum (99.999 \%, -100 mesh, Shanghai Quken New Material Technology Co., China), tungsten (99.999 \%, -100 mesh China Rhenium Co., China) and selenium (99.9999 \%, 2-4mm, Wuhan Xinrong New Materials Co., China) were placed in stochiometric ratio corresponding to 50 g in quartz ampoule ($50 \times 250$ mm) together with SeBr$_4$ (99 \%, Strem, USA) and 2 at.\% excess of selenium and melt sealed under high vacuum ($<1 \times 10^{-3}$ Pa using a oil diffusion pump with a liquid nitrogen cold trap) using oxygen-hydrogen welding torch. The ampoules were placed in a muffle furnace and heated at 450$^\circ$ C for 25 hours, at 500$^\circ$ C for 50 hours, at 600$^\circ$ C for 50 hours, and finally at 800$^\circ$ C for 50 hours. The heating and cooling rate were 1$^\circ$ C/min. Between each heating step the ampoule was mechanically homogenized for 5 minutes by shaking. For the CVT growth, the ampoule was placed in two zone horizontal tube furnace. First, the growth zone was heated at 1000$^\circ$ C and the source zone at 800$^\circ$ C. After 2 days the thermal gradient was reversed and the source zone was heated at 900$^\circ$ C and the growth zone at 800$^\circ$ C. Finally, the ampoule was cooled to room temperature. The growth zone was kept at 400$^\circ$ C for two additional hours to remove any excess of selenium and transport medium. The ampoule was opened in a argon filled glovebox.\\

\noindent
\textbf{Crystal elemental analysis}\\

The elemental composition of the bulk Mo$_{x}$W$_{1-x}$Se$_2$ crystals (from which the MLs were exfoliated) was examined by energy dispersive X-ray (EDX) analysis using a ZEISS-Sigma300 scanning electron microscope equipped with an Oxford Instruments X-Act 100 mm energy-dispersive spectrometer. The data were acquired with an acceleration voltage of 28 kV and analyzed with the INCA software. A copper sample has been used as reference for the elemental quantum calibration. The spatial resolution was in the 10-15 $\mu$m range.\\

\noindent
\textbf{Optical and magneto-optical experiments}\\

Photoluminescence (PL) experiments at zero magnetic field were performed using $\lambda= $515 nm (2.41 eV) radiation from diode lasers. 
The excitation laser beam was focused through a 50x long-working distance objective with a 0.55 numerical aperture, producing a spot of about 1 $\mu$m diameter.  
The signal was collected via the same microscope objective. 
The signal was sent through a 0.75 m monochromator and then detected by using a liquid nitrogen-cooled charge-coupled device (CCD) camera. 

Low-temperature micro-magneto-PL experiments were carried out in Faraday geometry, $i.e.$, with the magnetic field being oriented perpendicular to the ML plane.
Measurements (spatial resolution $\sim$1 $\mu$m) were performed with the aid of superconducting and resistive magnetic coils producing fields correspondingly up to 16~T and 30~T using a free-beam-optics arrangement. 
The sample was placed on top of a $x$-$y$-$z$ piezo-stage kept at $T$=10~K (superconducting magnet) or $T$=4.2~K (resistive magnet) and excited using a CW laser diode with a wavelength of 515 nm (2.41 eV photon energy). 
The emitted light was dispersed with a 0.5 m focal length monochromator and detected with a CCD camera. 
In the superconducting coil, the combination of a quarter-wave plate and a Wollaston prism was used to analyze and detect the circularly polarized light of opposite helicity ($\sigma^\pm$) at the same time.
In the resistive magnet, the combination of a quarter-wave plate and a linear polarizer was used to analyze the circular polarization of the signal (the measurements were performed in a fixed circular polarization configuration, whereas reversing the direction of magnetic
field yielded information corresponding to the other polarization component due to time-reversal symmetry).

Room-temperature Raman scattering measurements were performed using a micro-Raman LabRAM HR system equipped with a 100$\times$ objective lens and a 532~nm excitation laser. The scattered light was collected in a backscattering configuration and dispersed by a spectrometer with an 1800~grooves/mm diffraction grating blazed at 500~nm. The lateral resolution was estimated to be below 1~$\mu$m.

The excitation power focused on the sample was kept below 50 $\mu$W during all measurements to avoid local heating.\\

\noindent
\textbf{Theoretical calculations}\\

DFT calculations were conducted in the Vienna Ab initio Simulation Package (VASP)\cite{VASP} with the Projector Augmented Wave method \cite{PAW}. 
Perdew-Burke-Ernzerhof parametrization \cite{PBE} of the generalized gradients approximation to the exchange-correlation functional was used. 
The plane waves basis cut-off energy was set at 450 eV and a 12$\times$12$\times$1 (6$\times$6$\times$1) $\Gamma$-centered Monkhorst-Pack k-grid sampling was used for primitive (super-) cells. 
Atomic positions were optimized with $10^{-3}$~eV/\AA~criterion for the interatomic forces with Grimme's D3 correction\cite{D3}. 
The periodic images of the monolayers were separated by more than 20 \AA~of vacuum. 


The momentum operator matrix elements were calculated within Density Functional Perturbation Theory, as implemented in VASP\cite{Gajdos}. They were employed to find the optically active direct transition with $\sigma+$ polarization from the topmost valence band $v$ to a specific conduction band $c$ at the K$^+$ point. Their g-factors $g_v$ and $g_c$ were calculated from the bands-summation method developed in Ref.~\citenum{Wozniak2020}. 324/1296 bands were sufficient to converge $g_v$ and $g_c$ in primitive-/2$\times$2 supercells. The spin angular momenta at the K$^+$ point are $S=1$.
We evaluated the $g$-factor of X exciton as $g=2(g_c - g_v)$.

\clearpage
\newpage
\renewcommand{\thesection}{S\arabic{section}}
\setcounter{section}{0}
\setcounter{figure}{0}
\section{Elemental characterization of the crystals\label{S1}}

The Mo$_{x}$W$_{1-x}$Se$_2$ crystals from HQ Graphene used in this work were characterized by a nominal composition $x = 0.25, 0.50,$, $0.75$ and $0.85$. The self-grown sample was characterized by a nominal composition of 0.5. To verify the real composition of the samples and their uniformity over large (hundreds of $\mu$m) scales, we performed scanning electron microscopy (SEM) with energy dispersive X-ray analysis (EDX) (see Methods for details). 

First, to have precise information on the sample composition, we took highly resolved EDX spectra on 4-5 different points of mm-size crystals with a spatial resolution of about 10-15 $\mu$m. The EDX signals were integrated 240 s for each point. The cumulative spectra are shown in Figure \ref{fig:EDX_spectra}. From an averaged quantitative elemental analysis, we estimated the Mo concentration $x$ values displayed on the right (the spectra are ordered from the bottom to the top according to the estimated Mo content). The self-grown crystal is characterized by a Mo concentration $x = (49.40 \pm 0.81) \%$, in agreement within the uncertainty with the nominal value of 50\%. The other crystals are characterized by Mo concentrations of $(22.5 \pm 1.9) \%$, $(45.8 \pm 1.7) \%$, $(67.79 \pm 0.94) \%$ and $(85.41 \pm 0.63) \%$ Interestingly, the self-grown crystal (with $x \approx 0.49$) is characterized by a quite small standard deviation, indicating that the sample is  homogeneous, whilst the  commercial crystals are affected by a bit larger standard deviation (particularly for low Mo concentrations), accounting for some compositional inhomogeneity.

\begin{figure}[h!]

	\centering
	\includegraphics[width=0.7\linewidth]{S1_1.jpg}%
	\caption{
		 EDX high resolution spectra acquired on the five different Mo$_{x}$W$_{1-x}$Se$_2$ alloys. The spectra shown here are the normalized cumulative spectra obtained after measuring each crystal on multiple (4 or 6) points. From a quantitative analysis of the Mo, W and Se line intensity, the relative Mo concentration composition [Mo] (or $x$) is estimated, and displayed on the right. The spectra were ordered for increasing Mo concentration from the bottom to the top.}
	\label{fig:EDX_spectra}
\end{figure}

To attest the homogeneity over the whole size of the investigated mm-size crystals, we performed EDX elemental maps using the same experimental condition and integration time for all the samples.
Figure \ref{fig:EDX_map} shows the SEM images (first column on the left) of the investigated region for each sample. The other columns show the spatial distribution for the single elements (Se, W, and Mo), while the "Mix" map (last column on the right) combines the signal associated to the three identified elements (Se, W and Mo) in a RGB color figure (red: Se; green: W; blue: Mo). Indeed, a quite uniform intensity is found over the whole crystals, but for a Se segregation observed on the samples with the highest and lowest Mo concentration. Some shadows are present in the maps due to the presence of terrace and flake overlay on the crystal surface; as well as a signal intensity gradient from the lower-left angle to the upper-right one of the map due to an angle between the sample surface and the detector in the experimental set up.

\begin{figure}[h!b]

	\centering
	\includegraphics[width=0.9\linewidth]{S1_2.png}%
	\caption{ SEM-EDX maps of big pieces of Mo$_{x}$W$_{1-x}$Se$_2$ crystal with Mo concentration equals to $85\%$ (a), $68\%$ (b), $49\%$ (c), $46\%$ (d), and $23\%$ (e). Scale bar equals to 0.25 mm in each panel. The left column panels show the SEM image of the investigated region for each crystal, whilst the other column panels show the elemental map for each identified element (Se, W, and Mo) and their sum as a RGB image (red: Se; green: W; blue: Mo).}
	\label{fig:EDX_map}
\end{figure}

SEM/EDX measurements were also employed to characterize samples with a lower Mo concentration. In that case, however, the crystal quality was very low, the samples being highly inhomogeneous and presenting Se conglomerates.

\clearpage
\newpage
\setcounter{figure}{0}
\section{Temperature and power evolutions of the PL spectra\\ of M\lowercase{o}WS\lowercase{e}$_2$ monolayers\label{S2}}

This section is devoted to the analysis of the PL spectra of Mo$_{x}$W$_{1-x}$Se$_2$ MLs with four different Mo/W concentrations (Mo$_{0.68}$W$_{0.32}$Se$_2$, Mo$_{0.49}$W$_{0.51}$Se$_2$, Mo$_{0.85}$W$_{0.15}$Se$_2$ and Mo$_{0.23}$W$_{0.77}$Se$_2$) measured as a function of temperature and excitation power.

\begin{figure}[!h]

	\centering
	\includegraphics[width=0.9\linewidth]{S2_1.pdf}%
	\caption{(a) Temperature evolution of the PL spectra of a Mo$_{0.85}$W$_{0.15}$Se$_2$ ML encapsulated in hBN flakes measured with 2.41 eV laser light excitation and power of 10~$\mu$W. (b) Room temperature PL spectrum of the Mo$_{0.85}$W$_{0.15}$Se$_2$ ML. The colored Lorentzian curves (green and purple) display fits to the corresponding X and T lines. The blue area corresponds to the cumulative fit. (c-d) Temperature (c) and power dependence (d) of the integrated intensities of the X and T emission lines. The power evolution was measured at $T$=5~K. The black dashed line is a reference for a linear behavior.}
	\label{fig:temp_pow_085}
\end{figure}

\begin{figure}[!h]

	\centering
	\includegraphics[width=0.9\linewidth]{S2_2.pdf}
	\caption{ (a) Temperature evolution of the PL spectra of a Mo$_{0.68}$W$_{0.32}$Se$_2$ ML encapsulated in hBN flakes measured with 2.41 eV laser light excitation and power of 50~$\mu$W. (b) Room temperature PL spectrum of the Mo$_{0.68}$W$_{0.32}$Se$_2$ ML. The colored Lorentzian curves (green and purple) display fits to the corresponding X and T lines. The blue area corresponds to the cumulative fit. (c-d) Temperature (c) and power dependence (d) of the integrated intensities of the X and T emission lines. The power evolution was measured at $T$=5~K. The black dashed line is a reference for a linear behavior.}
	\label{fig:temp_pow_066}
\end{figure}

Fig.~\ref{fig:temp_pow_085}(a) and Fig.~\ref{fig:temp_pow_066}(a) show the temperature evolutions of the PL spectra measured on the Mo$_{0.85}$W$_{0.15}$Se$_2$ and Mo$_{0.68}$W$_{0.32}$Se$_2$ MLs encapsulated in hBN flakes, respectively.
As it can be seen in the figures, a temperature increase from 5 K up to room-temperature leads to a significant change in the PL spectra.
At low temperature ($T$=5~K), two emission lines ascribed to the neutral (X) and charged (T) excitons are well resolved, with the T intensity being greater than the X intensity.
When the temperature is increased, the trion emission vanishes much faster, and the neutral exciton emission dominates the spectra in the limit of high temperature. 
The room temperature PL spectrum of the Mo$_{0.85}$W$_{0.15}$Se$_2$ and Mo$_{0.68}$W$_{0.32}$Se$_2$ MLs are shown in Fig.~\ref{fig:temp_pow_085}(b) and Fig.~\ref{fig:temp_pow_066}(b), respectively.
The spectra can be well deconvoluted using two Lorentzian curves, which confirms the presence of the T emission at room temperature for both samples.
The temperature evolutions of the integrated intensities of the X and T emission lines are presented in Fig.~\ref{fig:temp_pow_085}(c) and Fig.~\ref{fig:temp_pow_066}(c), respectively.
It is seen that the T intensity decreases by three orders of magnitude with a temperature increase from 5~K to 300~K. 
On the other hand, the X intensity decreases by about 2 orders of magnitude when the temperature is increased from 5~K to 300~K.
Similar temperature dependences of the X and T intensities have been reported for MoSe$_2$ MLs~\cite{Jadczak2017, aroramose2}, which may be the indication of the bright nature of the ground excitonic state in the Mo$_{0.85}$W$_{0.15}$Se$_2$ and Mo$_{0.68}$W$_{0.32}$Se$_2$ MLs.
Finally, we investigated the X and T intensity evolution at 5 K as a function of the excitation power, as shown in
Fig.~\ref{fig:temp_pow_085}(d) and Fig.~\ref{fig:temp_pow_066}(d).
An almost linear dependence on the excitation power is found for both X and T, which is expected in the case of excitonic complexes composed of a single electron-hole ($e$-$h$) pair (in contrast to biexcitons, whose intensities increase quadratically with excitation power).~\cite{Klingshirn2012}

\begin{figure}[!h]

	\centering
	\includegraphics[width=0.9\linewidth]{S2_3.pdf}%
	\caption{ (a) Temperature evolution of the PL spectra of a Mo$_{0.49}$W$_{0.51}$Se$_2$ ML encapsulated in hBN flakes measured with 2.41 eV laser light excitation and power of 5~$\mu$W. (b) Room temperature PL spectrum of the Mo$_{0.49}$W$_{0.51}$Se$_2$ ML. The colored Lorentzian curves (green and purple) display fits to the corresponding X and T lines. The blue area corresponds to the cumulative fit. (c-d) Temperature (c) and power dependence (d) of the integrated intensities of the X and T emission lines. The power evolution was measured at $T$=5~K. The black dashed line is a reference for a linear behavior.}
	\label{fig:temp_pow_055}
\end{figure}

Fig.~\ref{fig:temp_pow_055}(a) shows the temperature evolution of the PL spectra measured on the Mo$_{0.49}$W$_{0.51}$Se$_2$ ML encapsulated in hBN flakes.
As it can be seen in the figure, the observed change in the PL spectra is very similar to the temperature dependences of the PL spectra measured on the Mo$_{0.68}$W$_{0.32}$Se$_2$ presented in Fig.~\ref{fig:temp_pow_066}(a).
At low temperature ($T$=5~K), two emission lines ascribed to the neutral (X) and charged (T) excitons are well resolved with comparable X and T intensities.
When the temperature is increased, the trion emission vanishes much faster, and the neutral exciton emission dominates the spectra in the limit of high (room) temperature. 
The room temperature PL spectrum of the Mo$_{0.49}$W$_{0.51}$Se$_2$ ML is shown in Fig.~\ref{fig:temp_pow_055}(b).
The spectrum can be well deconvoluted using two Lorentzian curves, which confirm the T emission at room temperature.
The temperature evolutions of the integrated intensities of the X and T emission lines are presented in Fig.\ref{fig:temp_pow_055}(c).
It is seen that the T intensity decreases 1000 times with temperature increased from 5~K to 300~K. 
In addition, the intensity of X decreases significantly about 100 times with temperature increased from 5~K to 300~K.
Similar temperature dependences of the X and T intensities have been reported for MoSe$_2$ MLs~\cite{Jadczak2017, aroramose2}, which may be the indication of the bright nature of the ground excitonic state in the Mo$_{0.49}$W$_{0.51}$Se$_2$ ML.
Finally, to confirm the origin of the aforementioned emission lines, we investigated their intensity evolutions as a function of the excitation power,  which are shown in
Fig.\ref{fig:temp_pow_055}(d).
The integrated intensities of the X and T emission lines are characterized by almost linear dependences on the excitation power, which is expected in the case of excitonic complexes composed of a single electron-hole ($e$-$h$) pair (in contrast to biexcitons, whose intensities increase quadratically with excitation power).~\cite{Klingshirn2012}

\begin{figure}[!h]

	\centering
	\includegraphics[width=0.9\linewidth]{S2_4.pdf}%
	\caption{ (a) Temperature evolution of the PL spectra of a Mo$_{0.23}$W$_{0.77}$Se$_2$ ML encapsulated in hBN flakes measured with 2.41 eV laser light excitation and power of 50~$\mu$W. (b) The temperature and (c) power dependencies of the integrated intensities of the X emission line. The power evolution was measured at $T$=5~K. The black dashed line is a reference for a linear behavior.}		\label{fig:temp_pow_022}
\end{figure}

Fig.~\ref{fig:temp_pow_022}(a) shows the temperature evolution of the PL spectra measured on the Mo$_{0.23}$W$_{0.77}$Se$_2$ ML encapsulated in hBN flakes.
As can be seen in the figure, the PL spectra are formed by a single emission line observed in the whole range of temperatures, which is ascribed to the neutral exciton (X).
The temperature evolution of the integrated intensity of the X emission lines is presented in Fig.~\ref{fig:temp_pow_022}(b).
The intensity of X increases twice with the temperature increased from 5~K to 300~K.
A similar temperature dependence of the X intensity has been reported for WSe$_2$ MLs~\cite{Zhang2015,Wang2015}, which may be an indication of the dark nature of the ground excitonic state in the Mo$_{0.23}$W$_{0.77}$Se$_2$ ML.
Finally, to confirm the origin of the aforementioned emission line, we investigated its intensity evolution as a function of the excitation power,  which is shown in Fig.\ref{fig:temp_pow_022}(d).
The integrated intensity of the X emission line is characterized by an almost linear dependence with the excitation power, which is expected in the case of excitonic complexes composed of a single electron-hole ($e$-$h$) pair (in contrast to biexcitons, whose intensities increase quadratically with the excitation power)~\cite{Klingshirn2012}.\\
\clearpage

\setcounter{figure}{0}
\section{Degree of circular polarization of the X and T emission lines\label{S3}}

As an exemplary case, this section is devoted to the analysis of the magnetic-field–induced valley polarization of the neutral (X) and charged (T) exciton emission lines measured on the hBN-
	encapsulated Mo$_{0.49}$W$_{0.51}$Se$_2$ monolayer, which magnetic field evolution is presented in Fig. 2 of the main text.

\begin{figure}[h!t]

	\centering
	\includegraphics[width=0.27\linewidth]{S3.pdf}
	\caption{  Magnetic-field dependences of the circular polarization degrees ($P_{circ.}$) for the X and T transitions measured on the hBN-
		encapsulated Mo$_{0.49}$W$_{0.51}$Se$_2$ monolayer.}
	\label{fig:pol}
\end{figure}

We can then quantify the degree of circular polarization of the X and T emission lines, defined by the formula:

\begin{equation}
	P_{\mathrm{circ.}} =
	\frac{I_{\sigma^+} - I_{\sigma^-}}{I_{\sigma^+} + I_{\sigma^-}},
	\label{eq:pol}
\end{equation}

 where $I_{\sigma^+}$ and $I_{\sigma^-}$ are the intensities of the corresponding circularly polarized components.
	As shown in Fig.~\ref{fig:pol}, the $P_{circ.}$ values are nearly zero in the absence of the field, as expected, and increase with the field, reaching values of 0.68 and 0.96 for charged and neutral excitons at 30~T, respectively. 
	We also find similar magnitudes of the polarization degree also for other Mo$_{x}$W$_{1-x}$Se$_2$ MLs, see Sec.~\ref{S5} for details.
	To our knowledge, these are the highest reported values of the magnetic-field-induced valley polarization for the neutral exciton in any S-TMD MLs~\cite{Mitioglu2105, Arora_MoTe2, Plechinger2016, Nagler2018, Koperski2019, Prando2021}, where values up to 0.8 were reported so far for MoSe$_2$ \cite{Koperski2019} and MoTe$_2$ MLs \cite{Arora_MoTe2}.

\setcounter{figure}{0}
\section{Magnetic-field evolutions of the PL spectra of M\lowercase{o}S\lowercase{e}$_2$ and WS\lowercase{e}$_2$ monolayers\label{S4} }

\begin{figure}[h!t]

	\centering
	\includegraphics[width=1\linewidth]{S4.pdf}
	\caption{ (a) Helicity-resolved PL spectra of an hBN-encapsulated MoSe$_2$ ML at $T$=10~K measured at selected values of the applied out-of-plane magnetic field. The red (blue) curves corresponds to the $\sigma^+$ ($\sigma^-$) polarized spectra. The spectra were measured under excitation energy of 2.41~eV and power of 10~$\mu$W. The spectra are vertically shifted for clarity. (b) Magnetic field evolutions of the energy difference between the two circularly polarized split components ($i.e.,$ the Zeeman splitting) of the X and T emission lines measured on the MoSe$_2$ ML. The solid lines represent fits according to the equation described in the main text. The spectra were measured under excitation energy of 2.41~eV and power of 10~$\mu$W. The g-factors estimated through the fits are displayed. (c) Low-temperature PL spectrum of an hBN-encapsulated WSe$_2$ monolayer at $T$=10~K measured at zero field. The spectrum was measured under excitation energy of 2.41~eV and power of 10~$\mu$W. Besides the neutral exciton X, several trions and biexcitons can be identified. (d) Magnetic field evolutions of the energy difference between the two circularly polarized split components of the X, T$^\textrm{S}$ (S: singlet), and T$^\textrm{T}$ (T: triplet) emission lines measured on the WSe$_2$ ML. The solid lines represent fits according to the equation described in the main text. The g-factors estimated through the fits are displayed.}
	\label{fig:mag_MoSe2WSe2}
\end{figure}

Fig.~\ref{fig:mag_MoSe2WSe2}(a) shows the magnetic field evolution of the PL spectra measured on a MoSe$_2$ ML encapsulated in hBN flakes.
The assignment of the emission lines to the neutral (X) and charged (T) excitons is based on the literature~\cite{Cadiz2017, Robert2020}.
As it can be seen in the figure, both the observed emission lines split into two $\sigma^\pm$ components due to the Zeeman effect.
The magnetic field dependencies of the energy differences between the $\sigma^\pm$ components ($i.e.$, the Zeeman splitting) of the X and T lines are presented in Fig.~\ref{fig:mag_MoSe2WSe2}(b).
The Zeeman splitting between $\sigma^\pm$ components ($\Delta E(B)=E_{\sigma^+}(B)-E_{\sigma^-}(B)$) is given by the formula: 
\begin{equation}
	\Delta E(B)=g \mu_\textrm{B} B, 
	\label{eq:Zeeman}
\end{equation}
where $g$ denotes the Land\'e $g$-factor of the considered transition and $\mu_B$ is the Bohr magneton. 
Using Equation~\ref{eq:Zeeman}, we determined the $g$-factors for both the X and T lines, which are red$-4.0 \pm 0.2$ and $-3.9 \pm 0.2$, respectively.
The found values are in very good agreement with those reported in the literature, with mean values equal to $-3.7 \pm 0.3$~\cite{Koperski2019} and $-4.2 \pm 0.3$~\cite{Robert2020,Kipczak_2023,Goryca2019}.

The magnetic field evolution of the PL spectra measured on the WSe$_2$ ML encapsulated in hBN flakes is shown in Fig.~\ref{fig:mag_MoSe2WSe2}(c).
The spectrum displays several emission lines with a characteristic pattern similar to that previously reported in several works on WSe$_2$ MLs encapsulated in hBN~\cite{Courtade2017,Chen2018, Li2019,Wang2017, Koperski2017, Koperski2019, Arora2020, Zinkiewicz2022, Barbone2018, Paur2019,He2020, Robert2021, Robert2021PRL, Arora2015W, Smolenski2016, Robert2017,Liu2020,Liu2019valley,LiuValley,Li2018, Lireplica2019, Li2019momentum, Molas2019}. 
According to these reports, the assignment of the observed emission lines is as follows:
X -- neutral exciton; XX -- neutral biexciton; 
T$^\textrm{S}$ and T$^\textrm{T}$ -- singlet (intravalley) and triplet (intervalley) negatively charged excitons, respectively; 
X$^\textrm{G}$ -- grey exciton;
XX$^-$ -- negatively charged biexciton; 
T$^\textrm{D}$ -- negatively charged dark exciton (dark trion);
X$^\textrm{G}_{\textrm{E}"(\Gamma)}$ -- phonon replica of the grey exciton;
T$^\textrm{D}_{\textrm{E}"(\Gamma)}$ -- phonon replica of the dark trion.
In our analysis, we focus only on three complexes: X, T$^\textrm{T}$, and T$^\textrm{S}$, as the investigated PL spectra of the Mo$_{x}$W$_{1-x}$Se$_2$ MLs are composed of two emission lines due to the neutral and charged excitons.
To determine the magnitude of the Zeeman splitting, we present the magnetic field dependences of the energy differences between the $\sigma^\pm$ components of the X, T$^\textrm{T}$, and T$^\textrm{S}$ lines in Fig.~\ref{fig:mag_MoSe2WSe2}(d).
Using Equation~\ref{eq:Zeeman}, we determined the $g$-factors for the X, T$^\textrm{T}$, and T$^\textrm{S}$ lines, which are $ -4.0 \pm 0.2 $, $-4.0 \pm 0.2$ and $-4.3 \pm 0.2$, respectively.
Such values are in very good agreement with those reported in the literature, whose corresponding average values are $-4.3 \pm 0.3$, $-4.5 \pm 0.2$ and $-4.8 \pm 0.2$ respectively ~\cite{LiuGate2019,Li2019,Ye2018,Förste2020}.


\setcounter{figure}{0}
\section{Magnetic-field evolutions of the PL spectra\\ of M\lowercase{o}$_{x}$W$_{1-x}$S\lowercase{e}$_2$ monolayers\label{S5} }

This section is devoted to the analysis of the PL spectra of MoWSe$_2$ MLs with four different Mo/W concentrations (Mo$_{0.85}$W$_{0.15}$Se$_2$, Mo$_{0.68}$W$_{0.32}$Se$_2$, Mo$_{0.49}$W$_{0.51}$Se$_2$, and Mo$_{0.23}$W$_{0.77}$Se$_2$) measured as a function of the out-of-plane magnetic field.

The magnetic field evolutions of the PL spectra measured on Mo$_{0.85}$W$_{0.15}$Se$_2$ ML are plotted in Fig.~\ref{fig:mag_085}, two samples of Mo$_{0.68}$W$_{0.32}$Se$_2$ are shown in Figs.~\ref{fig:mag_066}(a) and (d), while those of a second sample of Mo$_{0.49}$W$_{0.51}$Se$_2$ (the first sample's evolution is presented in the main text) and of a Mo$_{0.46}$W$_{0.54}$Se$_2$ MLs are presented in Figs.~\ref{fig:mag_046_055} (a) and (d), respectively. 
By analogy to the spectra of the so-called $"parents"$ MoSe$_2$ and WSe$_2$, we attributed the emission lines observed in these five samples to the neutral (X) and charged (T) excitons, correspondingly. As it can be seen in the figures, both the observed emission lines in the samples split into two $\sigma^\pm$ components due to the Zeeman effect. 
Similarly to the spectra of Mo$_{0.49}$W$_{0.51}$Se$_2$ shown in the main text in Fig.\ 2, the lowest energy transitions of the Zeeman split lines ($\sigma^+$) for both the neutral and charged excitons gain in intensity, while the highest energy transitions ($\sigma^-$) lose intensity with increasing magnetic field.

To determine the $g$-factor, we present the magnetic-field dependencies of the energy differences between the $\sigma^\pm$ components of the X and T lines in Fig.~\ref{fig:mag_066}(b) and (e) for two samples with relative concentrations of Mo/W equal to 68/32 and in Fig.~\ref{fig:mag_046_055} (b) and (e) for the samples with Mo/W of 49/51 and 46/54 as well as in Fig.~\ref{fig:mag_085} for the sample with a ratio of 85/15.

The Zeeman splitting between the $\sigma^\pm$ components, $\Delta E(B)$, is given by the expression introduced in the previous section (Sec.~S4). Using Eq.~\ref{eq:Zeeman}, we extracted the $g$-factors for the X and T transitions for all samples with various concentrations.
For Mo$_{0.85}$W$_{0.15}$Se$_2$, the extracted $g$-factors are identical 
	for both X and T transitions, with values of $-4.5 \pm 0.2$.
For the first sample of Mo$_{0.68}$W$_{0.32}$Se$_2$, the obtained values are $g_X = -6.0 \pm 0.2$ and $g_T = -3.6 \pm 0.2$, respectively.
For the second sample of Mo$_{0.68}$W$_{0.32}$Se$_2$, the $g$-factors are $-5.7 \pm 0.2$ for the X line and $-4.7 \pm 0.2$ for the T line. 
For the samples with a relative concentration of Mo/W atoms 49/51 and 46/54, the $g$-factors equal: $-7.1 \pm 0.2$ and $-7.3 \pm 0.2$ for X and $-4.5 \pm 0.2$ and $-4.7 \pm 0.2$ for T.
Although the $g$-factor for the trion is similar to those found for the MoSe$_2$ and WSe$_2$ MLs (about $-4.0$ and $-4.3$), the magnitude of the corresponding $g$-factor for the X line of all the alloys is larger than for the $"parent"$ MLs (around $-3.7$ and $-4.0$). To our knowledge, such high values have not been reported for 2D monolayers in the literature so far.

\begin{figure}[htbp]

	\centering
	\includegraphics[width=0.5\linewidth]{S5_1.pdf}
	\caption{(a) Helicity-resolved PL spectra of two hBN-encapsulated Mo$_{0.85}$W$_{0.15}$Se$_2$ MLs measured at $T$= 10~K, at selected values of the applied out-of-plane magnetic field. The red (blue) curves correspond to $\sigma^+$ ($\sigma^-$) polarized spectra. The spectra were measured under excitation energy of 2.41~eV and power of 2.5~$\mu$W. The $\sigma^+$-polarized spectra were normalized to the T intensity, while the $\sigma^-$-polarized spectra were multiplied by scaling factors to make them better visible. The spectra at different magnetic fields are vertically shifted for clarity. (b) Magnetic-field evolutions of the energy differences ($\Delta E$) between the two circularly polarized split components of the X and T emission lines measured on the Mo$_{0.85}$W$_{0.15}$Se$_2$ MLs. The solid lines represent fits according to Eq.~\ref{eq:Zeeman}. (c) Magnetic-field dependences of the circular polarization degrees ($P_{circ.}$) of the X and T transitions.} 
	\label{fig:mag_085}
\end{figure}

\begin{figure}[htbp]

	\centering
	\includegraphics[width=1\linewidth]{S5_2.pdf}
	\caption{ (a) and (d) Helicity-resolved PL spectra of two hBN-encapsulated Mo$_{0.68}$W$_{0.32}$Se$_2$ MLs (sample 1 and sample 2, respectively) measured at $T$=4.2~K and 10~K, respectively, at selected values of the applied out-of-plane magnetic field. The red (blue) curves correspond to $\sigma^+$ ($\sigma^-$) polarized spectra. The spectra were measured under excitation energy of 2.41~eV and power of 0.1~$\mu$W (sample 1) and 100~$\mu$W (sample 2). The $\sigma^+$-polarized spectra were normalized to the X intensity, while the $\sigma^-$-polarized spectra were multiplied by scaling factors to make them better visible. The spectra at different magnetic fields are vertically shifted for clarity. (b) and (e) Magnetic-field evolutions of the energy differences ($\Delta E$) between the two circularly polarized split components of the X and T emission lines measured on the Mo$_{0.68}$W$_{0.32}$Se$_2$ MLs. The solid lines represent fits according to Eq.~\ref{eq:Zeeman}. (c) and (f) Magnetic-field dependences of the circular polarization degrees ($P_{circ.}$) of the X and T transitions.}
	\label{fig:mag_066}
\end{figure}

\begin{figure}[htbp]

	\centering
	\includegraphics[width=1\linewidth]{S5_3.pdf}
	\caption{ (a) and (d) Helicity-resolved PL spectra of hBN-encapsulated Mo$_{0.49}$W$_{0.51}$Se$_2$ and Mo$_{0.46}$W$_{0.54}$Se$_2$ MLs, respectively, measured at $T$=4.2~K at selected values of the applied out-of-plane magnetic field. The red (blue) curves correspond to the $\sigma^+$ ($\sigma^-$) polarized spectra. The spectra were measured under excitation energy of 2.41~eV and power of 5~$\mu$W (Mo$_{0.49}$W$_{0.51}$Se$_2$) and 0.25~$\mu$W (Mo$_{0.46}$W$_{0.54}$Se$_2$). The $\sigma^+$-polarized spectra were normalized to the X intensity, while the $\sigma^-$-polarized spectra were multiplied by scaling factors to make them better visible. The spectra at different magnetic fields are vertically shifted for clarity. (b) and (e) Magnetic-field evolutions of the energy differences ($\Delta E$) between the two circularly polarized split components of the X and T emission lines measured on the Mo$_{0.49}$W$_{0.51}$Se$_2$ and Mo$_{0.46}$W$_{0.54}$Se$_2$ MLs. The solid lines represent fits according to Eq.~\ref{eq:Zeeman}. (c) and (f) Magnetic-field dependencies of the circular polarization degrees ($P_{circ.}$) of the X and T transitions. X$_{1}$ and T$_{1}$ in panel (c) correspond to the polarization degrees extracted from the helicity-resolved PL spectra of hBN-encapsulated Mo${0.49}$W$_{0.51}$Se$_2$ shown in Fig.~2 of the main text.}
	\label{fig:mag_046_055}
\end{figure}

Fig.~\ref{fig:mag_085}(c), Figs.~\ref{fig:mag_066}(c,f), as well as Figs.~\ref{fig:mag_046_055}(c,f) show the degree of circular polarization $P_{\mathrm{circ.}}$ defined in section S3. The $P_{\mathrm{circ.}}$ values are initially close to zero in the absence of a magnetic field and systematically increase with increasing field strength. For Mo$_{0.85}$W$_{0.15}$Se$_2$, the polarization degrees reach values of 0.85 for the X transition and 0.50 for the T transition. For the first sample of Mo$_{0.68}$W$_{0.32}$Se$_2$, $P_{\mathrm{circ.}}$ the polarization degrees increases up to 0.99 for the X transition and 0.68 for the T transition at 30~T. For the second sample with the same composition, the polarization degrees at 16~T are 0.71 for the X transition and 0.10 for the T transition. Finally, for the Mo/W ratios of 0.46/0.54 and 0.49/0.51 Fig.~\ref{fig:mag_046_055}(c,f), $P_{\mathrm{circ.}}$ at 30~T is nearly identical for both samples, reaching 0.96 for the X transition. For the T transition, the degree of circular polarization is 0.68 for the first sample and slightly lower, 0.62, for the second sample.

\begin{figure}[htbp]
	
	\centering
	\includegraphics[width=1\linewidth]{S5_4.pdf}
	\caption{ (a) and (d) Helicity-resolved PL spectra of two hBN-encapsulated Mo$_{0.23}$W$_{0.77}$Se$_2$ MLs (sample 1 and sample 2, respectively) measured at $T$=10~K, at selected values of the applied out-of-plane magnetic field. The red (blue) curves correspond to the $\sigma^+$ ($\sigma^-$) polarized spectra. The spectra were measured under excitation energy of 2.41~eV and power of 25~$\mu$W. The $\sigma^+$-polarized spectra were normalized to the X intensity, while the $\sigma^-$-polarized spectra were multiplied by scaling factors to make them better visible. The spectra at different magnetic fields are vertically shifted for clarity. (b) and (e) Magnetic field evolution of the energy difference ($\Delta E$) between the two circularly polarized split components of the X emission line measured on the Mo$_{0.23}$W$_{0.77}$Se$_2$ MLs. The solid line represents fit according to Eq.~\ref{eq:Zeeman}. (c) and (f) Magnetic-field dependence of the circular polarization degree ($P_{circ.}$) of the X transition.} 
	\label{fig:mag_022}
\end{figure}

\begin{figure}[htbp]

	\centering
	\includegraphics[width=0.5\linewidth]{S5_5.pdf}
	\caption{ (a) Helicity-resolved PL spectra of hBN-encapsulated Mo$_{0.23}$W$_{0.77}$Se$_2$ MLs (sample 3) measured at $T$=10~K, at selected values of the applied out-of-plane magnetic field. The red (blue) curves correspond to the $\sigma^+$ ($\sigma^-$) polarized spectra. The spectra were measured under excitation energy of 2.41~eV and power of 25~$\mu$W. The $\sigma^+$-polarized spectra were normalized to the X intensity, while the $\sigma^-$-polarized spectra were multiplied by scaling factors to make them better visible. The spectra at different magnetic fields are vertically shifted for clarity. (b) Magnetic field evolution of the energy difference ($\Delta E$) between the two circularly polarized split components of the X emission line measured on the Mo$_{0.23}$W$_{0.77}$Se$_2$ ML. The solid line represents fit according to Eq.~\ref{eq:Zeeman}. (c) Magnetic-field dependence of the circular polarization degree ($P_{circ.}$) of the X transition.}
	\label{fig:mag_022_3}
\end{figure}

The PL spectra of three samples of Mo$_{0.23}$W$_{0.77}$Se$_2$ MLs encapsulated in hBN (and deposited on different substrates) are presented in Fig.~\ref{fig:mag_022} panels (a,c)  and Fig.~\ref{fig:mag_022_3} (a). These spectra are different from those of the other samples with higher concentration of Mo atoms. Indeed, only the highest energy resonance attributed to neutral exciton (X) can be observed. The $g$-factors for this excitonic complex are equal to $-10.2 \pm 0.2$ , $-10.1 \pm 0.2$ and  $-10.7 \pm 0.2$  for the first, second and third sample respectively (see Fig.~\ref{fig:mag_022} and Fig.~\ref {fig:mag_022_3}(b)). It is worth mentioning the very similar $g$-factor values found on different samples indicating the robustness of the exceptionally high $g$-factor for this Mo concentration. This aspect is further addressed in Sec. S6.
The corresponding degrees of circular polarization are shown in Figs.~\ref {fig:mag_022} (c) and (f) and Fig.~\ref {fig:mag_022_3} (c). For the Mo$_{0.23}$W$_{0.77}$Se$_2$ MLs the polarization degree of the X transition at 12~T reaches 0.34, 0.05,  and 0.30 for samples 1, 2, and 3, respectively.
\newpage

\setcounter{figure}{0}
\section{Local strain and compositional variations in ML of M\lowercase{o}WS\lowercase{e}$_2$ encapsulated in hBN}
\label{strain} 

To verify the possible role of local strain and alloy disorder on the exciton $g$-factor, we performed a spatially resolved Raman scattering mapping of the Mo$_{0.23}$W$_{0.77}$Se$_2$ ML encapsulated in hBN. 
	This study was performed on the Mo$_{0.23}$W$_{0.77}$Se$_2$ MLs, since the largest linewidths of the exciton lines and the largest dispersion of their energies were observed for this alloy concentration (see Fig. 1 in the main article).

\begin{figure}[htpb]
	\centering    
	\includegraphics[width=0.6\linewidth]{S6_1.pdf}
	\caption{ (a) Optical image of the hBN-encapsulated Mo$_{0.23}$W$_{0.77}$Se$_2$ ML. The ML area is marked by white curve. 
			The room temperature ($T$=300 K) spatial maps of the (b) intensity and (c) frequency of the A$_{1g}$ peak.
			The red circle denotes the size of the  excitation/detection spot, whose diameter is about  1~$\mu$m.
			(d) Histogram of A$_{1g}$ frequencies extracted from the spatial maps shown in panel (c).}
	\label{fig:raman22_78}
\end{figure}

Fig.~\ref{fig:raman22_78}(a) displays the optical image of the Mo$_{0.23}$W$_{0.77}$Se$_2$ ML encapsulated in hBN. This ML was investigated using spatially resolved Raman scattering, whose map is presented in Fig.~\ref{fig:raman22_78}(b) and (c).  
	Note that the diameter of the excitation/detection spot is about 1~$\mu$m, which is similar to the spatial step in the mapping (1~$\mu$m).
	The intensities and frequencies of phonon modes are very sensitive to local strain in S-TMD MLs~\cite{BlundoAPR}.
	Indeed, the ML homogeneity is studied through the spatial distributions of the intensity and frequency of the $A_{1g}$ mode, shown correspondingly in Figs.~\ref{fig:raman22_78}(b) and (c).
	The evolution of the $A_{1g}$ intensity [Fig.~\ref{fig:raman22_78}(b)] varies between 1000 and 35000 counts, reflecting the spatial inhomogeneity of the Raman signal due to local deformation, $e.g.$, bubbles and/or edges~\cite{BlundoAPR}.
	The largest change in the $A_{1g}$ frequency [Fig.~\ref{fig:raman22_78}(c)] of about 1.5~cm$^{-1}$ is observed between the ML edges and center. Instead, the frequency of the $A_{1g}$ mode exhibits only minor variations within the center of the ML corresponding to the frequency range from 246.8 cm$^{-1}$ to 246.9 cm$^{-1}$. 
	To point out the reproducibility of the extracted $A_{1g}$ frequency, we plot the histogram of the $A_{1g}$ frequencies in Fig.~\ref{fig:raman22_78}(d).
	We found that most of the $A_{1g}$ frequencies (about 50 points) are located within the range 247.0$\pm$0.1~cm$^{-1}$, proving negligible strain and disorder effects in the central part of the ML, investigated in magneto-PL experiments.

\begin{figure}[htpb]
	\centering    
	\includegraphics[width=1\linewidth]{S6_2.pdf}
	\caption{ Helicity-resolved PL spectra of an hBN-encapsulated Mo$_{0.23}$W$_{0.77}$Se$_2$ ML at $T$=4.2~K measured at selected values of the applied out-of-plane magnetic field for three different samples: (a) sample 1, (b) sample 2 and (c) sample 3. The red (blue) color corresponds to the $\sigma^+$ ($\sigma^-$) polarized spectra. The measurements were performed  under excitation energy of 2.41~eV and power of 25~$\mu$W. The spectra are vertically shifted for clarity.  (d) Magnetic-field evolutions of the energy differences ($\Delta E$) between the two circularly polarized split components of the X transitions for all three samples. The solid lines represent fits according to the equation $\Delta E(B)=g \mu_\textrm{B} B$.}
	\label{PL_samples}
\end{figure}

\begin{figure}[htpb]
	\centering    
	\includegraphics[width=1\linewidth]{S6_3.pdf}
	\caption{ Helicity-resolved PL spectra of an hBN-encapsulated Mo$_{0.23}$W$_{0.77}$Se$_2$ ML at $T$=4.2~K measured at selected values of the applied out-of-plane magnetic field for three different positions on sample 1: (a) position 1, (b) position 2 and (c) position 3. The red (blue) color corresponds to the $\sigma^+$ ($\sigma^-$) polarized spectra. The measurements were performed  under excitation energy of 2.41~eV and power of 25~$\mu$W. The spectra are vertically shifted for clarity.  (d) Magnetic-field evolutions of the energy differences ($\Delta E$) between the two circularly polarized split components of the X transitions for all three position on sample 1. The solid lines represent fits according to the equation $\Delta E(B)=g \mu_\textrm{B} B$.}
	\label{PL_positions}
\end{figure}

To verify the influence of local strain and alloy disorder on the measured $g$-factor of excitons, we analyzed the magneto-optical properties of three independent hBN-encapsulated Mo$_{0.23}$W$_{0.77}$Se$_2$ MLs with the same nominal Mo/W ratio, and also three different spatial positions within the same Mo$_{0.23}$W$_{0.77}$Se$_2$ ML; compare Figs.~\ref{PL_samples} and \ref{PL_positions}.
	It is seen in Fig.~\ref{PL_samples}(a)-(c) that the energies of the neutral exciton (X) lines vary by up to 19 meV across three different samples, from 1.654 eV for sample 1, through 1.673 eV for sample 2, and to 1.657 eV for sample 3.
	Such a fluctuation of the X energy suggests its significant sensitivity to local environment, $e.g.$, small deviations of the Mo/W ratio.
	Despite these relatively large energy shifts, the extracted $g$-factors of the neutral exciton remain almost the same value of about -10, $i.e.$, $-10.2$, $-10.1$, and $-10.7$ for the three different samples [Fig.~\ref{PL_samples}(d)].
	An analogous change in X energy can also be found within the same Mo$_{0.23}$W$_{0.77}$Se$_2$ ML in various spatial locations, see Fig.~\ref{PL_positions}(a)-(c).
	The extracted X energies change by up to 7 meV at three different positions in the same ML, from 1.661 eV at position 1, through 1.656 eV at position 2, to 1.654 eV at position 3.
	In this case, the $g$-factors obtained for the neutral exciton also stay at almost the same value of about -10, $i.e.$, $-10.2$, $-10.1$, and $-10.2$ for the three different positions [Fig.~\ref{PL_positions}(d)].
	The insensitivity of the $g$-factor of the neutral exciton for three different MLs as well as three different positions in the same MLs, for which the X energies are characterized by a dispersion of several meV, makes it highly unlikely that all of them would coincidentally exhibit the same extrinsic strain conditions.
	Taken together, these results indicate that the observed enhancement of the exciton $g$-factor is an intrinsic genuine property of the alloy, arising from the alloy-induced $K$–$Q$ conduction band-mixing mechanism, and this conclusion can be extended to all Mo/W concentrations investigated in this study.


\setcounter{figure}{0}
\section{Band folding and mixing\label{theory} }

Figure \ref{fig:FBZs} depicts the first Brillouin zones of the primitive (1$\times$1) and 2$\times$2 supercells, evidencing the folding of the k-points. See details in the caption.

Figure \ref{unfolded} shows the electronic band structures of alloys with compositions with $x=0.25, 0.5, 0.75$ unfolded from $2\times2$ supercell Brillouin zones to primitive cell Brillouin zones. 

Figure \ref{fig:band-coupling} shows the orbital composition of the conduction bands for$a = 3.282$$\textrm{ \AA}$ and $a = 3.297$$\textrm{ \AA}$ MoSe$_2$ lattice constants, indicating the decoupling of the energy bands when Mo and W atoms are further apart. See details in the caption.

\begin{figure}[h!t]
	\centering
	\includegraphics[width=0.35\linewidth]{FigS7.1.pdf}
	\caption{ First Brillouin zones (FBZs) of the (1$\times$1) primitive cell of MoSe$_2$ and WSe$_2$ monolayers (dashed line hexagon) and the 2$\times$2 supercells of Mo$_{x}$W$_{1-x}$Se$_2$ alloys (solid line hexagons). The high-symmetry points with a ' refer to the 2$\times$2 supercells. The energy bands from the K valleys and Q valleys (green shades) in the primitive FBZ are folded to K valleys in the FBZ of the 2$\times$2 supercell. The blue arrows indicate 3 different directions that the Q valley can be approached, providing a geometrical explanation for the amount of conduction bands observed in the 2$\times$2 supercells.}
	\label{fig:FBZs}
\end{figure}

\begin{figure}[htpb]
	\centering    
	\includegraphics[width=1.0\linewidth]{FigS7.2.png}
	\caption{Unfolded band structures of alloys with $x=0.25, 0.5$ and 0.75. The size of dot is proportional to the unfolding weight, and the color represents the contribution of W and Mo orbitals.}
	\label{unfolded}
\end{figure}


\begin{figure}[h]
	\centering
	\includegraphics[width=\textwidth]{FigS7.3a.pdf}
	
	\includegraphics[width=\textwidth]{FigS7.3b.pdf}
	\caption{ Calculated orbital composition of conduction bands for $x$=0, 0.25, 0.5, 0.75 and 1 for Mo$_x$W$_{1-x}$Se$_2$ with  (a) $a = 3.282$$\textrm{ \AA}$ and (b) $a = 3.297$$\textrm{ \AA}$ MoSe$_2$ lattice constants. Red, cyan, yellow, and violet points depict the contribution of $d_0=d_{z^2}$,  $d_2=d_{x^2-y^2}+d_{xy}$,  $p_0=p_z$, and $p_1={p_x+p_y}$ orbitals of the Bloch states, respectively.}
	\label{fig:band-coupling}
\end{figure}

\clearpage

\end{document}